\documentstyle[12pt]{article}

\newcommand{\mbold}[1]{\mbox{\boldmath $ #1 $}}
\newcommand{\be}{\begin{equation}}
\newcommand{\ee}{\end{equation}}
\newcommand{\benn}{\begin{displaymath}}         % the same as 'equation' but
\newcommand{\eenn}{\end{displaymath}}           % without formula numbers
\newcommand{\ba}{\begin{eqnarray}}
\newcommand{\ea}{\end{eqnarray}}
\newcommand{\bann}{\begin{eqnarray*}}   % the same as 'eqnarray' but
\newcommand{\eann}{\end{eqnarray*}}              % without formula numbers

\newcounter{lr_tmp}

\newcommand{\lap}%
{\raisebox{-0.5ex}{$\stackrel{\scriptstyle <}{\scriptstyle 
\sim}$}}
\newcommand{\gap}%
{\raisebox{-0.5ex}{$\stackrel{\scriptstyle >}{\scriptstyle 
\sim}$}}
\def\opm{%
{\cal\rlap{$\scriptstyle\bigcirc$}%
\raisebox{0.05em}{$\scriptscriptstyle\kern0.1em\pm$}}}
\begin{document}
\hbadness=10000

\newpage
\title{BOSON INTERFEROMETRY \\ in \\ HIGH ENERGY
PHYSICS\thanks{To appear in Physics Reports, 1999}}
\author{{R. M. Weiner\thanks{E.Mail:
weiner@mailer.uni-marburg.de}
\date{Physics Department, University of Marburg,\\ Wieselacker
8, 35041 
Marburg, Germany \\and \\
Laboratoire de Physique Th\'eorique et Hautes 
\'Energies,
Univ. Paris-Sud,\\ 177 rue de Lourmel, 75015 Paris, France}}
}

\maketitle
\begin{abstract} 

Intensity interferometry and in particular that due to 
Bose Einstein correlations
(BEC)  constitutes at present the only direct
 experimental method for the determination of sizes and
lifetimes of sources in particle and nuclear physics. The
measurement of these is essential for an understanding of the
dynamics of strong interactions which are responsible for 
the existence and properties of atomic nuclei. Moreover a 
new state of matter, quark matter, 
in which the ultimate constituents of matter move freely,
is within the reach of
present accelerators or those under construction. The
confirmation of the existence of this new
state is intimately linked with the determination of its
space-time properties. Furthermore BEC provides information
about quantum coherence which lies at the basis of the
phenomenon of Bose-Einstein condensation seen in many
chapters of physics. Coherence and the associated classical
fields are essential ingredients in modern theories of
particle physics including the standard model. 
Last but not least 
besides this ``applicative" aspect
of BEC, this effect has implications for the 
foundations of quantum mechanics including the understanding
of the concept of ``identical particles".
Recent theoretical developments in BEC are reviewed and 
their application in
high energy particle and heavy-ion reactions is analyzed.  
The treated topics include: a) a comparison between the wave-function
approach and the space-time approach based on classical
currents, which predicts ``surprising" particle
anti-particle BEC b) the study of final state interactions
c) the use of hydrodynamics d) the relation between
correlations and multiplicity distributions.
\noindent
\end{abstract}

\newpage

\tableofcontents

\newpage

\section{Introduction} 
The method of photon intensity interferometry was 
invented in
the mid fifties by Hanbury-Brown and Twiss for the
measurement
 of stellar dimensions and is sometimes called the
HBT method. In 1959-1960 G.Goldhaber, 
S.Goldhaber,W.Lee and A.Pais discovered 
that identical charged pions produced in $\bar p-p$ 
annihilation are correlated (the GGLP effect). Both the HBT 
and the GGLP effects are based on Bose-Einstein correlations
(BEC). Subsequently also 
Fermi-Dirac correlations for nucleons have been observed. 
Loosely speaking
both these correlation effects can be viewed as a consequence
of the symmetry (antisymmetry) properties of the wave 
function with respect to permutation of two identical 
particles with integer (half-integer) spin and are thus 
intrinsic quantum phenomena.  At a higher level, these 
symmetry properties of identical particles are expressed by 
the commutation relations
of the creation and annihilation operators of
particles in the second quantization (quantum field theory).
The quantum field approach is the more general approach as
 it contains the possibility to deal with creation and 
annihilation of particles and certain phenomena like 
 the correlation between particles and antiparticles can
be properly described only within this formalism.
Furthermore, at high energies, because of the large number 
of particles produced, not all particles can be detected in 
a given reaction and therefore one measures usually only 
inclusive cross sections. For these reactions the wave 
function formalism is impractical.
Related to this is the fact that the second quantization
provides through the density matrix a transparent link
between correlations and multiplicity distributions. This
last topic has been in the center of interest of
multiparticle
dynamics for the last 20 years (we refer among other things
to Koba-Nielsen-Olesen (KNO) scaling and ``intermittency").
Furthermore one of the most important properties of
systems made of identical bosons and which is responsible
for the phenomenon of {\em lasing} is quantum statistical
coherence. This feature is also not accessible to a
theoretical
treatment except in field theory.

The present review is restricted to Bose Einstein
correlations which constitute by far the majority of 
correlations papers both of theoretical and experimental 
nature. This is due to the fact that BEC present important 
heuristic and methodological advantages as compared to
Fermi-Dirac correlations. Among the first we mention the
fact that quantum coherence appears only in BEC. Among the
second, one should recall that pions are the most 
abundantly produced secondaries in high
energy reactions.

In the last years there has been a considerable 
surge of interest in boson interferometry. This
can be judged by the fact that at 
present there is no
meeting on multiparticle production where numerous 
contributions to this subject are not presented. Moreover 
since 1990 \cite{camp} meetings dedicated exclusively to 
this topic are organized; this reflects the realization that
BEC are an important subject in its own. 
This development is
  due in part to the fact that 
at present intensity interferometry constitutes
the only
direct experimental method for the determination of sizes and
lifetimes of sources in particle and nuclear physics. 
Since soft strong interactions which are responsible for
multiparticle production processes cannot be treated by
perturbative QCD, phenomenological approaches have to be
used in this domain and space-time concepts are 
essential elements in these approaches. That is why  
intensity interferometry has become
an indispensable tool in the investigation of  
the dynamics of high energy reactions.  

However this alone could not yet explain the 
explosion of interest in BEC if it were not the search for
quark-gluon plasma which has mobilized the attention of 
most of the nuclear physics and of an appreciable
part of the particle physics community. For several reasons
the space-time properties of ``fireballs" produced in heavy
ion reactions are an essential key in the process of
undertanding whether quark matter has been formed.
 
The main emphasis of the present review will be on 
theoretical developments which took place
after 1989-1990. Experimental results will be mentioned 
only in so far as they
illustrate the theoretical concepts. 
For a review of older references cf. e.g.
the paper by Boal, Gelbke, and Jennings \cite{boal}. 
In the nineties  
the single most important 
theoretical event was in our view the space-time 
generalization of the
classical current formalism. For a detailed presentation of
this generalization and its applications up to 1993 
the reader is referred to ref.\cite{APW}. For a review of
experimental results in $e^+-e^-$ reactions cf. \cite{Hay}.
Finally a more pedagogical and more complete treatment of
the theory of BEC can be found in \cite{book}.

There are two categories of papers not mentioned:
(i) Those which the reviewer was unaware off; 
he apologizes to the authors of these papers for this. 
(ii) Those which he considered as irrelevant or as
 repetitions of previous work\footnote{As a matter of fact, 
``rediscoveries" are not rare in
this field and it would be desirable that journals and
referees should do a better job in trying to avoid such a
situation, not to mention the prevention of publication of 
misleading
statements and/or wrong results.}. The
large number of papers quoted despite these restrictions 
shows that an exhaustive listing of references on BEC is not
trivial.

\setcounter{equation}{0}

\section{The GGLP effect}

In the period 1954-1960 Hanbury-Brown and Twiss developped
the method of optical intensity interferometry for the 
determination of (angular sizes) of stars (cf. e.g.\cite{HBT}).
The particle physics equivalent of the Hanbury-Brown Twiss
(HBT) effect in optics is the Goldhaber, Goldhaber, Lee and 
Pais (GGLP) effect \cite{GGLP1}, \cite{GGLP} which we shall 
describe 
schematically below.   

However when going over from optics to particle physics 
the following point has to be considered:
In particle physics one does not measure distances $r$
in order to deduce 
(differences of) momenta $k$ and thus angular sizes, 
but one measures rather momenta in order to deduce 
 distances. 
This explains why GGLP were not aware of the HBT method
\footnote{For a comparison of optical and particle physics
intensity interferometry cf. e.g.\cite{book}.}.

Up to a certain point there are two ways of approaching 
the relationship between
optics and particle physics: the wave function approach
and the field theoretical approach. 
Although the first one is only a particular case of the
second
one it is still useful because it allows sometimes a more
intuitive
understanding of certain concepts and in particular that of 
the distinction between boson and fermion correlations. 
 We shall start therefore with the wave
function approach.

\subsection{The wave function approach}

Let us consider for the beginning 
 a source which consists of a number of
discrete emission points $i$ each of which is characterized 
by a probability amplitude  
\be
F_{i}({\bf r})=F_{i}\delta({\bf r}-{\bf r}_{i}).
\label{eq:emampl}
\ee

Let $\psi_{{\bf k}}({\bf r})$ be the wave function of an 
emitted particle. 
The total probability $P({\bf k})$ to observe the emission of
one particle with momentum ${\bf k}$ from this source is 
obtained by summing the contributions of all points $i$.
If this summation is done {\em incoherently}
 the single particle probability reads

\be
P({\bf k})=\sum_i|F_i\psi(r_i)|^2
\label{eq:incoh}
\ee
Instead of discrete emission points consider now a source
the emission points of which are continuously distributed 
in space and assume for simplicity that the wave functions 
are plane waves 
$\psi _{{\bf k}}({\bf r})\sim exp(i{\bf kr})$. 
The sum over $x_i$ will
now be replaced by an integral and we have 
\be
P({\bf k})=\int|F({\bf r})|^2d^3r.
\label{eq:Richard7}
\ee
Similarly the probability to observe two particles 
 with
momenta ${\bf k}_1,{\bf k}_2$ is
\begin{equation}
P({\bf k}_1{\bf k}_2)=\int|\psi_{1,2}|^2f({\bf
r}_1)f({\bf r}_2)d^3r_1d^3r_2
\label{eq:prob}
\end{equation} 
Here 
$\psi_{1,2}= \psi_{1,2}({\bf k}_1,{\bf k}_2;{\bf r}_1,
{\bf r}_2)$ is the two-particle wave function and we have
introduced the density distribution $f= |F|^2$.

Suppose now that the two particles are identical. 
Then the two-particle wave function has to be symmetrized or
antisymmetrized depending on whether we deal with 
identical bosons or fermions. Assuming plane waves we have  
\begin{equation} 
\psi_{1,2}=\frac{1}{\sqrt{2}}\left[e^{i({\bf k}_1{\bf r}_1
+{\bf k}_2{\bf r}_2)}\pm e^{i({\bf k}_1{\bf r}_2
+{\bf k}_2{\bf r}_1)}\right]
\label{eq:wf}
\end{equation}
with the plus sign for bosons and the minus sign for
fermions. 
With this wave function one obtains
\be
P({\bf k}_1,{\bf k}_2)=
|\tilde
f_I(0)|^2\pm|\tilde f_I({\bf k}_1-{\bf k}_2)|^2
\label{eq:P2inc}
\ee

The incoherent summation corresponds to random fluctuations
of the amplitudes $F_i$.
 
Following the GGLP experiment \cite{GGLP} 
consider 
the space points ${\bf r}_{1}$ and ${\bf r}_{2}$ within a 
source, so that each point emits two identical particles 
(equally charged pions in the case of GGLP)
with momenta ${\bf k}_{1}$ and ${\bf k}_{2}$. These 
particles are detected in the registration points $ x_1$ and 
$x_2$
so that in $x_1$ only particles of momentum $k_1$ and in
$x_2$ only particles of momentum $k_2$ are registered
(cf.Fig.1). Because of the identity of particles one cannot
decide which particle pair originates in $r_1$ and which
originates in $r_2$.

Assuming that the individual emission points of the source
act incoherently GGLP derived eq.(\ref{eq:P2inc}) which 
for bosons leads 
to the second order correlation 
function 
\be
C_{2}({\bf k}_1,{\bf k}_2)\equiv P_2({\bf k}_1,{\bf
k}_2)/P_1({\bf k}_1)P_1({\bf k}_2)
=1+|\tilde{f}({\bf q})|^2
\label{eq:Fourier}
\ee
where $\tilde{f}$ is the Fourier transform of $f$ and
${\bf q}$ the momentum difference ${\bf k}_{1}-{\bf k}_{2}$.

Eq.(\ref{eq:Fourier}) shows quite clearly how in particle 
physics momentum (correlation)
measurements  can yield information about
the space-time structure of the source 

In the case of coherent summation one gets instead
$C_2 = 1$ (cf. e.g.\cite{boal}). We retain herefrom (cf.
also below) that
coherence reduces the correlation and that a purely coherent
source has a correlation function which does not depend on
its geometry\footnote{Cf. also
\cite{MP2},\cite{Pratt2}
 for an attempt to approach the issue of coherence and chaos  
by using wave packets.}.    

\subsubsection{Newer correlation measurements in $\bar{p}-p $
annihilation} The original GGLP experiment \cite{GGLP1} 
measured the
correlation function in terms of the opening angle 
\footnote{It may be amusing to note that the use of the 
opening angle as a kinematical variable in BEC studies was
rediscovered in 1994 in ref. \cite{Chu} where it was
recommended as a tool for the investigation 
of final state interactions.} 
of a pion pair.

The GGLP experiment has been repeated in the last years 
 at LEAR (cf.e.g. \cite{Angel} where also older references 
 are quoted
\footnote{For a reanalysis of the results of an older 
annihilation experiment cf. \cite{Gaspero}.}). In these 
newer experiments the three momenta
of particles were measured, although one continued to use 
as a variable for the correlation function the invariant
momentum difference $Q$, as suggested already in \cite{GGLP}.

One of the remarkable observations made in all annihilation
reactions
is the fact
that the intercept of the second order correlation function
$C_2(k,k)$ appears consistently to exceed the canonical
value of $2$ reaching values up to $4$. (This effect was
possibly not seen in the original experiment \cite{GGLP1}
because of the averaging over the magnitudes of the
momenta.)

In \cite{Song} this effect was attributed to resonances
while subsequently \cite{Amado},\cite{Angel} a (non-chaotic)
Skyrmion type superposition of coherent states was proposed 
as an alternative explanation.

Another possible explanation of this intriguing observation 
may be that in annihilation processes squeezed states (cf.
section 2.2) are 
produced,
while this is not the case in other processes. Indeed,
it is known that squeezed states can lead to overbunching
effects (cf. below). Furthermore, as
shown in ref.(\cite{sqIgor}), squeezed states may be 
preferentially
 produced in rapid reactions. Or, according to 
 some authors \cite{Amado}, annihilation is a more 
rapid process than other reactions occuring at higher
energies.
Given the
importance of squeezed states, 
 further experimental and 
theoretical
studies of this issue are highly desirable
\footnote{In ref. \cite{sqIgor} the effect of chaotic 
superpositions
of squeezed states in BEC was also studied. It is shown that
such a superposition
 leads always to {\em overbunching}, while
pure squeezed states can lead also to antibunching.}. 

\subsubsection{Resonances, apparent coherence and other
experimental problems; the $\lambda$ factor}  
 
It is known that in multiparticle production processes,
an appreciable fraction of pions originate from resonance
decays. Resonances act in two opposite ways on the
correlation function of pion pairs. On one hand the
interference between pions originating from short lived
 resonances and
``directly" produced pions leads to a narrow peak in the
second order correlation function at small values of $q$ 
\cite{Grishin}, \cite{Grass}. On the other hand long 
lived resonances give
rise to pions which are beyond the range of detectors and
this leads to an apparent decrease of the intercept of the
second order correlation function $C_{2}(k,k)$. These 
modifications of the intercept are important among other 
things, because their understanding is
essential in the search for coherence through the intercept
criterion.
 As mentioned already
one of the most immediate consequences of coherence is the
decrease of the correlation function at small $q$.

Because of the large
number of different resonances produced in high energy
reactions   
a quantitative estimate 
of their effects is possible only via numerical techniques.
In sections 4.9.1 and 5.1.3 we will present a more
detailed discussion of the influence of resonances on BEC
within the Wigner function formalism. For older references 
 on this topic cf. also \cite{boal},
\cite{Ledn}, \cite{Thomas}.

Another experimental difficulty is that, because 
 of limited statistics or certain technical problems, 
sometimes not all
degrees of freedom can be measured in a given expriment.
This leads to an effective averaging over the non-measured
degrees of freedom and hence also to an effective reduction of
the correlation function. 
As a matter of fact 
   BEC experiments have shown from the very beginning
that
the extrapolation of the correlation function to $q=0$
almost never led to the maximum value $C_2(0)=2$ permitted
 by eq.(\ref{eq:Fourier}).

To take into account empirically this effect
experimentalists
introduced into the correlation function a correction 
factor $\lambda$.
Thus
eq.(\ref{eq:Fourier}) e.g. was modified into
\be
C_{2}({\bf k}_1,{\bf k}_2)=1+\lambda |\tilde{f}({\bf q})|^2    
\label{eq:lambda}
\ee

Because one initially thought that formally this 
generalization  offers also 
the possibility to describe
partially coherent sources, the corresponding parameter
$\lambda$ was postulated to be limited by (0,1) and 
called the ``incoherence" factor: indeed  
 $\lambda = 0$ leads to a totally coherent source and 
  $\lambda= 1$ to a totally chaotic one. 
Unfortunately this nomenclature
is not quite correct, as will be explained below. It should 
also be mentioned that there
exists a strong correlation between the empirical values of
$\lambda$ and and those of the ``radius" which enters
$\tilde{f}$ (cf. e.g. 
\cite{Peitz} for a special study of this issue.)

\subsubsection{The limitations of the wave function
formalism}

The wave function formalism presented in the previous
subsection has severe shortcomings: 

a) The correlation function (\ref{eq:Fourier})
depends only on the difference of momenta $q$ and not also
on the sum ${\bf k}_1+{\bf k}_2$, in contradiction with
experimental data. 
Although we will see in section 4.3 that this limitation 
disappears automatically in the quantum statistical 
(field theoretical) approach,
it can be remedied also within the first quantization 
formalism
 by using the Wigner function approach. In this approach 
the source function is from
the beginning a function both of coordinates and momenta and
therefore the correlation function depends on $k_1$ and
$k_2$ separately. There is of course a price to pay for this
procedure as it involves a semiclassical approximation 
\footnote{ 
 String models \cite{stringbowl}\cite{stringander} also use a
Wigner type formalism. Here it is postulated that
there exists a ``formation" time $\tau$ and therefore the 
particle production points are distributed around 
$t^2-x^2=\tau^2$.
This implies among other things a correlation between
particle production points and momenta.}.

b)   The wave function formalism may be
useful when exclusive reactions are considered as was the
case e.g. in the GGLP paper. Indeed $\psi_{1,2}$ in 
eq.(\ref{eq:wf}) is just the wave function of the two-boson
system i.e. the assumption is 
made that two and only two bosons are
produced. At low energies, i.e. low average multiplicities, 
this condition can be satisfied. However at
high energies the identification of all particles is very
difficult and up to now has not been done. Therefore one
measures practically always inclusive cross sections. This 
means that instead of (\ref{eq:wf}) one would have to use a 
wave function which describes the two bosons in the presence
of all other produced particles. To obtain such a wave 
function one would have to solve the Schr\"odinger equation 
of the many body, strongly interacting system, which is not 
a very practical proposal. Related to this is the difficulty
to treat  higher
order correlations within the wave function formalism.

c)  The correlation function $C_2$ in the wave function
formalism is independent of isospin
and is thus the same for charged and neutral particles.
We shall see in section 4.4 that
in a more correct quantum field theoretical approach, this
is not the case. This will affect among other things the
 bounds of the correlations and will lead to quantum
statistical particle-antiparticle correlations which are not
expected in the wave function formalism.

d) Coherence cannot be treated adequately (cf. below). 
 The correlation functions derived in
this subsection refer in general to incoherent sources and 
attempts to introduce coherence within the
wave function formalism are rather ad-hoc parametrizations. 
However coherence is the most characteristic and
important property of Bose-Einstein correlations among other
things because it is the basis of the phenomenon of Bose
Einstein
{\em condensates} found in many chapters of physics, like
superconductors, superfluids, lasers, and the  
recently discovered atomic condensates
\cite{Atomcond97}.   
It would be very surprising if coherence would not
be found
also in particle physics given the fact that the wave 
lengths of the
emitted particles are of the same order as that of the
sources. 
Furthermore as pointed out in this connection in \cite{FW}
modern particle physics is based on spontaneously broken
symmetries. The associated fields are coherent. That is why
one of the main motivations of BEC research should be the
measurement of the amount of coherence in strong
interactions. For this purpose the formalism of BEC has to 
be generalized to
include the presence of (partial) coherence and this again
can be done correctly only within quantum statistics i.e. 
quantum field theory.

We conclude this subsection with the observation that the
wave function or wave packet approach may nevertheless 
be useful in BEC for the investigation of final state 
interactions (cf. below) or for the construction of event 
generators, where phases
or quantum amplitudes are ignored anyway. Also for
correlations between fermions where coherence is
absent the wave function formalism may be an adequate
substitute, although
here too a field theoretical approach is possible.

\subsection{Quantum optical methods in BEC}
In high energy processes in which the pion multiplicity is
large, we may in general expect the methods of quantum
statistics (QS)
\footnote{By QS we understand in the following the density
matrix formalism within second quantization.}
to be useful.  
For BEC in particular they 
turn out to be indispensable. These methods have been applied
with great success particularly in quantum optics (QO),
superfluidity, supercoductivity etc. What distinguishes
optical phenomena from those in particle physics are
conservation laws and final state interactions which are
present in hadron physics. At high energies and high
multiplicities the first are unimportant. Neglecting for the
moment also the final state interactions, QS reduces then to
QO and we may take over the formalism of QO
to interpret the data on multipion production at high
energies, provided we consider identical pions. Given the
general validity of QS (or QO), it is then clear that any
model of multiparticle production must satisfy the laws of
quantum statistics and this has far reaching consequences,
independent of the particular dynamical mechanism which
governs the production process. The main tools in the QO
formalism are defined below \footnote{A review of the 
applications of quantum optical methods to multiparticle 
production up to 1988 can
be found in \cite{applic}.}.

\paragraph{Coherent states and squeezed coherent states} 
Coherent states $|\alpha>$ are
eigenstates of the (one-particle) annihilation operator $a$
\be     
a|\alpha>=\alpha|\alpha>.
\label{eq:coherent states}
\ee

Squeezed coherent states
are eigenstates $|\beta>_s$ of the two-particle annihilation
operator
\be
b=\mu a+\nu a^{\dagger}
\label{eq:2.66}
\ee  
with 
\be
|\mu|^2-|\nu|^2=1
\ee
so that
\be
b|\beta>_s=\beta|\beta>_s.
\label{eq:2.69}
\ee
One of the remarkable properties of these states which 
explains also their name is that for 
them the uncertainty in one variable can be
{\em squeezed} at the expense of the other so that
  
\be
(\Delta q)^2_{s}\le\frac{1}{2\omega},\;(\Delta
p)^2_{s}\ge\frac{\omega}{2}, 
\label{eq:Walls 2.54 sq}
\ee
or vice versa. The importance of this remarkable property
lies among other things in the possibility to reduce quantum
fluctuations and this explains the great expectations
associated with them 
in communication and measurement technology as well as 
their interest from
a heuristic point of view.

It has been found recently in \cite{sqIgor} that squeezed 
states appear naturally when one deals with rapid
phase transitions (explosions). Indeed consider the transition from a system $a$ to a system $b$
and assume that it proceeds rapidly enough so that the
relation
between the creation and annihilation operators and the
corresponding fields in the two ``phases" remains unchanged.
Mathematically this process will be
described by postulationg   
 at the moment of this transition
the following relations between the generalized coordinate $q$
and the generalized momentum $p$ of the field:
\ba
q &=& \frac{1}{\sqrt{2E_b}} (b^{\dagger} + b) = \frac{1}{\sqrt{2E_a}} (a^{\dagger} + a)
\nonumber\\
p &=& i \sqrt{\frac{E_b}{2}} (b^{\dagger} - b) = i \sqrt{\frac{E_a}{2}} (a^{\dagger} - a) 
\label{eq:qp}
\ea
$a^{\dagger},a$ are the free field creation and annihilation 
operators in the ``phase $a$" and $b^{\dagger},b$ the
corresponding operators in the ``phase $b$" . 
Eq. (\ref{eq:qp}) holds for each mode
$p$. Then we get immediately a connection between the $a$ and $b$ operators,
\ba
a &=& b \; \cosh \; r \; + \; b^{\dagger} \; \sinh \; 
r \; , \nonumber\\
a^{\dagger} &=& b \; \sinh \; r \; + \; b^{\dagger} \; \cosh \; r \; 
\label{eq:aa}
\ea
with
\be
r = r ({\bf p}) = \frac{1}{2} \, \log \, (E_a/E_b) \, .
\label{eq:rr}
\ee

The transformation (\ref{eq:aa}) is just the squeezing 
transformation (\ref{eq:2.66}) with 
\be
\mu = \cosh r; \qquad \nu = \sinh r
\label{eq:expl}
\ee
which proves the statement made above.

The observation of squeezed
states in BEC may thus serve as a signal for such rapid
transitions. Furthermore the existence of isospin induces in
hadronic BEC
certain effects which are specific for squeezed states. This
topic will be discussed in section 4.4. 

>From the point of 
view of BEC what distinguishes 
ordinary coherent states from squeezed states is the
following: for coherent states the intercept
$C_2(k,k) = 1$ while for squeezed states it can take
arbitrary values. In Fig.2 from ref.\cite{VW} one can see such an
example. 

\paragraph{Expansions in terms of coherent states}
Coherent states form an (over)complete set so that an
 arbitrary state $|f>$ can be expanded in a unique way  
in terms
of these states.

Of particular use is the expansion of the density matrix
$\rho$ in
terms of coherent states.  For a pure coherent state the 
density operator reads
\be
\rho=|\alpha><\alpha|.
\label{eq:7.1}
\ee

For an arbitrary density matrix case we have   
\be
\rho=\int{\cal P}(\alpha)|\alpha><\alpha|d^2\alpha.
\label{eq:7.6}
\ee
Here ${\cal P}$ is a weight function which usually, but not 
always, has the meaning of a probability.
The normalization condition for the density operator
translates
in terms of the ${\cal P}$ representation as
\be
tr\rho=\int{\cal P}(\alpha)d^2\alpha=1.
\label{eq:norm1}
\ee
Eq.(\ref{eq:7.6}) is
called also the Glauber-Sudarshan representation.

The knowledge of ${\cal P}(\alpha)$ is
(almost) equivalent to the knowledge of the density matrix.
However in most cases the exact form of ${\cal P}(\alpha)$
is not accessible and one has to content
oneself with certain approximations of it. Among these
approximations the
Gaussian form is priviledged because:

(i) one can prove that ${\cal P}(\alpha )$ is of Gaussian 
form for a certain physical situation which is frequently 
met in many body physics.

(ii) its use introduces an enormous mathematical 
simplification.

Proposition (i) is the subject of the central limit theorem
which states that if 

1. the number
of sources becomes large; 

2. they are stationary in the sense that their weight 
function ${\cal P}(\alpha)$ depends
only on the absolute value $|\alpha|$;

3. they act independently, 

then ${\cal P}(\alpha)$ is
Gaussian. These conditions are known to be fulfilled in most
cases of optics and presumably also in high energy physics.
Chaotic fields and in particular 
systems in thermal equilibrium are described by a Gaussian 
density matrix.
 
One of the reasons why the Gaussian form
for ${\cal P}$ plays such an important part in correlation 
studies is the fact that for a Gaussian ${\cal P}(\alpha)$ 
all higher order correlations can be expressed
in terms of the first two correlation functions.

On the other hand the coherent state representation is 
particularly important
for correlation studies because in this representation all
correlation functions can be expressed in terms of the
creation and annihilation operators $a$ and $a^{\dagger}$
of the fields
(particles). This follows from the Fourier
expansion of an arbitrary field in second quantization
\be
 \pi(x)=\sum_k 
[a_ke^{-ikx} +a^{\dagger}_ke^{ikx}].
\label{eq:2.12}
\ee

This property will be used extensively in section 4.2 within
the classical current formalism.

\paragraph{Correlation functions}

The first order correlation function  reads
\begin{equation}
G^{(1)}(x,x^{'})\equiv Tr[\rho\pi^{\dagger}(x)\pi(x^{'})]
\label{eq:corr1}
\end{equation}

Higher (n-th) order correlation functions are defined
analogously by
\begin{equation}
G^{(n)}(x_1...x_n,x_{n+1}...x_{2n})\equiv Tr[\rho\pi^
{\dagger}
(x_1)...\pi^{\dagger}(x_n)\pi(x_{n+1})...\pi(x_{2n})]
\label{eq:corrn}
\end{equation}

In quantum field theory because of the mathematical complexity
of the problem exact solutions of the field equations are
 available only in special cases. One such case will be
discussed later on. However for strong interactions
\footnote{Strong interactions are present not only in
hadronic physics but also in quantum optics.} even for this
case one has to use  
phenomenological parametrizations of  the
correlation functions and determine the parameters (which
have a definite physical meaning) by comparing with 
experiment. 
 
In
optics for
stationary chaotic fields two particular parametrizations are
used:\\

  1.Lorentzian spectrum:
\begin{equation}
G^{(1)}(x_1,x_2)=<n_{ch}>e^{-|x_1-x_2|/\xi}
\label{eq:Lorentz}
\end{equation}

  2.Gaussian spectrum:
\begin{equation}
G^{(1)}(x_1,x_2)=<n_{ch}>e^{-|x_1-x_2|^2/\xi^2}
\label{eq:Gaussdef}
\end{equation}

$\xi$ is the {\em coherence length} in x-space 
and $<n_{ch}>$ is the mean number of particles associated
with the
chaotic fields.

In \cite{FW} it was proposed to use the
analogy between time and rapidity in applying the methods of
quantum optics to particle
physics. Indeed in optics processes are usually stationary
in time while in particle physics the corresponding 
stationary variable (in the rapidity plateau region) is 
rapidity \footnote{This property does not hold
for other  variables.}.

Pure coherent or pure chaotic fields are just extreme cases.
In general one expects 
{\em partial coherence} i.e.
a superposition of coherent and chaotic fields
\be
\pi = \pi_{coherent} + \pi_{chaotic}.
\label{eq:supergen}
\ee

This leads for the Lorentzian case e.g.to a 
 second order correlation
function of the form
\begin{equation}
C_{2}(x,x')=1+2p(1-p)e^{-|x-x'|/\xi}+p^{2}e^{-2|x-x'|/\xi} 
\label{eq:super}
\end{equation}
where $p$ is the chaoticity, which varies between $0$ (for
purely coherent sources) and $1$ (for totally chaotic
sources). Eq. (\ref{eq:lambda}) is seen to be a particular
form of the above equation for $\lambda =p=1$ and it
is clear herefrom that $\lambda$ does not describe (partial) coherence
as its name would imply.
  The presence of coherence
introduces a new term into $C_2$. However it is remarkable 
that the number
of free parameters in eq.(\ref{eq:super}) is the same as in
eq.(\ref{eq:lambda}).
Formally it appears as if there would
act two sources rather than one, but the ``weights" and the
space-time characteristics of these
two sources are in a well defined relationship.  

This circumstance had been
 forgotten up to 1989 
\cite{revisit} both by experimentalists and theorists. The 
reason for this collective amnesia is the fact that during 
the eighties the wave
function formalism was dominating the BEC literature,
especially the experimental one. 

>From the foregoing discussion it should be clear that there
 are various reasons besides coherence why the bunching 
effect in BEC is
reduced. However it should also be clear that the
empirical description of this state of affairs through the
$\lambda$ factor is possible only for 
 totally chaotic sources. Since in an experiment
this is never known a priori, this implies that the fitting
of data with a formula of this type is misleading and should
be avoided, the more so that the correct formula
(\ref{eq:super}) does not contain more free parameters 
than eq.(\ref{eq:lambda}). 
In the example presented above
these free parameters are $p$ and $\xi$ for eq.
(\ref{eq:super}) and $\lambda$ and the effective 
radius (which enters in $\tilde{f}$) for eq.(\ref{eq:lambda})
respectively.
 
Using the rapidity-time analogy of ref.\cite{FW} 
for a partially chaotic field
the second order correlation function in rapidity is then
given by eq.(\ref{eq:super}),
with $x$ being replaced by rapidity $y$.

The fact that the last two terms in eq.(\ref{eq:super}) are
in a well defined relationship and depend in a
characteristic way on the two parameters $p$ and $\xi$   
 is a consequence of the superposition of the two {\em fields}
 (coherent
and chaotic) and distinguishes a partially coherent
source from a source which is a superposition of two independent {\em chaotic
 intensities}. Because of this the form (\ref{eq:super}) 
  was proposed in \cite{revisit} to be used
as a signal for detection of coherence in BEC
\footnote{Eq. (\ref{eq:super}) is a special case
of superposition of coherent and chaotic fields; it can be
considered as 
corresponding to point-like coherent and chaotic sources
 and a momentum
independent chaoticity; superpositions of more general
finite-size sources 
are considered in section 4.4 (eq.\ref{eq:c2plus1}) and
section 4.8.}. As a matter
of fact, an attempt in this direction was done in an 
experimental study 
by Kulka and L\"orstad \cite{KL}. In this analysis
BEC data from $pp$ and $\bar{p}p$ reactions at
$\sqrt{s}=53 GeV$ were used to compare various forms of 
correlation functions. Among other things one considered 
formulae of QO type for rapidity
  
\begin{eqnarray}
C_{2}=1+2p(1-p)e^{-|y_1-y_2|/\xi}+p^{2}e^
{-2|y_1-y_2|/\xi}
\label{eq:superlor}
\end{eqnarray}    
(corresponding to a Lorentzian spectrum)
and
\begin{eqnarray}
C_{2}=1+2p(1-p)e^{-|y_1-y_2|^{2}/\xi^{2}}+p^{2}e^
{-2|y_1-y_2|^{2}/\xi^2}
\label{eq:superpgauy}
\end{eqnarray}
(corresponding to a Gaussian spectrum)
as well as arbitrary superpositions of two chaotic sources
of exponential or Gaussian form respectively.

\begin{eqnarray}
C_{2}=1+\lambda_{1}e^{-|y_1-y_2|/\xi}+\lambda_{2}e^
{-2|y_1-y_2|/\xi}
\label{eq:superlory}
\end{eqnarray}

\begin{eqnarray}
C_{2}=1+\lambda_{1}e^{-|y_1-y_2|^2/\xi^2}+
\lambda_{2}e^{-2|y_1-y_2|^2/\xi^2}
\end{eqnarray}
Here $\lambda_{1}$ and $\lambda_{2}$ represent arbitrary weights
of the two chaotic sources.

Because of the limited
statistics no conclusion could be drawn as to the preference
of the QO form versus the two-source form.
Similar inconclusive results were obtained when one replaced
in the above
equations $y_1-y_2$ by the invariant momentum difference
$Q^2=(k_1-k_2)^2$.

\subsubsection{Higher order correlations}

We have mentioned above that a characteristic
 property of the Gaussian form of 
density matrix (not to be confused with the Gaussian form of
the correlator or the Gaussian form of the space-time 
distribution) is the
fact that all higher order correlation functions are 
determined just by the first two correlation functions. 
Since all BEC studies in particle
physics performed so far assume a Gaussian density matrix,
the reader may wonder why it is necessary to measure higher
order correlation functions.

There are at least three reasons for this:

(i) The conditions of the applicability of the above theorem
and in particular the postulate that the number of sources
is infinite and that they act independently can never be
fulfilled exactly. 

(ii) In the absence of
a theory which determines from first principles the
first two correlation functions,
 models for these quantities are used,
which are only approximations. The errors introduced by
these phenomenological parametrizations manifest themselves
differently in each order and thus violate the above theorem
even if (i) would not apply.

(iii) In experiments, because of limited statistics and
sometimes also because of theoretical biases not all
physical observables are determined, but rather 
averages over certain variables are performed, which again
introduce errors which propagate (and are amplified) from
lower to higher correlations.
  
Conversely, by comparing correlation functions of
different order one can test the applicability of the
theorem quoted above and pin down more precisely the 
 parameters which determine the first two correlation
functions (e.g. the chaoticity $p$ and the correlation
length $\xi$ in eqs. (\ref{eq:superlor}),
(\ref{eq:superpgauy})), which 
is essentially the 
purpose of particle interferometry.  
       
The phenomenological application of these considerations 
will be discussed in the following as well as in section 6.3
for the particular case 
of the invariant $Q$ variable, but the arguments
(i),(ii),and (iii) have general validity. It would be a
worthwhile research project to compare the deviations 
introduced in the relation between lower and higher 
correlation functions, due to (i) with those introduced by
(ii) and (iii).

The simplification brought by the variable $Q$ can be
enhanced by a further approximation proposed by Biyajima et
al. \cite{Biya90}. 
With the notation $Q_{ij}=k_i-k_j$ the analogue of eq.
(\ref{eq:superpgauy}) can be written
\be
C_2 = 1 + 2p(1-p)\exp(-R^{2}Q_{12}^{2}+p^{2}\exp(-2Q_{12}^
{2}R^{2})
\label{eq:C2Gauss}
\ee
 For the third order correlation function one obtains
\begin{eqnarray}
C_3 &=&
1+2p(1-p)[\exp\{-R^2Q^2_{12}\}+\exp\{-R^2Q^2_{13}\}
+\exp\{-R^2Q^2_{23}\}]\nonumber\\
& & +p^2[\exp\{-2R^2Q^2_{12}\}+\exp\{-2R^2Q^2_{13}\}+\exp
\{-2R^2Q^2_{23}\}]\nonumber\\
& & +2p^2(1-p)[\exp\{-R^2(Q^2_{12}+Q^2_{23})\}+\exp
\{-R^2(Q^2_{13}+Q^2_{23})\}\nonumber\\
& &+\exp\{-R^2(Q^2_{12}+Q^2_{13})\}]+2p^3[\exp\{-R^2(Q^2_{12}
+Q^2_{13}+Q^2_{23})\}]
\label{eq:Biya11}
\end{eqnarray}
 Ref.\cite{Biya90} proposed to use symmetrical
configurations for all two-particle momentum differences 
i.e. to consider $Q_{ij}$ independent of $(i,j)$. 
For $C_2$ this assumption does not of course introduce any
modifications. However for higher order the simplification
is important. 
 Thus e.g. for the
third order correlation with $Q_{12}=Q_{13}=Q_{23}$
and with the definition 
$Q^{2}_{three} = Q^{2}_{12}+Q^{2}_{13}+Q^{2}_{23}$ 
eq.(\ref{eq:Biya11}) becomes
\begin{eqnarray}
C_3 &=& 1+6p(1-p)\exp\left\{-\frac{1}{3}R^2
Q^2_{three}\right\}\nonumber\\
& & +3p^2(3-2p)\exp\left\{-\frac{2}{3}R^2Q^2_{three}\right\}
+2p^3\exp\{-R^2Q^2_{three}\}
\label{eq:Biya9}
\end{eqnarray} 

In ref.\cite{Biya90} similar expressions for $C_4$ and $C_5$, 
again for a Gaussian correlator, were given.  

These relations for higher order BEC were subjected to an
experimental test in ref.\cite{UA1}, using the UA1 data
for $\bar{p}p$ reactions at $\sqrt s = 630$ and $900 GeV$.  
For reasons which will become clear immediately,
we discuss here this topic in some detail.

The procedure used in \cite{UA1} for this test consisted in 
determining
$R$ and $p$ separately for each order $q$ of the correlation
and
comparing these values for different $q$. It was found that
a Gaussian correlator did not fit the data.  
Next in \cite{UA1} one tried
to replace the Gaussian correlator by an exponential (cf.
eq.(\ref{eq:superlor}).
To do
this one substituted simply in the expressions for the 
correlation functions of ref.\cite{Biya90} the factor
$\exp(-R^{2}Q^{2})$ with $\exp(-RQ)$. Such a procedure was at
hand given the fact that for $C_2$ the QO formulae
both for an exponential correlator and a Gaussian correlator
were known \cite{revisit} and their comparison suggested
just this substitution, as can be seen from 
eq.(\ref{eq:superpgauy}).    
 
In \cite{UA1} one used then for the exponential correlator
the relations

\be
C_{2}^{empirical} = 1 + 2p(1-p)\exp(-RQ_{12})+p^{2}\exp(-2Q_
{12}R)
\label{eq:C2exp}
\ee
and 

\begin{eqnarray}
C^{empirical}_{3} &=& 1+6p(1-p)\exp\left(-\frac{1}{3}
RQ_{three}\right)\nonumber\\
& &+3p^2(3-2p)\exp\left(-\frac{2}{3}RQ_{three}\right)+
2p^3\exp\left(-RQ_{three}\right)
\label{eq:C3wrong}
\end{eqnarray}

 With these modified formulae one still could not find 
in \cite{UA1} a unique set of values $p$ and $R$ 
for all orders of
correlation functions. However now a clearer picture of the
``disagreement" between the QO formalism and the data emerged.
It seemed that while the parameter $p$ was more or less
independent of $q$,  the radius $R$ increased with the 
order $q$ in a way which could be approximated by the
relation \footnote{This reminds one of Polonius's bewilderment in Shakespeare's 
Hamlet:
``Though this be madness, yet there is method in't".}

\be
R_q=R\sqrt {\frac{1}{2}q(q-1)}.
\label{eq:emp}
\ee

Indeed in \cite{mad} it was shown  that the findings of 
\cite{UA1} and in particular eq.(\ref{eq:emp}) not
only did not contradict QS but on the contrary constituted
a confirmation of it. 
While eq.(\ref{eq:C2exp}) for the second order correlation
function coincides with that derived in
quantum optics for an exponential spectrum, this
is not the case with the expressions for higher order 
correlations $C_q^{empirical}$ 
(eq.(\ref{eq:C3wrong})).   
 The formulae for $C_3$, $C_4$ and $C_5$
corresponding in QS to an exponential correlator and derived   
in \cite{mad} differ from the empirical ones used in
\cite{UA1}. 
As an example we quote 
\begin{eqnarray}
C_3 &=& 1+6p(1-p)\exp\left(-\frac{1}{\sqrt{3}}RQ_{three}
\right)\nonumber\\
& & +3p^2(3-2p)\exp\left(-\frac{2}{\sqrt{3}}RQ_{three}
\right)+2p^3\exp(-\sqrt{3}RQ_{three})
\label{eq:C3correct}
\end{eqnarray}
As observed in \cite{mad} and as one easily can check by
comparing eq.(\ref{eq:C3wrong}) with eq.(\ref{eq:C3correct}),
one
 can make coincide the 
empirical formulae for $C_q$ used in \cite{UA1} with the
correct ones, by replacing the parameter $R$ with a scaled
parameter $R_s$ and the relation between $R$ and $R_s$, is,
nothing else but $R_s=R_q$, where $R_q$ is given by
eq.(\ref{eq:emp}). The fact that this happens for three
different orders, i.e. for $C_3, C_4$, and $C_5$ makes a
coinicidence quite improbable.

 By empirically modifying
the formulae of higher order BEC for the exponential case, 
paper \cite{UA1} had explicitely violated QS and the
``phenomenological" relationship (\ref{eq:emp}) between $R$ 
and $q$ 
just compensated this violation. 

\begin{footnotesize} The fact that this
compensation and the final agreement between theory and
experiment was not perfect is not surprising and is
discussed in \cite{mad}. Besides the 
reasons (i), (ii), (iii) mentioned above, one has to take
into account that the QO formalism on which
eqs.(\ref{eq:C2exp}),(\ref{eq:Biya11})
 are based
assumes stationarity in $Q$, i.e. assumes that the
corrrelator depends only on the difference of momenta
$k_1-k_2$ and not also on their sum. As mentioned already 
 this condition is in general not fulfilled in BEC.
Furthermore, the parameter $p$, if it is related to
chaoticity, is in general momentum dependent (cf. section 
4.8). Also, the
symmetry assumption, $Q(i,j)$ independent of $i,j$, may be
too strong. Besides these theoretical caveats, there are
also experimental problems, related to the fact that the
$UA1$ experiment is not a dedicated BEC experiment and thus
suffers from specific diseases, which are common to almost
all particle physics BEC experiments performed so far. Among
other things, there is no identification of particles (only
$85\%$ of the tracks recorded are pions), and the 
normalization of
correlation functions is the ``conventional" one, i.e. not
based on the single inclusive cross sections as the
definition of correlation functions demands 
(cf.(\ref{eq:Fourier}) and
section 4.11), but rather uses an empirically determined
``background" ensemble.\end{footnotesize}

The very fact that QS was 
effectively ``reinvented" 
%\footnote {Imre Toth called this a ``boomerang" effect.}
by an unbiased experimental team
 is in our view overwhelmingly convincing. It may enter the
history of physics as a classical and quite unique example
of the intimate relationship between theory and experiment.  
Ref.\cite{mad} concludes that experiment \cite{UA1}
 supports QS in general and the standard 
form of the 
density
matrix in particular.
 From an epistemological
point of view, the message of this ``conspiracy" is highly
significant: {\em QS is robust enough to resist attempts
 of falsification}.

In the mean time further theoretical and experimental
developments took place. 

On the theoretical side a new space-time approach 
to BEC was
developped \cite{APW}, \cite{aw} which is more appropiate to particle 
physics and
which contains as a special case the QO formalism. In
particular the two exponential feature of the correlation
function is recovered. 

On the experimental side a new technique 
for the study of higher order correlations
was developed,
the method of correlation integrals which was applied \cite{UA1new} 
to a subset of the UA1 data in order to test the
above quoted QO fromalism. The fits were
restricted to second and third order cumulants only.
Again it was found that by extracting the parameters $p$ 
and $R$ from the second order data, the ``predicted" third
order correlation, this time by using a correct QO formula, 
differed significantly from the measured one. 

If confirmed, such a result could indicate that the QO
formalism provides only a rough description of the data and
that higher precision data demand also more realistic
theoretical tools. Such tools are the QS space-time approach
to BEC presented in section 4.3. A further, but more remote
possibility would be to look for deviations from the Gaussian
form of the density matrix.      
However it seems premature to speculate along these lines
given the fact that the procedure used to test the relation
between the second and third order correlation functions 
may have to be qualified. Indeed
in \cite{UA1new} one did not perform a {\em simultaneous} 
fit of
second and third order data to check the QO formalism.
Such a simultaneous fit appears necessary before drawing
conclusions, because
as
mentioned above (cf.(ii) and (iii)), the errors involved in
``guessing" the form of the correlator, and the fact that 
the variable $Q$ does not characterize completely the
two-particle correlation, limit the applicability of the
theorem which reduces higher order correlations to first and
second ones. As a matter of fact, it was found \cite{Nelly} 
(cf. also section 6.3)  in a comparison of the QS space-time
approach
with higher order correlation data, that the second order  
correlation data is quite insensitive to the values of the
parameters which enter the correlator, while once higher 
order data are used in a simultaneous fit, a
strong delimitation of the acceptable parameter values
results. Thus there are several possible solutions if one
restricts the fit to the second order correlation and the
correct one among those can be found 
only by fitting {\em simultaneously} all correlations. If by
accident one chooses in a lower correlation the wrong
parameter set, then the higher correlations cannot be fitted
anymore\footnote{The fact that in \cite{UA1new} the
correlations were normalized by mixing events rather than 
by comparing with the single inclusive cross sections in the
same event, as prescribed by the definition of the
correlation functions 
 may also influence the applicability of the
central limit theorem.}.

Before ending these phenomenological considerations in which
the variable $Q$ played a major part, a few remarks about
its use may be in order.

\paragraph{The invariant momentum difference $Q$} 

BEC studies in particle physics use often a priviledged 
variable namely  
the squared momentum difference
\be
Q^2=(k_1-k_2)^2 = ({\bf k}_1-{\bf k}_2)^2-(E_1-E_2)^2. 
\label{eq:Q}
\ee
It owes its special role to the fact that it is a 
relativistic invariant and was used already in the
pioneering paper by GGLP \cite{GGLP}. It has also the 
advantage that it involves all
four components of the momenta $k$ simultaneously 
so that the intercept of the correlation function 
$C_2({\bf k},{\bf k})$
 coincides with $C_2(Q=0)$. Thus by measuring
$C_2$ as a function of one single scalar quantity $Q$ one
gets automatically the intercept. This is not the case
with other single scalar quantities used in BEC like
$y_1-y_2$ or $k_{\bot,1}-k_{\bot,2}$ which characterize 
only partially the intercept.

On the other hand, $Q$ suffers from certain serious diseases
which make its use for practical interferometrical purposes
questionable.
 
The first and most important deficiency of $Q$ is 
the fact that it mixes time and space coordinates: the
associated quantity $R$ in the conventional parametrization
of the correlation function $C_2= 1+\lambda \exp(-R^2Q^2)$
is neither a radius nor a lifetime, but a combination of 
these, which cannot be easily disentangled. Another
deficiency of $Q$, which is common to all
  single scalar quantities is the circumstance that it does 
not fully
characterize the correlation function. Indeed the second
order correlation function $C_2$ is in general a function 
of six independent quantities 
 which cannot be
replaced by a single variable.

 An improvement on $Q$ was proposed by Cramer
\cite{Cramer91}
with the introduction of {\em coalescence} variables which
constitute a set of three boost invariant variables  
to replace, for a pair of identical particles, the 
single variable $Q$. 
They
  are related to $Q$ by 
$Q^2 = 2m^2(C^2_{L}+C^2_{T}+C^2_{R})$, where $C_L, C_T, C_R$
denote longitudinal, tranverse and radial coalescence 
respectively, and $m$ the mass of the particle. They have the
properties that $C_L=0$ when $y_1=y_2, C_T=0$ when
$\phi_1=\phi_2$ and $C_R=0$ when either $m_1=m_2$ or
$k_1=k_2$. Here $m_i$ is the transverse mass and $\phi$ the
azimuthal angle in the transverse plane. It is shown in
\cite{Cramer91} that with these new variables a Lorentz
invariant separation of the space-like and time-like
characteristics of the source is possible, within the
kinematical assumptions involved by the particular choice of
the coalescence variables. This separation  is however
rather involved. In \cite{Cramer91}
the coalescence variables are used for the introduction of
Coulomb corrections into second and higher order 
correlation functions.

Another way to compensate in part for the fact that one single
variable does not characterize completely the
two-particle system, but which maintains the use of $Q$  
 is, as was explained above, to consider higher order 
correlations.

\section{Final state 
interactions of hadronic bosons}

One of the most important differences between the HBT
effect in optics and the corresponding GGLP effect in
particle physics is the fact that in the first case we
deal with photons while in the second case with hadrons.

While
 photons in a first approximation do not interact,
 hadrons do interact.
 This interaction has two effects: (i) it influences the
correlation between identical hadrons and (ii) it leads to
correlations also between non-identical hadrons.

This review deals only with correlations due to the identity
of particles and in particular with Bose-Einstein
correlations. Therefore only effect (i) will be discussed
\footnote{The reader interested in correlations between
non-identical particles is refered to \cite{boal} for the
period up to 1990; for more recent literature cf.
\cite{Ard}, where correlations in low energy heavy ion 
reactions are reviewed.}.

 Effect (i) is usually described in terms of 
 final state interactions.
 In some 
theoretical studies (cf. e.g. \cite{AHR} and references
quoted there) 
 emission
of particles at different times is also treated as an
effective final state interaction.
>From the BEC point 
of view
the final states interaction constitutes in
general an unwanted background, which has to be substracted
in order to obtain the ``true" quantum statistical effect on
which interferometry measurements are based. That this   
is not always a trivial task will be shown in the following.  

 There are two types of
final state interactions in hadronic interferometry: 
 Electromagnetic, traded under the generic
name of Coulomb interactions, and strong. Furthermore one
distinguishes between one-body final state interactions 
and many body final state interactions. 
\subsection{Electromagnetic final state interactions}

The plane wave two-body function used in the 
considerations
above (cf. eq.(\ref{eq:wf})) applies of course only for 
 non-interacting particles. 

As a first step
towards a more general treatment consider charged particle
interferometry. As a matter of fact the vast majority of BEC
studies, both of experimental and theoretical nature, refer
to charged pions.  
For two-particle
correlations, we will have to consider the interaction of 
each member of
a pair with the charge of the source and the Coulomb 
interaction between the two particles constituting the pair.
The first effect will affect primarily the single particle
probabilities and is not expected to depend on the momentum
difference $q$. 
Attention has been paid so far mostly to the second
effect, i.e the modification of the two-particle wave
function due to the Coulomb interaction between the two
particles. While initially, having in mind the Gamow formula, 
it was assumed that this effect 
is (for small $q$ values) quite important, at present  
serious doubts about these estimates have arisen.
 The model dependence of corrections for this
effect makes it almost imperative that experimental data 
should be presented also without Coulomb corrections, so 
that it should be left to the reader the possibility of 
introducing (or not introducing) corrections according to 
her/his own prejudice.

\subsubsection{Coulomb correction and ``overcorrection"}
 As usual one separates the 
center of mass motion from the
relative motion. For the last one the scattering wave
function reads: 

\be
u({\bf r})\longrightarrow_{r \to \infty}e^{ikz}+r^{-1}g
(\theta,\phi)e^{ikr}
\label{eq:scatt}
\ee
where the relative position vector ${\bf r}$ has polar
coordinates $r, \theta,\phi$.

The form of the function $g$ depends on the scattering
potential.
In the Coulomb case the corresponding
 Schr\"odinger equation can be solved exactly and the
correction to the two-particle wave function and the
correlation function can be calculated.
So far three more and more sophisticated procedures were used 
for this purpose.
In \cite{GKW} the value of the square modulus of the 
wave function $u$ 
in the origin
$r=0$ was proposed as a correction term $G$ to the 
correlation function $C_2$. Up to non-interesting factors 
this is the Gamow factor which reads
\be
|u(0)|^2=\frac{2\pi\eta}{\exp(2\pi\eta)-1}=G(\eta).
\label{eq:Bowl2}
\ee
Here 
\be 
\eta=\alpha m_{\pi}/q
\label{eq:Gamov} 
\ee
and $=|k_{1} -k_{2}|$. $m_{\pi}$ is the pion mass.

However as pointed out by Bowler \cite{Bowler} and subsequently also
by others, in BEC this approximation may be questionable.
Indeed in a typical $e^{+}-e^{-}$ reaction e.g. 
 the source which gives rise to BEC has a size of the 
order of $1 fm$
which is a large number compared with a 
typical ``Coulomb
length" $r_{t}$ defined as the classical 
turning point where
the kinetic energy balances the potential Coulomb energy:
\be
\tilde{q}^{2}/2m_{red}=e^2/r_t
\label{eq:turn}
\ee
where $\tilde{q}=q/2$ and $m_{red}$ is the reduced mass of the pion pair.
 For a typical BEC momentum
difference of $q=100 MeV$, one gets from (\ref{eq:turn})
$r_{t}=0.08 fm$. This suggests that by taking the
value of the wave function at $r=0$ one overstimates the 
Coulomb correction.

More recently Biyajima and collaborators \cite{BMOW1} (cf.
also \cite{BMOW2}) have considered a further correction to
the correction proposed by Bowler, which decreases even
more the Coulomb effect and which is also of heuristic interest. 
In \cite{BMOW1} it is pointed out that the wave function 
(\ref{eq:scatt}) used by Bowler does not yet take into account
the symmetry of the two-particle system. It has to be
supplemented by an exchange term so that the rhs of
(\ref{eq:scatt}) becomes
\be
(e^{iqz}+e^{-iqz})+[f(\theta,\phi)+f(\pi-\theta,\phi+\pi)]
r^{-1}e^{iqr}
\label{eq:scattsym}
\ee

The corrections due to this new effect are of the same order
as those found in \cite{Bowler} and go in the same direction.

Another approach to the Coulomb correction in BEC
 has been
suggested by Baym and Braun-Munzinger\cite{Baym}. Starting 
from the observation of \cite{Bowler} about the classical 
turning point  
these authors propose the use for heavy ion reactions of a
{\em classical} Coulomb correction factor arising in the
assumption that the Coulomb effect of the pair is negligible 
for separations less than an initial radius $r_0$. 
This model is tested by comparing its results with
experimental data on $\pi^{+}\pi^{-}, \pi^{-}p,\pi^{+}p$
correlations in heavy ion collisions\footnote{The
measured $\pi^{+}\pi^{-}$
correlations were also used in two recent 
experimental papers \cite{1Na35}, \cite{2Na49} to estimate 
the Coulomb correction. Why such a
procedure is questionable is explained below.} (Au-Au at AGS
energies).
 The assumption behind this
comparison is that the observed correlations are {\em solely}
due to the Coulomb effect
\footnote{This assumption has to be qualified among
other things because 
 final state strong
interactions effects due to resonances can also influence
these correlations. 
 Furthermore there exists
also a quantum statistical correlation for the
$\pi^{+}\pi^{-}$ system (cf. sections 4.4, 4.7 and 4.8) 
which, however 
may be weak.}. Indeed qualitatively this seems to be
the case: thus the data for the $\pi^{+}\pi^{-}$ and $\pi^{-}p$
correlations show a bunching effect characteristic for an 
attractive interaction while the data for the $\pi^{+}p$
correlation show an antibunching effect, characteristic for
repulsion.
After this test
the authors compare their correction with the
Gamow correction and find that the last one is much
stronger. Herefrom they also conclude that the Gamow factor  
overestimates
quite appreciably the Coulomb effect 
\footnote{Cf. however also ref.\cite{Hardtke} where 
rescattering is added to the classical Coulomb effect and
where somewhat different results are obtained. It is unclear
whether the strong position-momentum correlations implicit in
this rescattering model do not violate quantum mechanics.}. 

An even stronger conclusion is reached by Merlitz and Pelte
\cite{MP3} from the solution of the time dependent
Schr\"odinger equation for two identical charged scalar bosons  
in terms of wave packets. These authors find that the
``expected" Coulomb effect in the correlation function is
obliterated by the dispersion of the localized states and
is thus unobservable. This makes the interpretation of
experimentally observed $\pi^{+}\pi^{-}$ correlations in
terms of Coulomb effects even more doubtful.

  The theoretical studies of the Coulomb effect in BEC
quoted so far are based on the
solution of the Schr\"odinger equation and apply in fact only
for the non-relativistic case. While one might argue that
the relative motion of two mesons in BEC is for small $q$
non-relativistic, this is not true for the single
particle distributions (cf. below). Therefore in principle
one should replace the solution of the Schr\"odinger
equation in the Coulomb field used above by the
corresponding solution of the Klein-Gordon equation. This
apparently has not yet been done, with the exception of
of a calculation by Barz\cite{Barz} who investigated 
 the influence of the Coulomb correction on the
measured values of radii. He found an 
important change of these radii due to the Coulomb field
only for momenta $\leq$ 200 MeV.

\begin{footnotesize}Finally, one must mention that the
corrections of the wave function described above do not take
into account the fact that the charge distribution of
a meson is not point-like, but has a finite extension of the
order of 1 fm. 
This means that in principle the
Schr\"odinger (or the Klein-Gordon) equation has to be
solved with a Coulomb potential modified by this finite size
effect. Given the great sensitivity of BEC on small
corrections in the wave function, this might be
a worthwile enterprise for future research. As a matter of
fact it is known from atomic physics (isotopic and isomeric shifts and hyperfine 
structure) that these finite size effects lead to observable
consequences.\end{footnotesize}

Besides the wave function effect which influences the BEC due to
directly produced particles or those originating from short
lived resonances, one has to consider \cite{Bowler} 
the Coulomb 
overcorrection applied to pairs of which one 
particle is a daughter of a long lived state.
This effect may bias the correction by up to 20\%.

\paragraph{Coulomb correction for higher order
correlations}
The Coulomb corrections discussed
above were limited to single and two-body interactions. In
present high energy heavy ion reactions we have already
events with  hundreds of particles and in the very near
future at the relativistic heavy ion collider at Brookhaven 
(RHIC) this number will increase by an order of magnitude. 
Then many-body final state interactions may become
important.
Unfortunately the theory of many-body interactions even for
such a ``simple" potential as the Coulomb one is apparently 
still unmanageable. The long-range
nature of the electromagetic interaction does not make this
task simpler. Bowler \cite{Bowler} sketched a scenario for
Coulomb screening based on the string model. In such a model
particles are ordered in space-time, so that e.g. at least one
$\pi^+$ must be situated between the members of a
$\pi^{-}\pi^{-}$ pair. While the net influence of this
effect on
$\pi^{+}\pi^{-}$ is expected to be small, for a 
$\pi^{-}\pi^{-}$
pair the situation is different, because instead of 
repulsion one obtains attraction. In the case of long range
interactions such an effect may become important if the
$\pi^{+}_{1}$ propagates together with the
$\pi^{-}_{2}\pi^{-}_3$ pair. This happens if 
$Q_{12}\sim Q_{13}\sim Q_{23}$. To take care of this  effect
Bowler suggests the replacement
\be
C(Q_{23})\to C(Q_{23})<C^{-1}(Q_{12})C^{-1}(Q_{13})>_{k3}
\label{eq:corrlr1}
\ee
for the Coulomb $\pi^{-}\pi^{-}$ correction and
\be
C(Q_{12})\to C(Q_{12})<C^{-1}(Q_{23})C(Q_{13})>_{k3}
\label{eq:corrlr2}
\ee
for the Coulomb $\pi^{+}\pi^{-}$ correction.
Here $< ...>_{k_3}$ symbolizes averaging with respect to 
the momentum of particle $3$. While on the
average
\be
<C^{-1}(Q_{23})C(Q_{13})>_{k3}
\label{eq:Bowle}
\ee
is unity, at small $Q$ the function $C(Q)$ oscillates
rapidly and therefore the factor
\be
<C^{-1}(Q_{12})C^{-1}(Q_{13})>_{k3}
\label{eq:Bowlf}
\ee
is sensitive to the distribution of $Q_{12}, Q_{13}$
associated with the local source. According to \cite{Bowler}
this last factor for
$\pi^{+}\pi^{-}$ pairs does not exceed $0.5\%$ but for like
sign pairs no estimate is provided.

\paragraph{Coulomb and resonance effect in single 
inclusive cross sections} 
At a first look one might 
be tempted to believe that for single inclusive cross
sections in heavy ion reactions the estimate of Coulomb 
effects is straightforward.
Unfortunately this is not the case and 
so far there is no reliable theoretical estimate of this 
effect. This is so because the produced charged secondaries
move not simply in the electromagnetic field of the
colliding nuclei but interact at the same time with all the
other secondaries. 
In view of this situation, recently an 
  attempt has been made to put in evidence experimentally 
the Coulomb effect in the single inclusive cross section of 
pions in heavy ion reactions \cite{Na44coul}. In this 
experiment an
excess of negative pions over positive pions in $Pb-Pb$
reactions at 158 AGeV was observed which the authors of
\cite{Na44coul} atrributed to the Coulomb interaction of
produced pions with the nuclear fireball. However this
interpretation has been challenged in \cite{Nellypm} where, 
in a detailed hydrodynamical simulation it
was shown that a similar excess in the $\pi^-/\pi^+$ ratio is
expected as a consequence of resonance (especially hyperon)
decays. This qualification goes in the same direction as
that mentioned above with respect to the exaggeration of 
the effects of Coulomb interaction in BEC. It also
illustrates the complexity of the many body problem of heavy
ion reactions even for weak interactions like the
electromagnetic one, which in principle are well known.

\subsection{Strong final state interactions.}   

This is a very complex problem because we are dealing with
non-perturbative aspects of quantum chromodynamics. That 
there is no fully satisfactory solution to
this problem can be seen from the very fact that we have at
least three different approaches to it. As will
become clear from the following, these different approaches
must not be used simultaneously, as this would constitute
double (or triple) counting. That is also why a fourth
solution proposed here and which is of heuristic nature 
will appear in many cases more appealing and more efficient.

\subsubsection{Final state interactions through resonances}

The majority of secondaries produced in high energy 
collisions are
pions out of which a large fraction (between 40 and 80$\%$) arise from
resonances\footnote{For
 experimental estimates 
of resonances cf. refs. \cite{expres}.}.
Since 
the resonances have finite lifetimes and momenta, their 
decay products are 
created in general outside the production region of the 
``direct'' pions 
(i.e., 
pions produced directly from the source) and that of the
resonances. As a consequence, the two-particle correlation 
function 
of pions reflects not only the geometry of the (primary) 
source but also the momentum spectra and lifetimes of 
resonances \cite{boal}. 
Kaons are much
less affected by this circumstance \cite{gpkaon}, however 
correlation
experiments with kaons are much more difficult because of the 
low 
statistics. For a more detailed discussion of kaon BEC cf.
section 5.1.3.

 The (known) resonances have been taken 
into account explicitely within the wave
function formalism (cf. e.g.\cite{Grishin},\cite{Pod72},\
\cite{Grass}), the string model (\cite{stringbowl}, 
\cite{stringander}) 
or within other variants of 
the Wigner function formalism (cf. e.g.
\cite{Gyu89}, \cite{Pra90}, \cite{Ledn}, \cite{Bolz}). 
The drawback of this explicit approach is that it is
rather complicated, it
 is usually applicable only at small
momenta differences $q$ and it presupposes a detailed 
knowledge of
resonance characteristics, including their weights, 
which, with few exceptions cannot be measured directly and have to
be obtained from event generators\footnote{That event
generators are not a reliable source of information for this
purpose was demonstrated in the case of $e^{+}-e^{-}$ reactions in
\cite{DeWolf}, \cite{Verb}.} or other models\footnote{In
ref.\cite{Bolz} the weights were determined
from thermodynamical considerations.}.

With these essential caveats in mind one finds that the distortion 
of the two-particle correlation function due to 
resonance decay leads  
to two obvious effects: (a) the effective radius of the 
source 
increases,
i.e., the width of the correlation function decreases, and (b) 
due
to the finite experimental resolution in the momentum difference
the presence of very 
long-lived resonances leads to an apparent decrease of the 
intercept 
of the correlation function.  
Effect (b) is particularly
important if one wants to draw conclusions from the intercept
about a possible contribution of a coherent component in 
multiparticle 
production.   

 In hydrodynamical studies of
multiparticle production processes one 
considers resonances within the Wigner function
approach (cf. section 5.1.3).  
 
\subsubsection{Density matrix approach}
In ref. 
\cite{Stelte} one describes the effect of strong final state 
interactions by constructing a density matrix based on 
 an effective Lagrangian of 
Landau-Ginzburg form as used in statistical physics and  
  quantum optics (cf. also 
\cite{Perugia}). This is
a more theoretical approach as it allows to study the effect
of the strength of the interaction $g$ on BEC, albeit in 
 an effective Lagrangian description. 

One writes the density matrix
\begin{equation}
\rho=Z^{-1}\int\delta\pi|\pi>e^{-F(\pi)}<\pi|\qquad;\qquad
Z=\int\delta\pi e^{-F(\pi)}
\label{eq:74}
\end{equation}
where $F$ is the analogue of the Landau-Ginzburg free energy
and the integrals are functional integrals over the field
$\pi$.  
This field is written as a superposition of coherent
$\pi_c$ and chaotic fields $\pi_{ch}$ 
\be
\pi  = \pi_{c}(y) + \pi_{ch}(y)
\label{eq:39}
\ee
The variable y refers in particular to rapidity.
 The total mean multiplicity $<n>$ is related to the 
 field $\pi$ by 
\be
<n>=<|\pi ^{2}|>;
\label{eq:mean}
\ee
 similar relations hold for the coherent and chaotic parts of
$\pi$.

One assumes stationarity in y, i.e. the field correlator
$G(y,y')=<\pi (y)\pi (y')>$
depends only on the difference $y-y'\equiv \Delta y$.
 One writes     
the Landau-Ginzburg form for $F$ as
\be
F(\pi)=\int^y_0dy\left[a\pi(y)+b\left|\frac{\partial\pi(y
)}{\partial
 y}\right|^2+g|\pi(y)|^4\right]
\label{eq:77}
\ee
where $a,b$ and $g$ are constants. The strong interaction
coupling is represented by $g$. The constants $a,b$ can be
expresed in term of $<n_{ch}>$ and the ``coherence length"
 $\xi$ which is defined through the correlator $G$ (cf.
eq.(\ref{eq:Lorentz}).

The main result of these rather involved calculations is 
the fact that while
the interaction does not play any significant role in
the value of the intercept $C_2(0)$  
 it plays an
important part in $(C_2(\Delta y\neq 0)$. This situation is
ilustrated in Figs.3,4.
Thus it is seen in Fig.3
 that all $C_2$ curves for various $g$ coincide in
the origin. The effect of the interaction in this approach
is similar to that of (short lived) resonances, i.e. it
leads to a decrease of the width of the correlation function. 
The sensitivity of $C_2$ on $g$ suggests that 
the shape of the correlation function can in principle
be used for the experimental determination of $g$.
One finds furthermore that there is no 
$g$-dependence
for purely chaotic or purely coherent sources.
 This
observation suggests that for a strongly coherent or chaotic
field the final state interaction does not manage to disturb
the correlation.

Note that the curves in Fig. 4 intersect at some $\Delta y$ 
which depends on the chaoticity. This is characteristic for 
correlations treated by quantum statistics, 
when one has a superposition of
coherent and chaotic fields,
and is 
a manifestation of the appearance of two (different) 
functions in the correlation function . 

We recall (cf. section 2.2) that for the particular case of 
a Lorentzian spectrum
one gets, {\em in the absence of final state interactions},
\begin{equation}
C_{2}(y,y')=1+2p(1-p)e^{-|y-y'|/\xi}+p^{2}e^{-2|y-y'|/\xi} 
\label{eq:41}
\end{equation}
instead of the empirical relation 
\be
C_{2}(y,y')= 1+ \lambda \exp(-|y-y'|/\xi).
\label{eq:Deutschmann}
\ee

As shown in Fig.5 the 
parametrization (\ref{eq:Deutschmann}) leads to 
parallel curves for various values
of $\lambda$, while the more correct parametrization of
ref.\cite{Stelte} leads to intersecting curves. This is due 
to the fact that the relation for $C_2$ derived in
\cite{Stelte}  contains as a particular
case (for $g=0$) 
eq.(\ref{eq:41}) and retains the
essential feature of eq.(\ref{eq:41}), which consists in the
superposition of two exponentials.

\subsubsection{Phase shifts} 
For charged pions the strong final state interactions can be
described also by
 phase shifts. It is known that  
for an isospin $I=2$ state (this the isopsin of a system of 
two identically charged 
pions) the corresponding strong interaction
 is repulsive. However it was suggested \cite{Bowl1}
  that
the range of strong interactions is smaller than the size
 of the hadronic source and therefore the correlation should
be
essentially unaffected by this effect\footnote{The
separation between resonances and phase shifts is of 
course not rigorous
because phase shifts reflect also the effect of resonances; 
however as long as phase shifts constitute a small effect, this should
not matter.}. 
  Even for particle
reactions like hadron-hadron or $e^{+}-e^{-}$
 the size of the source is of the order of $1 fm$ while 
 the range of interaction is only $0.2 fm$. 
On the other hand the effective size quoted above arises 
because of
the joint contribution to BEC of direct pairs and resonances.     
So it is interesting to analyze these two contributions 
separately.
  
Two identically charged pions are practically always
produced together with a third pion of opposite charge . 
Then according to Bowler \cite{Bowl1} one has also to
consider the $I=0$ attractive interaction between the
oppositely charged pions and this compensates largely the
$I=2$ state interaction. Thus it appears that also in particle
reactions  
only resonances play an important role in final state strong
interactions.

 The considerations about final state interactions made in
this
subsection treat separately Coulomb and strong
interactions. This is permitted as long as we deal with
small effects or when the ranges of the two types of
interactions do not overlap. 
For very small distances this
is not anymore the case. Furthermore the Gamow and the phase
shift corrections are 
based on the wave function formalism which 
ignores the possibility of creating particles. However when 
 entering the 
non-classical region the well known difficulties of the wave
function formalism become visible
(Klein paradox). 
 To consider this effect, in 
ref.(\cite{Anish} the joint contribution of the strong
interaction potential and the Coulomb potential are analyzed
in a version of the Bethe-Salpeter equation for spinless
particles. It is found that as expected also from the
considerations presented above that strong interaction
diminishes appreciably the Gamow correction.

\subsubsection{Effective currents}
As will be explained below the most
satisfactory approach to BEC is at present the classical
current approach, based on quantum field theory. 
There three types of source chracteristics appear: the
chaoticity, the
correlation lengths/times  and the space-time dimensions. 
It is obvious that these quantities contain already
information
about the nature of
the interaction and therefore it is quite natural to
consider them as effective parameters which describe 
 all effects of strong 
final state interactions.
 
This approach to strong final state interactions 
is probably the most
recommendable at the present stage for BEC in reactions
induced by hadrons or leptons 
because: 1. it is simpler; 
2. it avoids double counting; 3. it avoids the use of poorly known
resonance characteristics; 4. it avoids the use of the ill
defined concept of final state interactions for 
strong interactions. 
For heavy ion
reactions, when hydrodynamical methods are used
 the explicit consideration of resonances can be
practiced up to a certain point without major difficulties
 and then the strong final state interactions can be taken into
account through these resonances (cf. section 5.1.3). 
   
To conclude this discussion of final state interactions 
in BEC, it is interesting to note that in boson condensates
the final state interactions might be different than in 
normal
hadronic sources. In a condensate the Bose field becomes
long range in configuration space. This can be understood
as a consequence of the fact that in a condensate the
effective mass of the field carrier vanishes. Indeed a 
calculation \cite{2Wheeler} based on the chiral sigma model shows that   
the effective range of the pion field can increase several
times due to this effect.

\setcounter{equation}{0}

\section{Currents}
\subsection{Classical versus quantum currents}

In section 2.2 we were concerned mainly with the properties
of fields and did not ask the question where these fields
come from.
In the present chapter we shall put and try to answer 
this question.

We start by recalling the definition of correlation
functions within quantum field theory. 

Let $a^{\dagger}_{i} ({\bf k})$ and $a_{i}({\bf k})$ be the
creation and annihilation operator of a particle of momentum 
${\bf k}$, where the index $i$
labels internal degrees of freedom such as spin, isospin,
strangeness, etc. The $n$-particle inclusive distribution is

\begin{equation}
\frac{1}{\sigma}
\frac{d^n \sigma^{i_1 ... i_n}}{d\omega_1 ... d\omega_n}
\ =\ (2\pi)^{3n} \prod^n_{j=1} 2E_j  Tr \left( \rho_f 
\ a^\dagger_{i_1}
({\bf k}_1) ... a^\dagger_{i_n} ({\bf k}_n) a_{i_n}({\bf
k}_n) ... a_{i_1}
({\bf k}_1) \right) 
\label{eq:ae}
\end{equation}
where
\begin{equation}
d\omega_i = \frac{d^3k_i}{(2\pi)^3 2E_i}
\label{eq:di}
\end{equation}
is the invariant volume element in momentum space. 
With the notation
\begin{equation} 
G_n^{i_1 \cdots i_n} ({\bf k}_1, ... , {\bf k}_n) \ \equiv  \ 
\frac{1}{\sigma}\frac{d^n \sigma^{i_1 ... i_n}}
{d\omega_1 ... d\omega_n}
\end{equation}
the general $n$-particle correlation function is defined as
\begin{equation}
C_n^{i_1 ... i_n} ({\bf k}_1, ... ,{\bf k}_n ) \ =\ \frac{G_n
^{i_1 ... i_n} ({\bf k}_1 , ... ,{\bf k}_n)}
{G_1^{i_1} ({\bf k}_1) \cdot ... \cdot G_n^{i_n}({\bf k}_n)}
\label{eq:cln}
\end{equation}

The particles which the operators $a^{\dagger}$ and $a$ 
 are associated with are the quanta of 
the field (which we will denote in general by $\phi$);
these particles as well as 
the density matrix $\rho$ refer to the final state
where measurements take place. On the other hand we usually 
know
(or guess) the density matrix only in the initial state.
Therefore we will have to transform the above expression so
that eventually the density matrix in the final state
$\rho_{f}$
is replaced by the density matrix in the initial state 
$\rho_i$, while the fields will continue to refer to the 
final state.  
To emphasize
this we wrote in eq.(\ref{eq:ae}) 
$\rho_f$.
\be
\rho_{f}=S\rho_{i}S^{\dagger}  
\label{eq:S}
\ee
so that we have
\be
G_1(\mbold{k}) =
(2\pi)(2E_1)Tr\{\rho_i S^{\dagger} a^{\dagger}(\mbold{k}) 
a(\mbold{k})S\}\\
\ee
\ba
G_2(\mbold{k}_1,\mbold{k}_2)   
=(2\pi)^6(2E_1)(2E_2) Tr\{\rho_iS^{\dagger}a^{\dagger}
(\mbold{k}_2)
a^{\dagger}(\mbold{k}_1) a(\mbold{k}_1) a(\mbold{k}_2) S\}
\label{eq:(3+4)}
\ea 

Thus, if the initial conditions i.e. $\rho_i$ are given, 
in principle the knowledge of the S matrix suffices to
calculate the physical quantities of interest. 
In one case the $S$ matrix 
can even be derived without approximations. This happens when 
the currents are classical and we shall discuss this case 
in some detail in the present and the following section.  

Before doing this, we shall consider briefly the more general 
case when the currents are not necessarily classical.

 The $S$ matrix is given by the relation 
\be  
S\ =\ {\cal T}\exp\left\{i\int d^4x L^{int}(x)\right\}
\label{eq:S_def}
\ee
where the interaction lagrangian $L_{int}$ is a functional of
the fields $\phi$. $\cal T$ is the chronological time
ordering operator; we shall use below also the
antichronological time operator $\tilde {\cal T}$.

Consider for simplicity a scalar field produced by a current
$J$. Then
\be  
L^{int}(x)\ \equiv \ J(x)\phi(x).
\label{eq:L_def}
\ee

Equations (\ref{eq:S_def}) and (\ref{eq:L_def})
allow us now to calculate the correlations we are
interested in in terms of the currents after eliminating  
the fields. 
One obtains thus
\ba
P_1(\mbold{k})\;&=&\; Tr\left\{\rho_iJ_H^{\dagger}({\bf k})
 J_{H}({\bf k})\right\} \label{eq:P1JH} \\
P_2(\mbold{k}_1\,,\,\mbold{k}_2)\;&=&\;
Tr\left\{ \rho_i {\bf \tilde{\cal T}}
\left[ J_H^{\dagger}({\bf k}_1)J_H^{\dagger}({\bf k}_2)\right]
{\cal T}\left[ J_{H}({\bf k}_1)J_H({\bf k}_2)\right]\right\} 
\label{eq:P2JH}
\ea
where the label $H$ stands for the Heisenberg representation.
Now the cross sections depend only on the currents and the
density matrix in the initial state. The appearance
of the time ordering operators $\cal T$ and $\tilde {\cal T}$
in eqs.(\ref{eq:P1JH}),(\ref{eq:P2JH}) is a reminder of the fact
that the
current $J$ is here an operator.  
 
\subsection{Classical currents} 
Besides the fact that in this case an exact, analytical
solution  of the field equations is available and
that the limits of this approximation are quite clear, the
classical current has the important avantage that in it the
space-time characteristics of the source are clearly
exhibited and thus contact with approaches like the Wigner 
approach and
hydrodynamics are made possible. 

The assumption that the currents are classical implies that
$J$ is a $c$ number and then the order in eq.(\ref{eq:P2JH})
does not matter.
This approximation can be used 
in particle
physics
 when 
the currents are produced
by heavy particles (e.g. nucleons) and/or when the 
momentum transfer $q$ is small compared with the momentum $K$
of the emitting particles. 
\footnote{For multiparticle production processes this
implies also a 
constraint on the multiplicity and/or the  
momenta $k$ of
the produced particles}. In section 4.7 a new criterion for
the applicability of the classical current assumption in
terms of particle-antiparticle correlations will be presented.  

The classical current formalism was introduced
to the field of Bose-Einstein correlations in
\cite{shuryak,podgor,GKW}. In this approach, particle
sources are 
treated as external classical currents $J(x)$, the 
fluctuations of
which are described by a probability distribution $P\{ J \}$.
>From many points of view like e.g. the understanding the 
space-time properties of the
sources or the isotopic spin dependence of BEC this approach
is superior to any other approach.
This has become clear only in the last years \cite{aw}, 
\cite{APW} when a
systematic investigation of the independent physical
quantities which enter the dynamics of correlation functions   
has been made (cf.below).

 The 
classical current 
formalism in momentum space is mathematically identical 
with the coherent
state formalism used in quantum statistics 
and in particular in quantum optics (cf. section 2.2), the
classical currents in $k$-space $J(k)$ being proportional to
the eigenvalues
of the coherent states $|\alpha >$. This explains the
importance of the coherent state formalism for applications
in particle and nuclear physics.

 The density matrix is 
\begin{equation}
\rho = \int {\cal D} J \ {\cal P}\{ J \} \ |J><J|
\label{eq:what2}
\end{equation}
where the symbol ${\cal D}J$ denotes an integration over 
the space 
of functions $J(x)$, and the statistical weight ${\cal P}
\{ J \}$ is normalized to unity,
\begin{equation}
\int {\cal D}J \ {\cal P} \{ J \}\  =\  1.
\label{eq:dj}
\end{equation}
(The reader will recognize in eq.(\ref{eq:what2}) the $\cal P$-representation
introduced in section 2.2.)

 Expectation values of field 
operators 
can then be expressed as averages over the corresponding  
functionals of the currents, e.g.,
\begin{eqnarray}
&& Tr \left( \rho a^{\dagger}({\bf k}_1)\ ...\ a^{\dagger}
({\bf k}_n) a({\bf k}_{n+1})\ ... \ 
a({\bf k}_{n+m}) \right)\\
&& = \prod^n_{j=1} \frac{(-i)}{\sqrt{(2\pi)^3 \ 2 E_j}} 
\ \prod^{n+m}_{j=m+1}
\frac{i}{\sqrt{(2\pi)^3 2 E_j}} \ 
 <J^*({\bf k}_1)\ ...\ J^*({\bf k}_n) J({\bf k}_{n+1})\ 
...\ J({\bf k}_{n+m})>\nonumber
\label{eq:trw}
\end{eqnarray}

In the following we shall discuss the special case when the 
fluctuations
of the currents $J(x)$ are described by a  
Gaussian distribution ${\cal P}\{J\}$. The reasons for this
choice were given in section 2.2.

As in quantum optics we write
the current $J(x)$ as the sum of 
a chaotic and a coherent component,
\begin{equation}
J(x) \ = \ J_{chaotic}(x)\ + \ J_{coherent}(x)
\label{eq::jsum}
\end{equation}
with 
\begin{eqnarray}
J_{coherent}(x) & = & <J(x)>\\
J_{chaotic}(x)  & = & J(x) - <J(x)> 
\label{eq::jchao}
\end{eqnarray}
By definition, $<J_{chaotic}(x)>=0$.
The case $I(x) \not= 0$ corresponds 
to {\em single particle coherence}.

The Gaussian current 
distribution is completely determined by
specifying its first two moments: the first moment coincides,
because of eq.(\ref{eq::jchao}), with the coherent component,
\begin{equation}
I(x) \ \equiv \ <J(x)>
\label{eq:a1}
\end{equation}
and the second moment is given by the 2-current correlator,
\begin{eqnarray}
D(x,x') & \equiv & <J(x) \ J(x')> - <J(x)> \ <J(x')>\nonumber\\
  & = & <J_{chaotic}(x) \ J_{chaotic}(x')>
\label{eq:a2}
\end{eqnarray}

\subsection{Primordial correlator, correlation length and 
space-time distribution of the source}

We come now to a more recent development \cite{aw},
\cite{APW} of the current
formalism which has shed new light on both fundamental and
applicative aspects of BEC. 

There are two fundamental and in principle
independent aspects of physics which come together in the
phenomenon of BEC in particle and nuclear physics. One
refers to the {\em geometry} of the source and goes back to
the original Hanbury-Brown and Twiss interefence experiment 
in astronomy. The ``geometry" is characterized by the
size of the source e.g.the longitudinal and transverse radius
 $R_{\parallel}$ and $R_{\perp}$ respectively
and the lifetime of the source $R_0$.

The second aspect is related to the {\em dynamics} of
the source and is expressed through correlation lengths. 
In the following
we will use two correlation lengths  
$L_{\parallel}$, 
$L_{\perp}$ and a correlation time $L_0$ 
\footnote{We refer here to short range
correlations. Cf.
section 6.1.3 for a distinction between short range and long
range correlations.}.  
As a consequence
of the finite space-time size of sources in particle physics
 one cannot in general separate the geometry from the
dynamics in the second (and higher) order correlation 
function.   
This separation is possible (cf. below) only by using
simultaneously also the single inclusive cross section.
This remarkable fact 
\cite{aw} is not yet realized in
most of the theoretical and experimental papers on this 
subject\footnote{For systems in local equilibrium 
 Makhlin and Sinyukov \cite{MS1}     
introduced   
a length scale (called in \cite{MS} ``length
of homogeneity") which characterizes the 
hydrodynamical
expansion of the source and can be different from the size of the
system. Further references on this topic can be found e.g.in
\cite{Akk}, \cite{csorlor}}.
One of the main objects of the present review is to clarify
this situation.

Assuming Gaussian currents, one may take the point of view 
that the purpose of measuring $n$-particle distributions is 
to obtain information about the space-time form of the 
coherent component, $I(x)$, and of the correlator of the 
chaotic components of the current, $D(x,x')$. 
In practice, because of limited statistics it is necessary 
to consider simple parametrizations of these quantities, 
and to use the information extracted from experi\-mental 
data to determine the free parameters (such as radii, 
correlation lengths etc.). 

The approach considered in \cite{APW} 
 differs in 
some fundamental aspects from those applications of
the density matrix approach in particle physics which are 
performed in momentum
(rapidity) space and are limited usually to one dimension 
(cf. however \cite{prd44} where rapidity
{\em and} transverse momentum were considered).
The approach of ref.\cite{APW} is a  {\em space-time} approach
in which the parameters
refer to the space-time characteristics of the source. This 
approach has important heuristic advantages compared with the 
momentum (rapidity)
space approach as will be explained below. 

Among other things, in the quantum statistical (QS) space-time approach 
 the parameters of the source as
defined above can be considered as effective parameters 
which contain already
the entire information which one is interested in and which 
one could obtain from experiment, and thus distinguishing 
between directly produced particles and resonance decay 
pro\-ducts could amount to double counting.
 The apparent proliferation
of parameters brought about by the QS approach is
compensated by this heuristic and practical simplification.
Furthermore a new and essential feature of the 
approach of ref.\cite{APW} as compared with previous
applications of the current formalism \cite{podgor,GKW} 
which assume $L=0$, is the 
 {\em finite} correlation length (time) $L$. 
This fact has important theoretical and practical 
consequences. It leads 
 among other things to an effective correlation
between momenta and coordinates, so that e.g. the second
order correlation
functions does not depend only on the difference of momenta  
$q=k_1-k_2$ but also on the sum $k_1+k_2$.
This non-stationarity property, which is observed in experiment, is usually
associated with expanding sources and treated within the
Wigner function formalism. However from the 
considerations presented above it follows that
expansion is in general not a necessary condition for non-stationarity
in $q$. It will be shown how expanding sources can
be treated without the Wigner formalism, which restricts
unnecessarily the applicability of the results to small $q$
values  
 (cf. section 4.8).     
 
The distinction between correlation lengths and radii is
possible only in the current formalism; the Wigner formalism
provides just a length of homogeneity.

The results which follow from the space-time approach \cite{APW} 
include: 

(i) The existence of at least 10
independent parameters that enter into the correlation 
function; 

(ii new insights into the problem of partial coherence;

(iii) Isospin effects: $\pi^0\pi^0$ correlations are different
from $\pi^{\pm}\pi^{\pm}$
correlations; there exists a quantum statistical 
(anti)correlation
between particles and antiparticles ($\pi^+\pi^-$ in this 
case).
These effects are associated with the presence of squeezed 
states in the density matrix, which in itself is a
surprising and unexpected 
feature in conventional strong interaction phenomenology. 

It turns out that soft pions play an essential role in 
the experimental investigation of BEC, both with respect to 
the effect of particle-antiparticle
correlations predicted here as well as in the investigation 
of the coherence of the source.
Depending on the relative 
magnitude of the parameters of the coherent and chaotic
component, soft particles
can either enhance or suppress the coherence effect.

Consider first the case of an infinitely extended source. The 
correlation of currents at two space-time points $x$ and $y$ 
is described by a primordial correlator,
\begin{equation}
<J(x) \ J(y)>_0\  =\  C(x-y)
\label{eq:jx}
\end{equation}
Note that $C$ depends only on the difference $x-y$.
The correlator $C(x-y)$ contains some characteristic length (time) scales $L$, 
the so-called correlation lengths (times) \footnote{ 
These correspond to the ``coherence lengths" used in the 
quantum optical literature.}. In the current formalism used
in \cite{GKW} $L=0$. 

$C(x-y)$ is a real and even function of its argument.
 In the rest frame of the source, it is usually 
parametrized by an  exponential,
\begin{equation} 
C(x-y)\  =\  C_0 \ \exp \left[ -\frac{|x_0-y_0|}{L_0}- 
\frac{|{\bf x}-{\bf y}|}{L} \right]
\label{eq:cxy1}
\end{equation}
or by a Gaussian, 
\begin{equation}
C(x-y)\ = \  C_0 \ \exp \left[ -\frac{(x_0-y_0)^2}{2L^2_0}- 
\frac{({\bf x}-{\bf y})^2}
{2L^2} \right]
\label{eq:cxy}
\end{equation}
However, it should be clear that in principle any well 
behaved decreasing function of $|x-y|$ is a priori 
acceptable, and in practice it is usually
up to the experimenter to decide which particular form is more
appropriate.
The ans\"atze (\ref{eq:cxy1},\ref{eq:cxy}) need to be 
modified for
the case of an expanding source (cf. sections 4.8 and 4.9) 
where each source element is characterized not only by a 
correlation length $L^\mu$ but also by a four velocity 
$u^\mu$. 
 As a matter of
fact, the form of the function $C$ is
irrelevant as long as one is interested in the general 
statements of the theoretical quantum statistical (current) 
formalism. In practical applications, of course, in order to
obtain concrete information about the source and the
medium (i.e., about $L$) the form of $C$ has to be 
specified. In principle, a full
dynamical theory is expected to determine the functional 
form of the correlation function, 
 however at present this
``fundamentalist" approach is not yet applicable 
\footnote{Indeed one may hope that for strongly 
interacting systems 
 lattice QCD may provide in future the
corrrelation length $L$.}. One uses instead a 
phenomenological
approach like that reflected in eqs. (\ref{eq:cxy1},
\ref{eq:cxy}). 

Effects of the {\em geometry} of the source 
are taken into account by introducing
the space-time distributions of the chaotic and of the 
coherent component,
$f_{ch}(x)$ and $f_c(x)$, re\-spec\-ti\-ve\-ly. 
The expectation values of the currents, $I(x)$ and
$D(x,x')$, take nonzero values only in space-time 
regions where $f_c$ and $f_{ch}$ are nonzero. Thus, one may 
write 
\begin{eqnarray}
I(x) &=& f_c(x)\\
D(x,x') &=& f_{ch}(x) \  C(x-x') \ f_{ch}(x')
\label{eq:ixh}
\end{eqnarray}

We turn now to an important new aspect of the
current approach

\subsection{Production of an isospin multiplet}

Following \cite{APW} we generalize the previous results to the 
case of an isospin multiplet and derive explicit
expressions for the single inclusive distributions and 
correlation functions of particles that form an isotriplet
(such as the $\pi^+$,$ \pi^-$ and 
$\pi^0$-mesons). For the sake of definiteness, we 
will refer to pions in the discussion below, but it should 
be understood that the formalism is applicable to an 
arbitrary isomultiplet.

Consider the production of charged and neutral pions,
$\pi^+, \pi^- $ and $\pi^0$, by random currents. 
The initial interaction Lagrangian is written
\begin{eqnarray}
{\cal L}_{int} &=& J_+(x) \pi^-(x) + J_-(x)\pi^+(x) + J_0(x)\pi^0(x) .
\label{eq:lint}
\end{eqnarray}
 The 
current distribution is completely characterized by its 
first two moments. They read
for the case of an isotriplet

\begin{eqnarray}
I_i(x) & \equiv &  <J_i(x)> \qquad i,i'\ =\ +,-,0\\
D_{ii'}(x,x') & \equiv & <J_i(x) \ J_{i'}(x')> - <J_i(x)> \ 
<J_{i'}(x')>\nonumber
\label{eq:b1}
\end{eqnarray}

Invariance of the chaotic 2-current correlator under 
rotation in isospin space implies 
\begin{eqnarray}
D_{00}(x,x') & = & D_{+-}(x,x') \ \equiv  D(x,x')\\
D_{++}(x,x') & = & D_{--}(x,x') \ =\ 
D_{+0}(x,x') \ =\ D_{0-}(x,x') \ =\ 0.\nonumber
\label{eq:b35}
\end{eqnarray}
The corresponding current distribution is 
\begin{equation}
P \{ J_i \} \ = \ \frac{1}{{\cal N}}\  \exp[-{\cal A} 
\{ J_i \}]
\label{eq:b4}
\end{equation}
where in coordinate representation
\begin{equation}
{\cal N} \ = \ \int \int \int {\cal D}J_+ {\cal D}J_- 
{\cal D}J_0
 \ \exp[-{\cal A} \{ J_i \}]  
\label{eq:b45}
\end{equation}
and
\begin{eqnarray}
{\cal A} \{ J_i \} & = & 
\left.
\int\int \ d^4x \ d^4y \right[ (J_+(x)-I_+(x)) M(x,y) 
(J_-(y)- I_-(y))\nonumber\\
& & \qquad \qquad + \left.\frac{1}{2}
(J_0(x)-I_0(x)) M(x,y) (J_0(y)-I_0(y))\right]  
\label{eq:b5}
\end{eqnarray}
with 
\begin{equation}
M(x,x') \ = \ D^{-1}(x,x')
\label{eq:fd2}
\end{equation}

For the general case of a partially coherent source,
the single inclusive distributions of pions of charge $i$ 
$(i = +,-,0)$ can be expressed as the sum of a 
chaotic component and a coherent component,
\begin{equation}
\frac{1}{\sigma} \frac{d\sigma^i}{d\omega} = 
\left.\frac{1}{\sigma} \frac{d\sigma^i}
{d\omega}\right|_{chaotic} + \left.\frac{1}{\sigma} 
\frac{d\sigma^i}{d\omega}
\right|_{coherent}
\label{eq:sig1}
\end{equation}
with
\begin{equation}
\left.\frac{1}{\sigma} \frac{d\sigma^i}
{d\omega}\right|_{chaotic} \ =\ D(k,k)
\label{eq:ncha}
\end{equation}
and
\begin{equation}
\left.\frac{1}{\sigma} 
\frac{d\sigma^i}{d\omega}
\right|_{coherent}
\ =\  |I(k)|^2
\label{eq:nco}
\end{equation}

\noindent
In general, the chaoticity parameter $p$ will be momentum 
dependent,
\begin{equation}
p(k) \ =\ \frac{D(k,k)}{D(k,k)+|I(k)|^2}
\label{eq:fk}
\end{equation}

To write down the correlation functions in a concise form, 
one introduces the normalized current correlators,
\begin{equation}
d_{rs} = \frac{D(k_r,k_s)}{[D(k_r,k_r)\cdot 
D(k_s,k_s)]^{\frac{1}{2}}},
\quad
\tilde{d}_{rs} = \frac{D(k_r,-k_s)}{[D(k_r,k_r)\cdot
D(k_s,k_s)]^{\frac{1}{2}}},
\label{eq:drs}
\end{equation}
where the indices $r,s$ label the particles. 
Note that $\tilde{d}$ is the {\em same} function as $d$ 
but of the
change of sign of one of its variables. 
 One may express the correlation functions
in terms of the magnitudes and the phases,
\begin{eqnarray}
T_{rs} &\equiv & T(k_r,k_s) \ =\ |d(k_r,k_s)|\nonumber\\
\tilde{T}_{rs} &\equiv& \tilde{T}(k_r,k_s) \ = \ 
| \tilde{d}(k_r,k_s)| \nonumber \\
\phi^{ch}_{rs} &\equiv& \phi^{ch}(k_r,k_s)\ =\
 {\rm Arg} \  d(k_r,k_s)\\
\tilde{\phi}^{ch}_{rs} &\equiv& \tilde{\phi}^{ch}(k_r,k_s) 
\ =\  {\rm Arg} \ \tilde{d}(k_r,k_s) \nonumber
\label{eq:trs}
\end{eqnarray}
and the phase of the coherent component,
\begin{equation}
\phi^c_r  \ \equiv \ \phi^c(k_r) \ =\ {\rm Arg}\  I(k_r)
\nonumber \, .
\label{eq::phic}
\end{equation}
The same notation will be used for the chaoticity 
parameter,
\begin{equation}
p_r \ \equiv \  p(k_r)
\label{eq:pr}
\end{equation}
 The two-particle
correlation function is
\begin{equation}
C^{++}_2({\bf k}_1,{\bf k}_2)\ =\ 1 +\ 2 \sqrt{p_1(1-p_1)
\cdot p_2(1-p_2)}
\ T_{12}\  \cos(\phi^{ch}_{12} - \phi^c_1+\phi^c_2)\ 
+\  p_1p_2 \ T^2_{12}
\label{eq:c2plus1}
\end{equation}

In \cite{APW} higher order correlation functions up to and
including order 5 are given.

In the absence of single particle coherence 
the two-particle correlation functions for different pairs of 
$\pi^+,\pi^-, \pi^0$ mesons read
\begin{eqnarray}
C_2^{++}({\bf k_1},{\bf k_2}) &=& 1 + |d_{12}|^2,\nonumber\\
C_2^{+-}({\bf k_1},{\bf k_2}) &=& 1 + |\tilde{d}_{12}|^2,
\nonumber\\
C_2^{+0}({\bf k_1},{\bf k_2}) &=& 1, \label{eq:c4}\\
C_2^{00}({\bf k_1},{\bf k_2}) &=& 1 + |d_{12}|^2
+ |\tilde{d}_{12}|^2 \nonumber
\end{eqnarray}

These results \cite{apw} were surprising in that they 
disagreed with some of the pre\-con\-cei\-ved notions on 
Bose-Einstein correlations.
For instance, it was commonly assumed that without taking into
account final state interactions
 and in the absence of coherence, the maximum of the 
two-particle correlation  of identical pions is 2
(for ${\bf k_1} = {\bf k_2}$). It was also assumed that 
there are no correlation effects among different 
kinds of pions because these particles are not identical.
(This last assumption is even sometimes used in normalizing the 
experimental data on $C_2^{\pm\pm}$ with respect to 
$C_2^{+-}$.) 

The results (\ref{eq:c4}) 
show that these assertions are not necessarily true. 
In particular, looking at the two pion correlations one can 
see that in addition to the familiar correlations of 
identical particles 
(the terms $|d_{12}|^2$) there are particle-antiparticle 
-- in this case, $\pi^+\pi^-$  -- correlations 
(the terms $|\tilde{d}_{12}|^2$). The $\pi^0$ has both
terms, as it is identical with its antiparticle. 
Essentially this last fact is the explanation for the 
appearance of
the ``surprising" effects. Obviously it is a specific
quantum field effect.
It will be shown in section 4.8, that for soft pions and for small lifetimes of the 
source the terms
$|\tilde{d}_{12}|^2$ can in principle become comparable with
the conventional terms $|d_{12}|^2$.
This implies  that the distribution $C_2^{00}({\bf k_1},
{\bf k_2})$ of two neutral pions can be as large as 3, and 
the maximum value of $C_2^{+-}$ is 2 
(instead of 1). The corresponding limit of $C_3^{000}$ is 15
(instead of 6),  and that of $C_3^{++-}$ is 6 (instead of 2).
However, it should be noted that these are merely upper 
limits, which for massive particles are not reached except for sources
of infinitesimally small lifetimes. For
soft photons however the situation is different
(cf. below). 
 
The ``new" terms, proportional to 
$<a_\ell(k_1) a_\ell(k_2)>$ are due to the 
non-stationarity (in $k$ space) of the source. 
While in quantum optics time-stationarity is the rule, 
in  particle physics this
is not the case because of the finite lifetime and finite 
radius of the sources.
The existence of a non-vanishing expectation value of the 
products $a(k_1) a(k_2)$ is what one would expect (cf. 
section 2.2) from two-particle coherence 
(squeezing), just as $<a(k)>\not= 0$ follows from ordinary 
(one-particle) coherence (note that the latter has not been 
assumed here).  
The fact that squeezing which, as mentioned above, is quite 
an exceptional situation in optics,
discovered only recently, is a natural consequence of the 
formalism for present particle physics,
is possibly one of the most startling
results obtained recently in BEC.
A characteristic feature of squeezed states is the fact
that for static sources they lead anti-correlations.
Finally, it should be pointed out again that the above 
results -- in particular, the fact that $C_2^{00}$ in 
general differs from $C_2^{--}$ -- are consistent with 
isospin symmetry. 

In closing this section one should note that the existence of
particle-antiparticle correlations is not restricted to
pions but applies also to other systems, like e.g. neutral
kaons. In principle thus there exist also $K_0\bar{K}_0$
quantum statistical correlations. However since $K_0$ particles cannot be
observed except in linear combinations with $\bar{K}_0$ in the
form of $K_s$ and $K_l$ the QS particle-antiparticle correlation
effect has to be disentangled from the $K_sK_s$ or $K_lK_l$
 Bose-Einstein correlation ($K_s$ and $K_l$ are of course bosons and thus subject
to BEC) which exist also in the ``old fashioned"
wave function formalism which ignores the intrinsic
``new" $K_0\bar{K}_0$ correlation. As a matter of fact
$K_sK_s$ correlations have been observed experimentally,
however no attempt has been made so far to extract from them
the ``surprising" effects
\footnote{Because of their larger mass as compared with that
of pions, these effects may be even more quenched than in
the case of pions, except for sources of very short lifetime
(cf. section 4.8).} (For more recent experiments cf. 
\cite{Opal}) and for a theoretical analysis
of the ``old" effects and their possible application in CP
violation phenomena cf. \cite{Lipkin}).  
 
\footnote{The analysis in the preceding subsections referred to the
production of an isotriplet assuming just symmetry between
the isospin components (cf. eq.(\ref{eq:lint})) 
 of the current.
In principle, for strong
interactions the conservation of isospin $I$ has also to be
considered. While the chaotic part is not affected by this
condition, the coherent component is influenced by
conservation of isospin\cite{AB}. In particular this can lead to an
additive positive term in the correlation function and thus to an
increase of the bounds of BEC for pions. 
%For $I=0$,
%where the effect is maximum, this additive term is $1/5$ for
%charged pions and $4/5$ for neutral pions. 
It remains to be
seen whether this effect can be distinguished from the 
effect of long range correlations (cf. section 6.1.3).
Moreover in hadronic reactions and in particular those involving 
nuclei such an effect would be strongly suppressed
 because of the following circumstance.
The initial state has to be averaged over all components of
isospin $I$  which is a first ``diluting" factor. Furthermore
the effect is appreciable only for low total isospin.
This total isospin has to be shared by the chaotic component
$I_{ch}$ and the
coherent component $I_c$: $I=I_c+I_{ch}$.
The first one arises mostly from
resonances with different isospin values, so that even if
the total isospin takes its minimum value
($I=0$), $I_{ch}$ and therefore $I_c$ can take larger
values.}

\subsubsection{An illustrative model of uncorrelated point-like 
random sources}

To clarify the origin of different terms in the functions 
$C_2^{ab}(k_1,k_2)$, let us consider, a 
source consisting of $N$ point-like random sources

\begin{equation}
J_a(x) = \sum^N_{i=1} \ j_a(x_i) \ \delta(t-t_i) \ \delta^3
({\bf x}-{\bf x}_i),\quad a = +,-,0
\label{eq:e1}
\end{equation}

and assume that the currents $j_a(x_i)$ at different points 
$x_i$ are mutually
independent and have the same statistical properties, i.e.

\begin{equation}
<j_a^*(x_i) \ j_b(x_j)> = \delta_{ij}<j_a^* j_b>.
\label{eq:e2}
\end{equation}

We also assume in this section

\begin{equation}
<j_a> = 0,
\label{eq:e3}
\end{equation}
i.e. ignore a possible coherent component $<j_a>$ to make 
the presentation more transparent.

Now the one-particle distribution is
\begin{equation}
<J_a^*({\bf k}) \ J_a({\bf k})> = N \cdot <j_a^* j_a>
\label{eq:e4}
\end{equation}

and the two-particle distribution takes the form
\begin{eqnarray}
&<&J^*_a(k_1) \ J_a(k_1) \ J^*_b(k_2) \ J_b(k_2)> = 
\nonumber\\
&=&\sum^N_{i=1} < j^*_a(x_i)j_a(x_i) j^*_b(x_i)j_b(x_i)> 
\nonumber\\
&+&\sum^N_{i \not= j} <j^*_a(x_i)j_a(x_i)> \cdot <j^*_b(x_j)
j_b(x_j)>\nonumber\\
&+&\sum^N_{i \not= j} <j^*_b(x_i)j_a(x_i)> \cdot <j^*_a(x_j)
j_b(x_j)> \cdot e^{i(k_1-k_2)(x_i-x_j)} \nonumber\\
&+&\sum^N_{i \not= j} <j^*_a(x_j)j^*_b(x_j)> \cdot <j_a(x_i)
j_b(x_i)> \cdot e^{i(k_1+k_2)(x_i-x_j)} \nonumber\\
\label{eq:e5}
\end{eqnarray}

Let us consider separately the four different terms on the 
right-hand side of eq. (\ref{eq:e5}). The first term 
corresponds to two particles being emitted
from a single point (Fig. 6a); it is proportional to the 
number of emitting points.

The second term describes an independent emission of two 
particles from different points (Fig. 6b).

The third term, being non-zero for $a=b$, describes an 
interference effect of direct and exchange diagrams, 
characteristic for identical particles, emitted
from different points (Fig. 6c). This is the usual 
BE-correlation effect.

The fourth term describes an interference of two-particle 
emissions from different points (Fig. 6d) (``two-particle 
sources"). It appears to be non-zero for real currents with 
$a=b \ (\pi^0\pi^0)$ and for
complex currents with $J^*_a = J_b \ (\pi^+\pi^-)$, that is 
for particle-antiparticle associative emission. In this 
simple model, it represents the ``surprising" effects discussed 
above.

This diagramatic illustration of the ``surprising" 
BE-correlations is due to Bowler \cite{bowler}, who found 
that the ``surprising" effects 
derived for the first time in ref. \cite{apw} can be 
understood in terms of these more qualitative considerations.
Bowler derived the ``new" effects from the string model.

\subsection{Photon interferometry. Upper bounds of BEC.}

The advantage of photon BEC resides in the fact that photons 
 are not influenced by final state interactions.
Photons present also an interesting subject of theoretical
research from the general BEC point of view, since they are
spin-one bosons while pions and kaons used in hadronic BEC
are scalar particles. We shall see below that this 
supplementary degree of freedom has specific implications 
in BEC. Last but not least photon correlations are for various reasons to be discussed
below of particular interest
for the search of quark matter.  We present some of 
the results of
\cite{Leo}, which contain as a special case those of
\cite{Neu} and where these topics are discussed.

 Consider a heavy ion reaction where photons are produced
through bremsstrahlung from protons in independent
proton-neutron collisions\footnote{Photon emission from
proton-proton collisions is suppressed because it is of
quadrupole form.}. The corresponding
elementary dipole currents are 

\be     
j^{\lambda}(k)=\frac{ie}{mk^0}{\bf p}.{\bf
\epsilon}_{\lambda}(k)
\label{eq:Leo1}
\ee
where ${\bf p} = {\bf p}_i-{\bf p}_f$ is the difference
between the initial and the final momentum of the proton,
${\bf \epsilon}_{\lambda}$ is the vector of linear 
polarization
and $k$ the photon momentum; $e$ and $m$ are the charge and 
mass of the proton respectively. The total current is written 

\be
J^{\lambda}(k)=\sum^N_{n=1}e^{ikx_n}j_n^{\lambda}(k).
\label{eq:Leo2}
\ee
For simplicity we will discuss in the following only the 
case of pure chaotic currents\\ $<J^{\lambda}(k)> = 0$ and 
refer for coherence effects to the original 
literature \cite{Leo}, \cite{Neu}.  
In analogy to the considerations of the previous subsection
the index $n$ labels the independent nucleon collisions
which take place at different space-time points $x_n$. These
points are assumed to be randomly distributed in the
space-time volume of the source with a distribution function
$f(x)$ for each elementary collision. The current correlator
 is proportional to products of
the form 
\be
<J^{\lambda_1}(k_1)J^{\lambda_2}(-k_2)>={\bf
{\epsilon}}^{i}_{\lambda_1}(k_1)\left(\sum^N_{n=1}<{\bf
p}^{i}_n{\bf p}^j_n>\right){\bf {\epsilon}}^j_{\lambda_2}(k_2)
\label{eq:Leo10}
\ee

 For central collisions due to the axial symmetry around the beam direction one
has for the momenta the tensor
decomposition
\be
<{\bf p}^{i}_n{\bf p}^j_n>=\frac{1}{3}\sigma_n\delta^{ij}+
\delta_nl^{i}l^j,
\label{eq:Leo11}
\ee
where $l$ is the unit vector in the beam direction and
$\sigma _n, \delta _n$ are real positive constants. In
\cite{Neu} an isotropic distribution of the momenta was
assumed. This corresponds to the particular case 
$\delta _n = 0$. 
The generalization to the form (\ref{eq:Leo11}) is due to
\cite{Leo}. The summation over polarization indexes is
performed by using the relations

\be
<({\bf \epsilon}^{i}.{\bf p}_l)({\bf \epsilon}^{j}.{\bf p}_
{l^{'}})>=\frac{1}{3}({\bf \epsilon}^{i}.{\bf \epsilon}^{j})
\delta_{ll'}
\label{eq:sum}
\ee

and

\be
\sum^2_{\lambda=1}{\bf \epsilon}^{i}_{\lambda}(k){\bf
\epsilon}^j_{\lambda}(k)=\delta^{ij}-{\bf n}^{i}{\bf n}^j,
\label{eq:Leo21}
\ee
where ${\bf n} = {\bf k}/ {|{\bf k}|}$.
 
We write below the results for
the second order correlation function
\be
C_2(k_1,k_2)=\frac{\rho_2(k_1,k_2)}{\rho_1(k_1)\rho_1(k_2)}
\label{eq:Leo24}
\ee
 for two extreme cases:

(1) Uncorrelated elementary currents (isotropy) ($\sigma \gg
\delta$)

\be
C_2(k_1,k_2;\;\sigma\neq0,\;\delta=0)=
1+\frac{1}{4}[1+({\bf n}_{1}.{\bf n}_{2})^2]\left[|\tilde{f}
(k_1-k_2)|^2+|\tilde{f}(k_1+k_2)|^2\right],
\label{eq:Leo25}
\ee
leading to an intercept 
\be
C_2(k,k)=\frac{3}{2}+\frac{1}{2}|\tilde{f}(2k)|^2
\label{eq:Leo26}
\ee
limited by the values (3/2,2).  

(2) Strong anisotropy
($ \sigma  \ll \delta $)
\be
C_2(k_1,k_2;\;\sigma=0,\;\delta\neq0)=
1+|\tilde{f}(k_1-k_2)|^2+|\tilde{f}(k_1+k_2)|^2
\label{eq:Leo27}
\ee
with an intercept
\be
C_2(k,k)=2+|\tilde{f}(2k)|^2
\label{eq:Leo28}
\ee
limited this time by the values (2,3). 
 Note that due to the form of the photon-current
interaction (\ref{eq:Leo1}) in this (strong anisotropy) case 
 the photons emerge practically completely polarized 
so that the summation over polarizations does not affect
the correlation.

 These results are remarkable among other things because they
illustrate the specific effects of photon spin on BEC. Thus
while for (pseudo-)scalar pions the intercept is a constant
(2 for charged pions and 3 for neutral ones) even for
 unpolarized photons the intercept is a function of k.

For a graphical illustration and explanation of this fact
cf. Fig. 7 from ref.\cite{Leo}). It is seen that to perform
the summation over polarization implied by eq.(\ref{eq:sum})
only one direction of the linear polarization can be chosen
equal for both photons, while the other
polarization direction differs by the angle $\theta$ between
the two momenta ${\bf k}_1, {\bf k}_2$.

One thus finds that, while for a system of charged pions (i.e. a mixture of 
$50\%$ 
positive and $50\%$ negative) the maximum value of this 
intercept $MaxC_2(k,k)$ is 1.5, for
photons $MaxC_2(k,k)$ exceeds this value and this excess 
reflects the space-time properties (represented by
$\tilde{f}(k)$)
, the degree of (an)isotropy of the source represented by 
the quantities $\sigma$ and $\delta$, and the supplementary
degree of freedom represented by the photon spin. 

As a consequence of 
the fact that $\tilde{f}$ is a decreasing function of its 
argument, in eqs.(\ref{eq:Leo25}),(\ref{eq:Leo27}) 
the terms with $\tilde{f}(k_1+k_2)$ are in general
smaller than the terms with $\tilde{f}(k_1-k_2)$, except for
small momenta $k$.

The fact that 
the differences between charged pions and photons
 are enhanced for soft photons reminds us of a similar
effect found with neutral pions (cf. section 4.4). 
Neutral pions are in general more bunched than identically 
charged ones and this difference is more pronounced for soft
pions.
This
similarity is 
not accidental, because photons as well as
$\pi^0$ particles are neutral and this circumstance has
 quantum field theoretical implications which will be
mentioned also below.

We see thus that photon BEC can provide
information both about the space-time form of the source
represented by $f$ and
the dynamics which are represented by $\delta$
\footnote{Cf. \cite{Marques},\cite{Badala} where
 photon correlation experiments in 
 low energy (100 MeV/nucleon) heavy ion reactions are
reported. For a theoretical discussion of these experiments 
cf.\cite{Barz96}.} \footnote{The relation between
photon interferometry and the formation length of photons 
is discussed in \cite{PPT}.}. 

The results on photon correlations presented above refer to
the case that the sources are ``static" i.e. not expanding.
Expanding sources were considered in \cite{Feld} within a 
covariant formalism. 

 Some of the results above, in particular eqs.
(\ref{eq:Leo25},\ref{eq:Leo26}),
which had been initially derived by Neuhauser \cite{Neu}, were
challenged by Slotta and Heinz \cite{Heinz}. 
Among other things, these authors claim that for photon
correlations due to a chaotic
source ``the only change relative to 2-pion interferometry
is a statistical factor $\frac{1}{2}$ for the overall
strength of the correlation which results from the
experimental averaging over the photon spin". In \cite{Heinz}
an intercept $\frac{3}{2}$ is derived which is 
in contradiction with the results presented above
and in particular with eq.(\ref{eq:Leo26}) where besides the 
factor $\frac{3}{2}$ there appears also the $k$ dependent 
function $\frac{1}{2}{|\tilde{f}(2k)|^2}$. Similar
statements can be found in previous papers \cite{kap},
\cite{kap1}, \cite{kap2}, \cite{Sriv} where 
more detailed
applications concerning heavy ion reactions based on this
assertion of \cite{Heinz} are presented. 
 
Some of the papers quoted above were criticized
 immediately after their publication in \cite{bounds}, 
\cite{Axel} \cite{Feld}
and 
the paper \cite{Heinz} was written with the intention 
to ``settle" this ``controversy". 

It should be pointed out here that the reason for
the difference between the results of \cite{Neu},\cite{Leo}
on the one hand and those of ref.\cite{Heinz} on the other
is mainly due to the fact that in 
\cite{Heinz} a formalism was used which is less general
than that used in \cite{Neu} and \cite{Leo} and which is 
inadequate for the present problem. This implies among 
other things that unpolarized photons cannot be treated in
the way proposed in \cite{Heinz} and that the results
of \cite{Neu} and \cite{Leo} are correct, while the results
of \cite{Heinz} are not.

 In 
\cite{Heinz} 
the following formula for the second order correlation 
function is used: 
\be
C({\bf k}_1,{\bf k}_2)=1+\frac{\tilde {g}_{\mu \nu}(
{\bf q,K})\tilde {g}^{\nu \mu}({\bf -q,K})}
{\tilde {g}^{\mu}_{\mu}({\bf 0,k}_1)\tilde {g}^{\mu}_{\mu}(
{\bf 0,k}_2)}
\label{eq:Heinz23}
\ee
Here $\tilde{g}$ is the Fourier transform of a source function, ${\bf q}={\bf k}_1-
{\bf k}_2$ and $ {\bf K}=\frac{1}{2}({\bf k}_1+{\bf k}_2)$. 
This formula is a particular case of a more general formula
for the second order correlation function derived
 by Shuryak \cite{shuryak}
using a model of uncorrelated sources, when 
 emission of particles from the same space-time point is
negligible (cf. section 4.4.1).

Since this equation is sometimes used in the recent literature
without giving the reader the possibility of evaluating the
 approximations used in its derivation, we will sketch this
derivation in the following.

In \cite{shuryak} one starts with the current correlator
\be
<J^{*}_i(x_1)J_j(x_2)>=\delta_{ij}J_i(x,\Delta x),
\label{eq:shuryak1}
\ee
where $J_i(x)$ is the current emitted by point $x$ and
\ba
x=(x_1+x_2)/2,\qquad \Delta x=x_1-x_2.
\ea

Eq. (\ref{eq:shuryak1}) assumes that the individual currents
$(i \neq j$) are uncorrelated. 
With the notation $\tilde{I}_{i}(q,K)$ for the Fourier transform of
$I_{i}(x, \Delta x)$ and
$\tilde{I}(q,K)\equiv\sum_i\tilde{I}(q,K)$
the inclusive single particle distribution reads
\be
W(k)=\left<\left|\sum_j\int e^{ikx}J_j(x)d^4x\right|\right>=
\sum_i\tilde{I}_i(0,k)=\tilde{I}(0,k)
\label{eq:shuryak3}
\ee
and the two particle distribution is given by
\be
W(k_1k_2)=\left<\left|\sum_{i,j}\int(e^{ik_1x_1+ik_2x_2}+
e^{ik_1x_2+ik_2x_1})J_i(x_1)J_j(x_2)dx_1dx_2\right|^2\right>
\label{eq:shuryak4}
\ee
 
or finally
\begin{eqnarray}
W(k_1,k_2) &=& \tilde{I}(0,k_1)\tilde{I}(0,k_2)+
|\tilde{I}(q,K)|^2\nonumber\\
& & +\sum_{i}[<J^{*}_{i}(k_1)J^{*}_{i}(k_2)J_{i}(k_1)J_{i}
(k_2)>
\nonumber\\
& & -\tilde{I}_i(0,k_1)\tilde{I}_i(0,k_2)-
|\tilde{I}_i(q,K)|^2]
\label{eq:shuryak5}
\end{eqnarray}

The first term on the rhs of eq.(\ref{eq:shuryak5}) is the
product of one-particle distributions and the second term
is the conventional interference term, corresponding to
figure 6c. By going over to the Wigner source function 
(cf. also section 4.9) 
\be
g_{\mu\nu}(x,K)=\int
d^4ye^{-iKy}<J^{*}_{\mu}(x+\frac{1}{2}y)J_{\nu}(x-\frac{1}{2}
y)>
\label{eq:heinz21}
\ee
these first two terms result in eq.(\ref{eq:Heinz23}) of 
\cite{Heinz}. However as is clear from eq.(\ref{eq:shuryak5})
there exists also a third term, neglected in
eq.(\ref{eq:Heinz23}) and which corresponds to the 
simultaneous  emission of two particles from a single point
($x_i$) as indicated in figure 6d.  While for massive
particles this term is in general suppressed, this is not
true for massless particles and in particular for soft
photons. In \cite{Neu} and \cite{Leo} this additional
term had not been neglected as it was done in \cite{Heinz}  
and therefore it is not surprising that
ref.\cite{Heinz} could not recover the results of
refs.\cite{Neu} and \cite{Leo}. The neglect of the term
corresponding to emission of two particles from the same
space-time point 
 is not permitted in the present case.
 As mentioned in section 4.4.1, in a model of
uncorrelated point-like random sources like the present one,
emission of particles from the same space-time point  
corresponds in a first approximation to
particle-antiparticle correlations and this type of effect
leads also to the difference between BEC for identical 
charged pions and the BEC for neutral pions.
This is so because neutral particles coincide with the
corresponding antiparticles. (As a consequence of this
circumstance e.g.
while for charged pions the maximum of the intercept is 2, for neutral pions
it is 3 (cf. section 4.4)). Photons being neutral particles,
similar effects like those observed for $\pi^0$-s are
expected and indeed found (cf. above).

 This misapplication of the current formalism  
 invalidates completely the conclusions of 
ref.\cite{Heinz} and confirms and strengthens the criticism 
expressed in  \cite{bounds}, \cite{Axel}, \cite{Feld} of the
 papers \cite{kap}-\cite{Sriv}. 
 The fact that for unpolarized
photons  $MaxC_2(k,k)$ is 2 and not 1.5 as stated in
\cite{Heinz}, can be understood by realizing that  
a system of 
unpolarized photons consists on the average of $50\%$
photons with the same helicities and $50\%$ photons with
opposite helicities. The first ones contribute to the
maximum intercept (of the unpolarized system) 
with a factor of $3$ and the last ones with a factor of 1
(coresponding to unidentical particles).

For the sake of clarification it must be mentioned that ref.
\cite{Heinz} contains also other incorrect statements. Thus   
the claim in \cite{Heinz} that the approach by Neuhauser
``does not correctly take into account the constraints from
current conservation" is completely unfounded
 as can be seen from eq.(\ref{eq:sum}) which is 
an obvious consequence of current conservation (cf.e.g.
eq.(7.61) in \cite{BD}). Last but not least the statement 
that because the tensor structure in eq.(20) of
ref.\cite{Feld} is parametrized
in terms of $k_1$ and $k_2$ separately ``instead of only in
terms of $K$, leading to spurious terms in the tensor
structure which eventually result in their spurious
momentum-dependent prefactor" has also to be qualified. As
mentioned above, eq.(\ref{eq:Heinz23}) to which this
observation about the $K$ dependence of \cite{Heinz} refers 
is not general enough for the problem of photon 
interferometry.

\subsection{Coherence and lower bounds of Bose-Einstein 
correlations} 

We mentioned in the previous subsections that the intercepts
of the second and higher order correlation functions
can deviate from the canonical values derived within the
wave function formalism. This effect is important for at
least three reasons:

(i) it illustrates the limitations of the wave function 
approach 

(ii) it can in principle  
(provided other effects like final
state interactions are taken into account) be used for 
the determination of the degree of coherence.

(iii) it can serve  
as a test of models of BEC, since the value of the
intercept follows from very general quantum statistical
considerations, in particular the Gaussian nature of the
density matrix. 

In most BEC models the intercept is
identical to the maximum of the correlation function
and therefore it can be studied by limiting the discussion
to chaotic sources as was done in the previous subsection 
where the upper bounds of correlation functions were
investigated.
On the other hand the minimum of the correlation functions
is determined both by the form of the density matrix and 
the amount of coherence (cf.ref.(\cite{bounds})) because
coherence leads to a decrease of the correlation function. 
This will be illustrated below by discussing the lower 
bounds of this function.
We will show among other things (cf. \cite{APW}) that in the quite general 
case of a Gaussian density matrix, for
  a purely chaotic system the two-particle 
correlation function must always be greater than one.  
On the other hand, in the presence
of a coherent component the correlation function may 
take values 
below unity. Some implications for experimental and 
theoretical results 
found in the literature will be discussed here as well as in
section 5.1.6.

We have seen in section 4.4 that 
for identically charged bosons (e.g., $\pi^+$) the two-
particle correlation function reads 
%\footnote{The effects of conservation of isospin on the
%coherent component (cf. section 4.4) are neglected here.}
\begin{equation}
C^{++}_2({\bf k}_1,{\bf k}_2)\ =\ 1 +\ 2 \sqrt{p_1(1-p_1)
\cdot p_2(1-p_2)}
\ T_{12}\  \cos(\phi^{ch}_{12} - \phi^c_1+\phi^c_2)\ 
+\  p_1p_2 \ T^2_{12}
\label{eq:c2plus}
\end{equation}

For neutral bosons like photons, or $\pi^0$'s,  the terms 
$\tilde{d}(k_r,k_s)$ also appear
(cf. eq:(\ref{eq:drs})
in the BEC function:
\begin{eqnarray}
C^{00}_2({\bf k}_1,{\bf k}_2) & = &  1  +  
2 \sqrt{p_1(1-p_1)\cdot p_2(1-p_2)}
\ T_{12}\  \cos(\phi^{ch}_{12} - \phi^c_1+\phi^c_2)\ 
+\   p_1p_2 \ T^2_{12}\nonumber\\
& & + 2 \sqrt{p_1(1-p_1)\cdot p_2(1-p_2)}
\ \tilde{T}_{12}\  \cos(\tilde{\phi}^{ch}_{12} - \phi^c_1-
\phi^c_2)
\ + \  p_1p_2 \  \tilde{T}^2_{12}\nonumber\\
&&
\label{eq:cnulnul}
\end{eqnarray}

Let us first consider the case of a purely chaotic source. 
Insertion   of  $p(k) \equiv 1$ in eqs. (\ref{eq:c2plus}) 
and (\ref{eq:cnulnul})
immediately yields $C_2({\bf k}_1,{\bf k}_2)\geq 1$. 
In the case of
partial coherence, 
the terms containing cosines come into play 
and 
consequently $C_2$ may take values below unity. 
Eqs. (\ref{eq:c2plus},\ref{eq:cnulnul}) imply that 
$C^{--}_2({\bf k}_1,{\bf k}_2)\geq 2/3$ and 
$C^{00}_2({\bf k}_1,{\bf k}_2)\geq 1/3$.
Because of the 
cosine functions 
in (\ref{eq:c2plus},\ref{eq:cnulnul})
one would expect
$C_2$ as a function of the momentum difference $q$ to 
oscillate between values above and below $1$. 
 
\begin{footnotesize}Such a 
behaviour of the Bose-Einstein correlation function has been
observed in high 
energy $e^{+}-e^{-}$ collision experiments (cf., e.g., ref. 
\cite{osc}), 
but apparently not in hadronic reactions. This observation 
was interpreted as a consequence of final state interactions
in ref. \cite{bowler}. If final state interactions determine
this effect, it is unclear why the effect is not seen in 
hadronic reactions. On the other hand, if coherence is 
responsible for it, this would be 
easier to understand. Indeed multiplicity distributions of 
secondaries in $e^{+}e^{-}$ reactions are much narrower (almost 
Poisson-like) than in $pp$ reactions, which is consistent 
with the statement that 
hadronic reactions are more chaotic than $e^{+}-e^{-}$ reactions 
\cite{fried}\footnote{This statemant is not necessarily in
contradiction with the empirical observation that the
$\lambda$ factor in $e^{+}-e^{-}$ reactions appears in general
to be larger than in $p-p$ reactions, given the fact that
$\lambda$ is not a true measure of coherence.}.
\end{footnotesize}
 
So far, two methods have been proposed for the detection of 
coherence in BEC: the intercept criterion \cite{gnf} 
($C_2({\bf k},{\bf k})<2$) and the 
two exponent structure of $C_2$ \cite{revisit}. Both these 
methods have their difficulties because of statistics 
problems or other  
effects. The observation
of $C_2({\bf k}_1,{\bf k}_2)<1$  could constitute a third 
criterion for
coherence.

 In \cite{kap},\cite{kap1} the two-particle correlation 
function has been 
calculated for photons emitted from a longitudinally expanding
system of hot and dense hadronic matter created in 
ultra\-rela\-ti\-vist\-ic nuclear collisions . 
For such a system, the particles are 
emitted from a large number of independent source elements 
(fluid elements), and consequently one would expect the 
multiparticle 
final state to be described by a Gaussian density matrix. 
However, although the system is assumed to be purely chaotic 
the 
correlation function calculated in \cite{kap} is 
found to take values significantly below unity. 
Clearly, this is in contradiction with the general result
derived above from quantum statistics 
($C_2\geq 1$ for a chaotic system). 

 The reason for this violation of the general bounds 
derived for a purely chaotic source is in this concrete case
the use of an inadequate approximation in the evaluation of 
the space-time integrals. However  
as pointed out in \cite{Axel} the  
expression for the two-particle inclusive distribution
used in ref. \cite{kap} (equation (3) of that paper), 
which in our notation
takes the form 
\begin{equation}
P_2({\bf k}_1,{\bf k}_2) \ =\ \int d^4x_1 \int d^4x_2\ 
g\left(x_1,k_1\right)\ 
g\left(x_2,k_2\right)\ [1\ +\ \cos((k_1-k_2)(x_1-x_2))],  
\label{eq:3}
\end{equation}
is also unsatisfactory 
\footnote{This formula appears apparently for the first time
in \cite{Yano} and was criticized (for other reasons) 
already in \cite{PaGyGa}. It is nevertheless used in certain
event generators for heavy ion reactions (cf. section 4.10).}
because for certain physical 
situations it can lead to 
values below unity for the two-particle
correlation function even if the integrations are performed
exactly.  To see this, consider, e.g., the 
simple ansatz
\begin{equation}
g(x,k)\ =\ const. \ 
\exp[-\alpha({\bf x}-\beta {\bf k})^2]\ \delta(t-t_0) 
\end{equation}
where $\alpha$ and $\beta$ are free parameters.
The expression for $P_2({\bf k}_1,{\bf k}_2)$ used in 
ref.\cite{kap} then yields
\begin{equation}
C_2({\bf k}_1,{\bf k}_2) \ =\ 1 \ +\ \exp\left[-\frac
{{\bf q}^{\ 2}}
{2\alpha}\right] \ \cos[\beta {\bf q}^{\ 2}]
\end{equation}
Clearly, if $\beta$ exceeds $\alpha^{-1}$ the above 
expression will 
oscillate and take values below unity. On the other 
hand, in the current formalism (cf. below) one obtains with 
the same ansatz for $g$
\begin{equation}
C_2({\bf k}_1,{\bf k}_2) \ =\ 1 \ +\ \exp\left[-\frac
{{\bf q}^{\ 2}}{2\alpha}\right] \ \geq \ 1.
\end{equation}

Thus, eq. (\ref{eq:3}) can lead to values $C_2<1$ if there 
is a strong correlation between the momentum of a particle 
and the space-time coordinate of the source element from 
which it is emitted. 
Such correlations between $x$ and $k$ can occur in the case
of an expanding source. The reason for this pathological
behaviour is the fact that the simultaneous specification of
coordinates and momentum as implied by eq.(\ref{eq:3}) is
constrained in quantum mechanics by the Heisenberg
incertainty relation and any violation of this constraint
leads necessarily to a violation of quantum mechanics. This
violation manifests itself sometimes, as in the present case,
 through a violation of
the conservation of probability. This phenomenon is also met
when using the Wigner function, which for this reason cannot
always be associated with a bona fide probability amplitude.
We will
discuss this problem also in section 5.1.6.

The above considerations concerning bounds 
for the BEC functions refer to the case of a Gaussian 
density matrix. In general, a different form of the 
density matrix may yield correlation functions that 
are not constrained by the bounds derived here. For 
instance we have seen in section 2.2  that for a
 system of squeezed states $C_2$ can take 
arbitrary positive values. Moreover, for 
particles produced in high energy hadronic or nuclear
collisions, the fluctuations of quantities such as impact 
parameter or inelasticity may introduce additional 
correlations which may also affect the bounds of the BEC 
functions.

\subsection{Quantum currents}

The results derived in the previous section, in particular
the isospin dependence of BEC, were obtained in the
assumption that the currents
were classical. 
The question arises up to what point these conclusions
survive in a fully quantum treatment of the problem. It
would also be important to get a more precise criterium for 
the
phenomenological applicabilbity of the classical assumption,
besides the
no-recoil prescription. This question was discussed in 
\cite{LeoQFT} where it was found that the ``surprising"
effects not only persist when the currents are quantum, but
that they can serve as an experimental estimate of the
size of the quantum corrections. We shall sketch briefly 
in the following the results of ref.(\cite{LeoQFT}).

As in the classical current case one starts with the
interaction Lagrangian    

\be 
L_{int}(x)\ \equiv \ J_{(+)}(x)\pi^{(-)}(x)\ +\ J_{(-)}(x)\pi^{(+)}(x)
\ +\ J_0(x)\pi^0(x)
\label{eq:L_def1}
\ee
The currents $J^{(+)},\; J^{(-)},\; J^{0}$ are 
{\em operators} which we assume again for simplicity not to
depend on the $\pi^{(\pm)}$  and $\pi^0$ 
fields \mbox{$([J,\pi]=0)$}. 
Taking into account the different isospin components in
eqs.(\ref{eq:P1JH}) and (\ref{eq:P2JH}) we find 
that as in the classical current case the single and double inclusive
cross sections depend on these components, 
e.g.\footnote{For reasons of notational simplicity we have
replaced $J^{\dagger}(k)$ by $J(-k)$.}

\ba
G^{(-)}_1(\mbold{k})\;&=&\; Tr\left\{\rho_iJ_H^{(+)}(-k)
 J_H^{(-)}(k)\right\} \label{eq:P1JH_m} \\
G^{(-+)}_2(\mbold{k}_1\,,\,\mbold{k}_2)\;&=&\;
Tr\left\{ \rho_i {\bf \tilde{\cal T}}
\left[ J_H^{(+)}(-k_1)J_H^{(-)}(-k_2)\right]
{\cal T}\left[ J_H^{(-)}(k_1)J_H^{(+)}(k_2)\right]\right\} 
\label{P2JH_mp}
\ea

>From now on we shall omit the label $H$ and assume that all
operators are written in the Heisenberg representation.
Assuming a Gaussian density matrix one gets
\ba
G^{(0)}_1(\mbold{k})\ &=&\  F^n(k,k)
                                                                                        \label{P0:F} \\
G^{(-)}_1(\mbold{k})\ &=&\ G^{(+)}_1(\mbold{k})\ =\ F^{ch}(k,k)
                                                                                        \label{P-:F}\\
G^{(--)}_2(\mbold{k}_1,\mbold{k}_2)\ &=&\
G^{(-)}_1(\mbold{k}_1) G^{(-)}_1(\mbold{k}_2) + |F^{ch}(k_1,k_2)|^2
                                                                                        \label{P--:F}\\
G^{(00)}_2(\mbold{k}_1,\mbold{k}_2)\ &=&\
G^{(0)}_1(\mbold{k}_1) G^{(0)}_1(\mbold{k}_2) + |F^{n}(k_1,k_2)|^2
+ |\Phi^{n}(k_1,k_2)|^2
                                                                                        \label{P00:F}\\
G^{(-+)}_2(\mbold{k}_1,\mbold{k}_2)\ &=&\
G^{(-)}_1(\mbold{k}_1) G^{(+)}_1(\mbold{k}_2) +
|\Phi^{ch}(-k_1,k_2)|^2
                                                                                        \label{P-+:F}
\ea
where the functions $F$ and $\Phi$ are defined for charged particles (upper index
$ch$) and for neutral ones (upper index $n$) as follows:
\ba
F^{ch}(k_1,k_2)\ &\equiv&\ \ll\!\! {\cal T}_c
\left\{ J^{(+)}_{\oplus}(-k_1)J^{(-)}_{\ominus}(k_2) \right\} \!\!\gg
\ =\ \ll\!\! J^{(+)}(-k_1)J^{(-)}(k_2) \!\!\gg
                                                \label{Fch} \\
\Phi^{ch}(k_1,k_2)\ &\equiv&\ \ll\!\! {\cal T}_c
\left\{ J^{(+)}_{\oplus}(-k_1)J^{(-)}_{\oplus}(k_2) \right\} \!\!\gg
\ =\ \ll\!\! {\bf \tilde{\cal T}}
\left\{ J^{(+)}(-k_1)J^{(-)}(k_2) \right\} \!\!\gg
                                                                                                \label{Fch:t} \\
F^n(k_1,k_2)\ &\equiv&\ \ll\!\! {\cal T}_c
\left\{ J^{(0)}_{\oplus}(-k_1)J^{(0)}_{\ominus}(k_2) \right\} \!\!\gg
\ =\ \ll\!\! J^{(0)}(-k_1)J^{(0)}(k_2) \!\!\gg
                                                \label{Fn} \\
\Phi^{n}(k_1,k_2)\ &\equiv&\ \ll\!\! {\cal T}_c
\left\{ J^{(0)}_{\oplus}(-k_1)J^{(0)}_{\oplus}(k_2) \right\} \!\!\gg
\ =\ \ll\!\! {\bf \tilde{\cal T}}
\left\{ J^{(0)}(-k_1)J^{(0)}(k_2) \right\} \!\!\gg
                                                                                                \label{Fn:t}
\ea
In contrast to the classical current approach 
\cite{APW} which
deals with only one type of two-current correlator we have here two
different kinds of two-current correlators depending on their ordering
prescriptions. Moreover and most remarkably, the difference
between these two correlators is reflected by the difference
between $--$ and $+-$ correlations. It thus follows from
\cite{LeoQFT} that 
the ``surprising'' effects found in \cite{apw}
i.e. the presence of particle-antiparticle Bose-Einstein type
correlations and
a new term in the Bose-Einstein correlation function for neutral
particles
are reobtained, but under a more general form which contains
also the quantum corrections.
These equations also prove that the above effects 
 are not an artifact of the classical current
formalism but have general validity.

Morover and most remarkably, from the above equations
follows that {\em the difference between the effects of the
classical and quantum currents resides in just these ``new
effects"  and in particular in the difference between $00$
and $--$ correlations i.e. in the $+-$ correlations}.   
This result can serve as an estimate of the importance of
quantum corrections to the classical current formalism of
BEC. Since $+-$ correlations are in general small, it
follows that the classical current approach is a good
approximation, except for very short lived sources, where
the $+-$ correlations become comparable to the $--$
correlations. 
It also follows that the
experimental
measurement of $+-$ correlations is a highly rewarding task,
since they are a rather unique tool for the investigation of
two very interesting effects in BEC, namely 
squeezed states and quantum corrections.

\subsection{Space-time form of sources in the classical 
current formalism}

In \cite{APW} two types of sources were considered, 
a ``static" one which corresponds to a source in rest
and an expanding one. We will present below some of the
results, as they exemplify certain important features of the
space-time approach within the classical current formalism.

\paragraph{A static source}
The space-time distributions of  static sources, as well 
as the primordial correlator, are parametrized as Gaussians:
\begin{eqnarray}
f_{ch}(x) & = & \exp\left(-\frac{x_0^2}{R_{ch,0}^2}
-\frac{x_{\parallel}^2}{R_{ch,\parallel}^2}
-\frac{x_{\perp}^2}{R_{ch,\perp}^2}\right)
\label{eq:fstat1}\\
f_{c}(x) & = & \exp\left(-\frac{x_0^2}{R_{c,0}^2}
-\frac{x_{\parallel}^2}{R_{c,\parallel}^2}
-\frac{x_{\perp}^2}{R_{c,\perp}^2}\right)
\label{eq:fstat}
\end{eqnarray}
and
\begin{equation}
C(x-y)\ =\ \exp\left[-\frac{(x_0-y_0)^2}{2L_0^2}
-\frac{(x_{\parallel}-y_{\parallel})^2}{2L_{\parallel}^2}
-\frac{({\bf x}_\perp-{\bf y}_\perp)^2}{2L_{\perp}^2}\right]
\label{eq:cxmy}
\end{equation}

Note that the term {\em static} here does
not imply time independence but
rather a specific time dependence defined by eqs. 
(\ref{eq:fstat1}),
(\ref{eq:fstat}) corresponding to source elements being at 
rest. This is to be contrasted to the {\em expanding} source,
discussed in the next section,
which explicitly contains velocities of source elements.

The main justification for this particular form of 
parametrization is mathematical convenience, because, as 
will be shown below, for this case the correlation functions
in momentum space can be calculated analytically and
the physical implications can be read off immediately.

In eqs.(\ref{eq:fstat1}-\ref{eq:cxmy}), 
$R_{ch,\alpha}$ and $R_{c,\alpha} (\alpha = 0,\perp, 
\parallel)$
are the lifetimes, transverse radii and longitudinal radii of
the chaotic source and of the coherent source, respectively,
and $L_{\alpha} (\alpha = 0,\perp, \parallel)$ are the 
correlation time and the corresponding correlation lengths in
transverse and in longitudinal direction.
The relative contributions of the chaotic and the coherent
component are determined by fixing the value of the (momentum
dependent) chaoticity parameter $p$ at some arbitrary scale 
(in this case, at $k=0$): 
\begin{equation}
p_0 \ \equiv \ p(k=0)
\label{eq:pop}
\end{equation}
The model contains 10 independent parameters: the radii and lifetimes of the chaotic and of the coherent 
source, the correlation lengths in space and time, and the 
chaoticity $p_0$. In \cite{APW} it is assumed that 
$L_{\parallel} = L_{\perp}\equiv L$, i.e., that the medium is
isotropic, which leaves us with 9 independent parameters.

With the definitions
\begin{equation}
R_{\alpha L}^2\ =\ \frac{R_{ch,\alpha}^2 L_{\alpha}^2}
{R_{ch,\alpha}^2 + L_{\alpha}^2} \qquad (\alpha = 0, \perp ,
\parallel )
\label{eq:rperp}
\end{equation}
one may write the single inclusive distribution in the form
\begin{equation}
E \frac{1}{\sigma} \frac{d^3 \sigma}{d^3k}\ = \ \left. E\frac{1}{\sigma}\frac
{d^3\ \sigma}{d^3 k}  \right| _{k=0} 
\left( p_0\ s_{ch}(k)\ +\ (1-p_0)\ s_c(k) \right)
\label{eq:eins}
\end{equation}
where
\begin{equation}
s_{ch}(k) \ = \ \exp \left[-\frac{E^2 R^2_{0L}}{2}
-\frac{k_{\parallel}^2 R^2_{\parallel L}}{2} 
- \frac{{\bf k}_{\perp}^2 R^2_{\perp L}}{2}\right]
\label{eq:schak}
\end{equation}
and
\begin{equation}
s_{c}(k) \ = \ \exp \left[-\frac{E^2 R^2_{c,0}}{2}
-\frac{k_{\parallel}^2 R^2_{c,\parallel}}{2} 
- \frac{{\bf k}_{\perp}^2 R^2_{c,\perp}}{2}\right]
\label{eq:schuk}
\end{equation}
The scales which determine the mean 
energy-momentum of the coherently produced particles
are given by the inverse lifetime and radii, $R^{-1}_{c,
\alpha}$, of the coherent source. For the chaotically 
produced particles, these scales are given by the inverse of
a combination of correlation lengths and dimensions of the 
chaotic source, $R^{-1}_{\alpha L}$. Eq.(\ref{eq:rperp}) 
implies that
$R_{\alpha L} \leq R_{ch,\alpha}$. 
The radius
of the chaotic source enters the single inclusive distribution only in
combination with the correlation length
 $L$. This feature which 
 occurs also for higher order correlations leads to the important
consequence that experimental measurements of BEC do not provide 
separately information about radii (lifetimes) of sources, nor about correlation-lengths
(-times), but rather about the combination of these quantities as given by
eq. (\ref{eq:rperp}). On the other hand, by measuring both the single
and the double inclusive distribution one can determine radii and
correlation lengths separately.

It follows from eqs.(\ref{eq:eins},
\ref{eq:schak},\ref{eq:schuk}) that in the presence of partial coherence 
in general (i.e., unless $R_{\alpha L} = R_{c,\alpha}$)
the single inclusive distribution is a superposition of two Gaussians 
of different widths.
If the geometry of the coherent source is the same as that of the chaotic 
source, one has $R_{c, \alpha} = R_{ch, \alpha} > R_{L \alpha}$, which would
imply that coherently produced particles can be observed predominantly in
the soft regime. However, if the coherent radii are small compared to the
chaotic ones, this situation is reversed.

As a next step, consider the correlation functions. 
The correlation function 
of two negatively charged pions is 
\begin{equation}
C^{--}_2(k_1,k_2)\ =\ 1 +\ 2\sqrt{p_1(1-p_1) \cdot p_2(1-p_2)}
\ T_{12}\  \cos(\phi^{ch}_{12} - \phi^c_1+\phi^c_2)\ 
+\ p_1p_2\ T^2_{12}
\label{eq:non}
\end{equation}
For the Gaussian parametrizations all phases in the second
order correlation function disappear
\footnote{This is not the case anymore for an expanding
source e.g. (cf. below) or in general for higher order 
correlations},
\begin{equation}
\phi_{12}^{ch}\  =\ \tilde{\phi}_{12}^{ch}\ =\ 
 \phi_{j}^{c}\ =\ \tilde{\phi}_{j}^{c}\ =\ 0 
\label{eq:fis}
\end{equation}
and
\begin{eqnarray}
T_{12} & = & \exp \left[-\frac{(E_1-E_2)^2(R^2_{ch,0}-R^2_{0L})}{8}
-\frac{(k_{1,\parallel}-k_{2,\parallel})^2(R_{ch,\parallel}^2-R^2_{\parallel L}
)}
{8}\right.\nonumber\\
&& \qquad  \left. 
- \frac{({\bf k}_{1\perp}-{\bf
k}_{2\perp})^2(R^2_{ch,\perp}-
R^2_{\perp L})}{8}\right]
\label{eq:fiss}
\end{eqnarray}
\begin{eqnarray}
\tilde{T}_{12}  & = & \exp \left[-\frac{(E_1+E_2)^2(R^2_
{ch,0}-R^2_{0L})}{8}
-\frac{(k_{1,\parallel}+k_{2,\parallel})^2(R_{ch,\parallel}^2-
R^2_{\parallel L})
}{8} \right. \nonumber\\
&& \qquad \left. - \frac{({\bf k}_{1\perp}+{\bf k}_{2\perp})^2
(R^2_{ch,\perp}-R^2_{\perp L})}{8}\right]
\label{eq:fiat}
\end{eqnarray}
The two particle correlation function $C_2^{--}$ is the sum
of a purely chaotic term $(\propto \ T^2_{12})$ and a mixed term
$(\propto \ T_{12})$. The momentum dependence of the chaoticity parameter,
$p=p(k)$, implies a momentum dependence of the contribution of the mixed
term relative to that of the purely chaotic term. To see how this 
affects the interplay between the two terms (i.e., the interplay between 
the two Gaussians), it is useful to explicitly insert the momentum 
dependence of the chaoticity parameter by writing
\begin{equation}
p_r \ =\  p(k_r)\ =\ \frac{p_0}{A_r} \qquad (r=1,2)
\label{eq:pr1}
\end{equation}
with
\begin{equation}
A_r \  \equiv \ A(k_r) \ =\  p_0 + (1-p_0) \ S_{rr} \qquad (r=1,2)
\label{eq:akr}
\end{equation}
and
\begin{eqnarray}
S_{rs} & = & \exp \left[-\frac{(E_r^2+E_{s}^2)(R^2_{c,0}-R^2_{0L})}{4}
-\frac{(k_{r\parallel}^2+k_{s\parallel}^2)
(R_{c,\parallel}^2-R^2_{\parallel L}
)}
{4}\right. \nonumber\\
&& \qquad  \left. - \frac{({\bf k}_{r\perp}^2+{\bf k}_
{s \perp}^2)(R^2_{c,\perp}-R^2_{\perp L})}{4}\right]
\label{eq:swelve}
\end{eqnarray}
With this, $C_2^{--}$ takes the form
\begin{equation}
C^{--}_2(k_1,k_2)\ =\ 1 +\ \frac{2p_0(1-p_0)S_{12}}{A_1 A_2}
\ T_{12}\  
+\  \frac{p_0^2}{A_1 A_2}\ T^2_{12}
\label{eq:cett}
\end{equation}

The momentum-dependence of the relative contributions of the purely chaotic
and of the mixed term is reflected in the factor $S_{12}$. Depending on
the sign of the combinations $R^2_{c,\alpha} - R^2_{\alpha L}$, 
$\ S_{12}$ may act
either as a suppression factor or as an enhancement factor of the mixed term
relative to the chaotic term. This is a consequence of the
fact that, in 
contrast with the case of the correlation function $C_2$ derived
within the wave function formalism, where the $C_2$ depends
only on the difference of momenta $k_1-k_2$ 
now the correlation function depends also on $k_1+k_2$.
\footnote{Up to recently this desirable physical property, which is
observed in most experimental data on BEC, was considered to
be a consequence of the {\em expansion} of the source and
used to be derived within the Wigner function formalism, 
which is also a particular case of the 
classical current formalism.) As
shown in the example treated above (cf.\cite{aw})  
it can be considered also a consequence of the (partial) 
coherence of a non-expanding source.} 

It is instructive to discuss the tilde terms that give rise to the 
particle-antiparticle correlations for the parametrization  
(\ref{eq:fstat1}-\ref{eq:cxmy}) of a 
static source. For the sake of transparency, consider only the purely 
chaotic case, $p_0 = 0$. The correlation functions of like and unlike 
charged pions then take the form
\begin{eqnarray}
C_2^{--} (k_1,k_2) &=& 1 + T^2_{12}\\
C_2^{+-} (k_1,k_2) &=& 1 + \tilde{T}^2_{12}
\label{eq:tills}
\end{eqnarray}
>From (\ref{eq:fiss}) and (\ref{eq:fiat}) it can be seen that
the ``new" $\tilde{T}_{12}$ terms that appear in 
the particle-antiparticle correlations are in general small 
compared to the ``ordinary" $T_{12}$ terms that
determine the particle-particle correlations.
The term $\tilde{T}_{12}$ gives rise to an anticorrelation 
effect due to the factor in eq. (\ref{eq:fiat}) containing 
the sum ${\bf k}_1 + {\bf k}_2$, if the first factor, 
containing $E_1+E_2$, is not too small. The latter is 
possible, if the time duration of the 
pion emission process and/or the pion energies are small 
enough. We thus 
expect an enhanced contribution of the ``new" terms for 
 soft pions. The appearance of anti-correlations 
\footnote{Some authors have recently called them
``back to back" correlations.}
is, as mentioned above, a general property of 
squeezed states, which are present in
the space-time formalism of \cite{APW}. 
The tilde terms arise as a consequence of the 
non-stationarity
of the source. In the limit of a stationary source, $R_0 
\rightarrow \infty$, and $\tilde{T}_{12} \rightarrow 0$. An 
upper limit is given by
\begin{equation}
C_2^{+-} \ = \ 1 + |\tilde{T}_{12}|^2 
\ \leq \ 1 + \exp\left[ -(R_0^2 - R^2_{0L}) m^2_\pi\right]
\label{eq:rol}
\end{equation}
In the limit $L_0 >> R_0$ on the other hand, $R_0 \simeq 
R_{0L}$ and $C_2^{+-}$ reaches its maximum value
$(C_2^{+-})_{max} = 2$.
We observe in the above equation that the contribution of the 
``tilde" terms increases with decreasing mass of the
particles
and reaches its maximum of $2$ for massless particles at 
fixed and non-vanishing $R_0-R_{0L}$. This is related to
the observation made in the case of photon BEC 
where we saw
that for unpolarized photons the maximum of $C_2$ is also
$2$. Indeed the role of the charge degrees of freedom
($(+,-$) for pions is for unpolarized photons taken over by
the spin. From this mass dependence or more general from the
energy dependence of the
anticorrelation follows that if the factor $E_1+E_2$
in eq.(\ref{eq:fiat}) could be decreased, an enhancement of
the ``new" terms would emerge. A possible mechanism for this
could be the sudden transition mechanism considered in
\cite{sqIgor}. Indeed as shown in this
reference, for a chaotic source the correlator 
$<a({\bf k}_1)a({\bf k}_2)>$ characteristic for the ``new" 
terms turns out to be an increasing function of the parameter 
$r=\frac{1}{2}\log(E_a/E_b)$ where $E_a, E_b$ are the
energies of the particle in the vacuum and medium 
respectively. 
Thus by
allowing for medium effects, which in a certain sense is
equivalent to an effective change of mass
\footnote{This particular possibility was suggested in 
\cite{Asakawa1}.
 Andreev \cite{IgorMatra} suggested 
a time 
evolution scenario for the medium effect, which involves two
modes $k$ and $-k$.}, one can possibly
enhance the anticorrelation effect  \footnote{Anticorrelations in disoriented chiral
condensates are considered in \cite{Hideaki}.}in BEC.

\paragraph{Expanding source}
 High energy multiparticle dynamics suggests
that the sources of produced particles are expanding. This
property is reflected in particular in hydrodynamical models
and also in string models. In terms of the current formalism
this means that the correlators are velocity dependent.
While many of the studies of Bose-Einstein correlations 
for expanding sources 
have followed, with slight variations, 
a Wigner function type of approach, 
 the use of the Wigner approach is in general
too restrictive and is recommendable only in the case when a
full-fledged hydrodynamical description of the system is
performed. 
In the present section, following \cite{APW}, we shall 
therefore start with a more general 
discussion of the expanding source which is based on the
space-time current correlator and the space-time form of the
coherent component and which is not affected by the 
semi-classical and small $q$ approximations inherent in the 
Wigner function approach.

We introduce the variables $\tau$, $\eta$ and  $x_{\parallel}$, with
\begin{equation}
\tau \ =\ \sqrt{x_0^2-x_{\parallel}^2}, \qquad \eta \ =\ 
\frac{1}{2} \ln \frac{x_0+x_{\parallel}}{x_0-x_{\parallel}}
\end{equation}
Here $\tau$ is the proper time, $x_{\parallel}$ the coordinate
in the longitudinal direction (e.g. the collision axis in p-p
reactions or the jet axis in $e^{+}-e^{-}$ reactions) and $\eta$ 
the space-time rapidity. An ansatz 
which is invariant under boosts of the coordinate frame in
longitudinal direction will be considered (Fig.8).
Physically this ansatz is motivated
by the prejudice that the single inclusive distribution in 
rapidity is flat.
The space-time distributions of the chaotic and of the 
coherent source and the correlator are then parametrized as
\begin{eqnarray}
f_{ch}(x) & \sim & \exp\left(-\frac{(\tau-\tau_{0,ch})^2}
{(\delta \tau_{ch})^2}\right)
 \ \exp\left(
-\frac{x_{\perp}^2}{R_{ch}^2}\right)
\label{eq:chaos1}\\
f_{c}(x) & \sim & \exp\left(-\frac{(\tau-\tau_{0,c})^2}
{(\delta \tau_{c})^2}\right)
 \ \exp\left(
-\frac{x_{\perp}^2}{R_{c}^2}\right)
\label{eq:coh1}
\end{eqnarray}

\begin{eqnarray}
C(\tau_1-\tau_2,\eta_1-\eta_2,x_{\perp,1}-
x_{\perp,2}) & = & \exp  \left[-\frac{(\tau_{1}-\tau_{2})^2}
{2L_{\tau}^2}
 -\frac{2 \tau_1 \tau_2}{L_{\eta}^2}
\sinh^2\left(\frac{\eta_1-\eta_2}
{2}\right)
\right. \nonumber\\
& & \ \qquad \left.-\frac{({\bf x}_{\perp,1}-{\bf x}_{\perp,2})^2}
{2L_{\perp}^2}\right]
\end{eqnarray}
 
The model contains again 10 independent parameters:
The proper time coordinates of the chaotic and the 
coherent source, $\tau_{0,ch}$, $\tau_{0,c}$,
their widths in proper time, $\delta \tau_{ch}$ 
and $\delta \tau_{c}$, the transverse radii,
 $R_{ch}$ and  $R_{c}$, the correlation
lengths $L_\tau$, $L_{\perp}$ and $L_{\eta}$,
and the chaoticity parameter $p_0$.

In order to be able to obtain explicit expressions for the 
single inclusive distribution and the correlation functions, 
a further simplifying assumption is made,
 namely, that $\delta \tau_{ch} = \delta \tau_{c} =0$. 
Eqs.(\ref{eq:chaos1},\ref{eq:coh1}) 
then take the form
\begin{eqnarray}
f_{ch}(x) & \sim &  \delta(\tau-\tau_{0,ch})\ \exp\left(
-\frac{x_{\perp}^2}{R_{ch}^2}\right)\\
f_{c}(x) & \sim & \delta(\tau-\tau_{0,c})\
 \exp\left(
-\frac{x_{\perp}^2}{R_{c}^2}\right)
\end{eqnarray} 
Now the results no longer depend on the correlation length
$L_\tau$, and one is left with 7 independent parameters: 
$\tau_{0,ch}$, $\tau_{0,c}$,  $R_{ch}$, $R_{c}$, $L_{\perp}$, 
$L_{\eta}$ and $p_0$.

The Fourier integrations necessary to obtain $D(k_1,k_2)$ and 
$I(k)$ can be performed by doing a saddle point expansion; this
should provide a good approximation if
\begin{equation}
a_i \ \equiv \ \frac{m_{i\perp} \tau_{0,ch}}{2} \gg 1, \qquad
b \ \equiv \ \frac{\tau_{0,ch}^2}{2L_{\eta}^2} \gg 1
\label{eq:ab}
\end{equation}

With the definitions:
\begin{equation}
R_{L}^2\ \equiv\ \frac{R_{ch}^2 L_{\perp}^2}
{R_{ch}^2 + L_{\perp}^2} 
\label{eq:rlexp}
\end{equation}
\begin{equation}
\gamma_{12} \ \equiv \ \frac{\tau_{0,ch}(m_{1\perp}-m_{2\perp})}
{L_{\eta}^2 m_{1\perp}m_{2\perp}}
\qquad \tilde{\gamma}_{12} \ \equiv \ \frac{\tau_{0,ch}(m_{1\perp}+m_{2\perp})}
{L_{\eta}^2 m_{1\perp}m_{2\perp}}
\label{eq:gamma12}
\end{equation}
the single inclusive distribution can be 
written as the sum of a chaotic and a coherent term,
\begin{equation}
E \frac{1}{\sigma}  \frac{d^3 \sigma}{d^3k}\ = 
\left( p_0\ s_{ch}(k)\ +\ (1-p_0)\ s_c(k) \right)
\left. E \frac{1}{\sigma}\frac{d^3 \sigma}{d^3 k} \right|_{k=0} 
\label{eq:eins1}
\end{equation}
with
\begin{eqnarray}
s_{ch}(k) & = & \frac{m_{\pi}}{m_{\perp}} \ \exp \left[
- \frac{{\bf k}_{\perp}^2 R^2_{L}}{2}\right]\\
s_{c}(k) & = & \frac{m_{\pi}}{m_{\perp}} \ \exp \left[
- \frac{{\bf k}_{\perp}^2 R^2_{c}}{2}\right]
\end{eqnarray}
where $m_{\perp}$ is the transverse mass of the pions emitted.
The momentum dependence of the chaoticity parameter takes
the form 
\begin{equation}
p_r \ =\  p(k_r)\ =\ \frac{p_0}{A_r} \qquad (r=1,2)
\label{eq:pr11}
\end{equation}
with
\begin{equation}
A_r \  \equiv \ A(k_r) \ =\  p_0 + (1-p_0) \ S_{rr} \qquad (r=1,2)
\label{eq:akr1}
\end{equation}
and
\begin{equation}
S_{rs} \ = \ \exp \left[- \frac{({\bf k}_{r \perp}^2+{\bf
k}_{s \perp}^2)
(R^2_{c}-R^2_{L})}{4}\right]
\label{eq:sxp12}
\end{equation}
Unless $R_c=R_L$, the transverse momentum distribution is a 
superposition of two Gaussians of different widths.  
The rapidity distribution is uniform, 
$dN/dy =const.$, as a result of boost-invariance. 
In opposition to what is assumed usually in simplified
pseudo-hydrodynamical treatments, the transverse
radius of the chaotic source, $R_{ch}$, cannot be determined 
independently by measuring only the single inclusive 
distribution, as the quantity $R_L$ which sets the scale for
the mean transverse momentum of the chaotically produced 
particles is a combination of $R_{ch}$ and the correlation 
length $L_{\perp}$.

We recall that the second order correlation functions 
\begin{equation}
C^{++}_2({\bf k}_1,{\bf k}_2)\ =\ 1 +\ 2 \sqrt{p_1(1-p_1)
\cdot p_2(1-p_2)}
\ T_{12}\  \cos(\phi^{ch}_{12} - \phi^c_1+\phi^c_2)\ 
+\  p_1p_2 \ T^2_{12}
\label{eq:c2plus11}
\end{equation}
are
defined in terms of the 
magnitudes and phases of $d_{rs}$ and $\tilde{d}_{re}$,
$T_{rs}$, $\tilde{T}_{rs}$, $\phi^{ch}_{rs}$
and $\tilde{\phi}^{ch}_{rs}$ of the chaotic source as well
the phases of the coherent
component, $\phi^{c}_{r}$. The expressions for these
quantities read for an expanding source:
\begin{eqnarray}
T_{12} & = & \left(1+\gamma_{12}^2\right)^{-1/4}
\exp \left[-\frac{b}{1+\gamma_{12}^2}(y_1-y_2)^2
- \frac{({\bf k}_{1\perp}-{\bf k}_{2\perp})^2(R^2_{ch}-R^2_
{L})}{8}\right]\nonumber\\
\tilde{T}_{12} & = & \left(1+\tilde{\gamma}_{12}^2\right)^
{-1/4}\exp \left[-\frac{b}{1+\tilde{\gamma}
_{12}^2}(y_1-y_2)^2
- \frac{({\bf k}_{1\perp}-{\bf k}_{2\perp})^2(R^2_{ch}-R^2_
{L})}{8}\right]
\end{eqnarray}
and
\begin{eqnarray}
\phi_{12}^{ch} &  = & \frac{b \gamma_{12}}
{1+\gamma_{12}^2}(y_1-y_2)^2\
-\ \tau_{0,ch} (m_{1\perp}-m_{2\perp})\ -
\ \frac{1}{2}\arctan \gamma_{12}\\
\tilde{\phi}_{12}^{ch} &  = & 
\frac{b \tilde{\gamma}_{12}}{1+\tilde{\gamma}_{12}^2}(y_1-y_2)^2\
-\ \tau_{0,ch} (m_{1\perp}+m_{2\perp})\ -
\ \frac{1}{2}\arctan \tilde{\gamma}_{12}\\
 \phi_{j}^{c}   &  = & 
-\ \tau_{0,c} m_{j\perp}
\end{eqnarray}
One thus finds again that the correlation functions do not
depend separately on the geometrical radii $R$ or on the
correlation lengths $L$ but rather on the combination $R_L$
defined in (\ref{eq:rlexp}). This expression reduces  
 in the limit $R_{ch}\gg  L$ to $L$
and in the limit. 
 $R_{ch} \ll L$ to $R$. The model considered in \cite{GKW}
is thus a particular case of the space-time approach
\cite{APW} for $L=0$.

As in the static case the tilde terms give rise to the 
particle-antiparticle correlations.
 For a purely chaotic system the intercept of the 
$\pi^+\pi^-$  correlation function is  
\begin{equation}
C^{+-}_2(k,k) \ = \  1 \ + 
\ \left(1+4\left(\frac{b}{a}\right)^2\right)^{-1/2}
\ = \  1 \ + 
\ \left(1+4\left(\frac{\tau_{0,ch}}
{m_{\perp}L_{\eta}^2}\right)^2\right)^{-1/2}
\label{eq:rol1}
\end{equation}

We conclude this section 
 with the observation that
in \cite{APW} a correspondence between the 
correlation length $L$ in the primordial correlator $C(x-y)$ and 
the temperature $T$ for a pion source that exhibits thermal 
equilibrium was established. In the limit of large volume $V\propto R^3$ and 
lifetime $R_0$ of the system, it reads 
\begin{equation}
L \ \sim \ T^{-1}
\label{eq:A.11}
\end{equation}

\subsection{The Wigner function approach}
 
As mentioned previously, the experimental observation of the
fact that the two particle correlation function depends not
only on the difference of momenta $q=k_1-k_2$ but also on 
the sum $k_1+k_2$ led to the introduction and the use
\cite{Pratt} of a ``source" 
function within the well known Wigner function formalism of
quantum mechanics 
\footnote{An attempt to consider the
correlation between coordinates and momentum was also
performed earlier within the ordinary wave function 
formalism by
Yano and Koonin \cite{Yano} who proposed a formula for the
second order correlation function of form (\ref{eq:3}).
However this form turned out subsequently to have 
pathological features as it
leads in some cases to a violation of the lower bounds of the
correlation function (cf. section 5.1.6). The reason for this
mis-behaviour was mentioned in section 4.6 and will also be
discussed in the following.}.   
While it turned out later that this
property of the correlation function can be derived within
the current formalism without the approximations involved by
the Wigner formalism, this formalism is still useful when
applied within a hydrodynamical context. 
On the other hand it should be clear that there is no
justification for using this formalism without a full
hydrodynamical apparatus (cf. also below). This message has
apparently not yet come through, as many phenomenological and
experimental papers have continued to use this restrictive
formalism.
  
The Wigner function approach for BEC was proposed in a
non-relativistic form in 
Ref.\cite{Pratt} and subsequently generalized in \cite{Schlei},
\cite{APW}(cf. also \cite{shuryak},\cite{PaGyGa}). 
The Wigner function called also source function, 
$g(x,k)$, may be 
regarded as the quantum analogue of the density of particles
of momentum $k$ at space-time point $x$ in classical statistical
physics. It is defined within the wave function formalism as

\begin {eqnarray}
g({\bf x},{\bf k},t)& =&\int d^3x^{'}\psi^*\left({\bf
x}+\frac{1}{2}{\bf x^{'}},t\right)\psi\left({\bf x}-\frac{1}{2}{\bf
x^{'}},t\right)e^{i{\bf kx{'}}}\nonumber\\
&=&\int d^3k^{'}\psi^*\left({\bf k}+\frac{1}{2}{\bf k{'}},t\right)
\psi\left({\bf k}-\frac{1}{2}{\bf k{'}},t\right)e^{-i{\bf k{'}x}}
\label{eq:Wigner}
\end{eqnarray}  
and is related to the coordinate and momentum densities
 by the relations
\be
n({\bf x},t)=\int d^3kg({\bf x},{\bf k},t)
\label{eq:(B)}
\ee
and
\be
n({\bf k},t)=\int d^3xg({\bf x},{\bf k},t)
\label{eq:(C)}
\ee
respectively.

Due to its quantum nature the function $g(x,k)$ takes 
real but not necessarily 
positive values. Although eq.(\ref{eq:Wigner}) is nothing
but a definition which does not imply any approximation, its
form suggests that it might be useful when 
simultaneous information about coordinates and momenta are
desirable, provided of course that
the limits imposed by
uncertainty relations are not violated. 
As a matter of fact as will be shown below, the Wigner
function is useful for BEC only if a more stringent
condition is fulfilled, namely that the difference of
momenta $q$ of the pair is small, as compared with the
individual momenta of the produced particles. It is thus clear that its
applicability  is more restricted
than that of the classical current approach, where only the
``no recoil" condition, i.e. small total momentum of produced
particles, as compared with the momentum of incident
particles, must be respected.   
This circumstance 
is often overlooked when comparing theoretical predictions 
based on the Wigner approach with experimental data. 
In particular herefrom also follows that the application of the
Wigner formalism to data has necessarily to take into account
from  the beginning resonances which dominate the small 
$q$ region. 
  It turns out that
the use of the Wigner function for BEC
is justified only in special cases as e.g. when a coherent 
hydrodynamical study is
performed, i.e. when the observables are related to an
equation of state and when simultaneously single and higher
order inclusive dstributions are investigated. 
Unfortunately only very few papers, where the Wigner function
formalism is used, are bona fide hydrodynamical studies. The
majority of  ``theoretical" papers in this context are
``pseudo-hydrodynamical" (cf. sections 5.1.5 and 5.1.6) in 
the sense that the form of the
source function is expressed in terms of {\em effective} 
physical variables
like temperature or velocity, which are not related by an
equation of state. In this case 
 the application of the Wigner
approach is a ``luxury" which is not justified. This is a
fortiori true since, as will be shown in the following, the
Wigner approach is mathematically not simpler that the 
classical current
approach, of which it is a particular case. Thus the
space-time
model \cite{APW}
presented above (cf. section 4.8) is
more general than the Wigner approach, albeit it is not more
complicated and has not more independent parameters

In second quantization
$g(x,k)$ is defined in terms of the correlator 
$< a^\dagger ( {\bf k}_i ) a ( {\bf k}_j )>$ 
by the relation
\begin{equation}
< a^\dagger ( {\bf k}_i ) a ( {\bf k}_j )  > =
\int d^4 x \exp [- i x_\mu ( k_i{}^\mu  - k_j{}^\mu ) ] \cdot
g [ x , \textstyle{\frac{1}{2}} ( k_i + k_j ) ]
\label{eq:wigdef}
\end{equation}

This is a natural generalization of (\ref{eq:(C)}) to which
 it reduces in the limit $k_{i}=k_{j}$.

Accordingly, for the second order correlation function one
writes
\be
P_2({\bf k}_1,{\bf k}_2) \ =\ \int d^4x_1 \int d^4x_2
\left[g\left(x_1,k_1\right)
g\left(x_2,k_2\right)\ +\ 
g\left(x_1,K\right)
g\left(x_2,K\right)
\ \exp\left[iq_\mu(x^\mu_1-x^\mu_2)\right] \right]
\label{eq:P2}
\end{equation}
where $K^\mu=(k_1^\mu+k_2^\mu)/2$ and $q^\mu=k_1^\mu-k_2^\mu$
are the mean momentum and momentum difference of the pair
\footnote{For neutral particles, there are additional 
contributions to $P_2({\bf k}_1,{\bf k}_2)$ which play a role
for soft particles and which will be neglected here.}. 
 
The relation between this Wigner approach and the classical
current approach is established by expressing the rhs of 
eq.(\ref{eq:wigdef}) in terms of the currents. One has
\begin{equation}
g(x,k)\ =\ \frac{1}{2 \sqrt{E_{i}E_{j}}(2\pi)^{3}} \ 
 \int d^4z \ <J\left(x+\frac{z}{2}\right) 
J\left(x-\frac{z}{2}\right)> \ \exp\left[-ik^\mu z_\mu\right]
\label{eq:wigcur}
\end{equation}
The derivation of the Wigner formalism from the classical
current formalism has the important advantage that it avoids
violations of quantum mechanical bounds as those mentioned
previously.

  Note that in the rhs of eq.(\ref{eq:P2}) enters the
off-mass shell average momentum $\frac{1}{2}(k_1+k_2)$ which is not equal
to the on-mass shell average $K=\frac{1}{2}\sqrt{E^2-m_{1}^2+m_{2}^2}$
where $E$ is the total energy of the pair (1,2). This means
among other things that
in this approach it is not
enough to postulate the source function $g$
in order to
determine the second (and higher order) correlation
function
$C_2$, but further assumptions are necessary. Usually 
oe neglects the off-mass shellness i.e. one
approximates $E$ by the sum $E_1+E_2$ where $E_i$ are the
on-shell energies of the particles (1,2), which means that
one neglects quantum corrections
\footnote{That these corrections can be important has
also been shown in \cite{Bertsch94}.} 
which is permitted as long as $k_1-k_2=q$ is small 
\footnote{It is sometimes argued that the
relevant $q$ range in BEC is given by $D^{-1}$ where $D$ is a
typical length scale of the source and therefore 
for heavy ion reactions this should be allowed. 
This is not quite correct,
because the {\em shape} of the correlation function from which one 
determines the physical parameters of the source is not
given just by the values of the correlation function near 
the origin, but depends also
on its values at large $q$.}.

As mentioned already, the use of the Wigner formalism
 is worthwhile within a true
hydrodynamical approach when the relation with the
equation of state is exploited. In this case
the probability to produce a particle of momentum $k$
from the space-time point $x$ then depends on the fluid velocity, 
$u^\mu(x)$, and the temperature, $T(x)$, at this point, and one
has  
\begin{eqnarray}
&&\sqrt{E_i E_j} < a^\dagger ( {\bf k}_i ) a ( {\bf k}_j ) >
 \nonumber\\
&&= \frac{1}{(2\pi)^3} \int\limits_\Sigma
\frac{\textstyle{\frac{1}{2}} (k_i{}^\mu + k_j{}^\mu)
d \sigma_\mu (x_\mu)}{\exp \left[\displaystyle{
\frac{\textstyle{\frac{1}{2}} (k_i{}^\mu + k_j{}^\mu)
u_\mu (x_\mu)}{T_f (x_\mu)} } \right] - 1
} \cdot \exp [- i x_\mu (k_i{}^\mu - k_j{}^\mu)] 
\label{eq:hydro}
\end{eqnarray}
Here, $d\sigma^\mu$ is the volume element on the 
freeze-out hypersurface $\Sigma$ where the final state 
particles are produced. We will discuss applications of this
approach in section 5. 

\subsubsection{Resonances in the Wigner Formalism}

For a purely chaotic source, the formalism to 
take into
account the effects of resonance decays on the Bose-Einstein 
correlation function can be found, e.g. in  
\cite{Grass,Gyu89}. An
extension of this approach is due to \cite{Bolz} and 
\cite{Schlei} which allows to consider also the effect of
coherence
and provides rather detailed
and subsequently, apparently, confirmed
predictions for heavy ion reactions. It is based on
 the Wigner function formalism.  
   
The correlation function of two identical particles
of momenta ${\bf k}_1$ and ${\bf k}_2$ can be written as
\begin{equation}
C_2( {\bf k}_1 , {\bf k}_2 )
\:=\: 1 + \frac{A_{12}\:A_{21}}{A_{11}\:A_{22}}
\label{eq:001}
\end{equation}

where the matrix elements $A_{ij}$ are given in terms of source
functions $g(x,k)$ as follows
\begin{equation}
A_{ij}\:=\:\sqrt{E_i E_j}<a^\dagger({\bf k}_i) a({\bf k}_j)>
\:=\:\int d^4x\: g(x_\mu,k^\mu)\: e^{\textstyle{i q^\mu x_\mu}}
\label{eq:002}
\end{equation}
A typical source function reads
\begin{equation}
g(x_\mu,p^\mu)\: =\: g_\pi^{dir}(x_\mu,p^\mu)
\:+ \sum_{res=\rho,\omega,\eta,...} 
g_{res \rightarrow \pi}(x_\mu,p^\mu)
\label{eq:003}
\end{equation}

where the labels $dir$ and $res \rightarrow \pi$ refer to 
direct pions 
and to pions which are produced through the decay of resonances
(such as $\rho$, $\omega$, $\eta$, ... etc.), respectively.

The contribution from a particular resonance decay is estimated in
\cite{Bolz},\cite{Schlei} using kinematical and phase space
considerations as well as the source function of that
resonance. 
The source distribution for 
the direct
production of pions and resonances is calculated assuming
local thermodynamcal and chemical equilibrium as is appropiate for a
hydrodynamical treatment. 

\begin{equation}
g_{\alpha}^{dir}(x_\mu,p^\mu)\: =\:
\frac{2J+1}{(2\pi)^3}
\int_{\Sigma} \:
\frac{p^\mu d\sigma_\mu(x'_\mu) \:  \delta^4(x_\mu-x'_\mu)}
{\exp \left[ \displaystyle{\frac{p^\mu u_\mu(x'_\mu)
-B_{\alpha}\mu_B(x'_\mu)-S_{\alpha}\mu_S(x'_\mu)}
{T_f(x'_\mu)}} \right] - 1}
\label{eq:00x}
\end{equation}
Here $\alpha$ denotes the particular resonance and 
 $d\sigma^\mu$ is the differential volume element and 
the integration is performed over the freeze-out hypersurface
$\Sigma$. $u^\mu(x)$ and $T_f$ are the four velocity of 
the fluid
element at point $x$ and the freeze-out temperature, 
respectively. 
$B$ and $S$ are the baryon number and the strangeness of the 
particle
species labeled $\alpha$, 
respectively, and $\mu_B$ and $\mu_S$ are the corresponding
chemical potentials. $J$ is the spin of the particle.

This approach is then extended \cite{Schlei} to include
also a coherent component rsulting in
 a second order correlation function of 
the form

\begin{equation}
C_2( {\bf k}_1 , {\bf k}_2 )= 1 + 2\:p_{eff}\:(1-p_{eff})\: 
Re\:d_{12} 
+ p^2_{eff}\:|d_{12}|^2
\label{eq:def}
\end{equation}
where $p_{eff}$ is an effective chaoticity
related to the true chaoticity $p_{dir}$
\footnote{It is assumed that only directly produced
particles have a coherent component.}
via 
\begin{equation}
p_{eff}\: =\: p_{dir} \:(1 - f^{res})\: +\: f^{res}.
\label{eq:017}
\end{equation}
$f^{res}$ is the fraction of particles arising from resonances. 
The form of this
equation us the same as that derived previously for a
partially coherent source within the current formalism and
manifests the characteristic two component structure.

The sensitivity of the correlation function on the 
chaoticity parameter
$p_{dir}$ can be estimated e.g. from the intercept
(cf. eq. (\ref{eq:def}))
\begin{equation}
I_o\: =\: C_2({\bf k},{\bf k})\: =\: 1 + 2\: 
p_{eff} - p_{eff}^2
\label{eq:018}
\end{equation}

Defining the fractions of pions produced directly (chaotically 
and coherently) and from resonances,  
\begin{equation}
f_{ch}^{dir} = \frac{p_{dir}\: A_{ii}^\pi}{A_{ii}}\:, \quad
f_{co}^{dir} = \frac{(1-p_{dir})\:A_{ii}^\pi}{A_{ii}}\:, \quad
f^{res} = \frac{\Sigma_{res=\rho,\omega,\eta,...}\: 
A_{ii}^{res}}{A_{ii}}
\label{eq:014}
\end{equation}
with
\begin{eqnarray}
f_{ch}^{dir} + f_{co}^{dir} + f^{res} = 1
\nonumber
\label{eq:015}
\end{eqnarray}

In fig. 9 the intercept of the correlation function is shown
as a function of $p_{dir}$ and $f^{res}$.
In order to read off the fraction of {\it direct}
chaotically produced particles, $p_{dir}$, from the intercept
of the correlation function, one has to extract the effective 
chaoticity $p_{eff}$ according to eq. (\ref{eq:018}) and then 
correct for the fraction of pions from resonance decays. 
Note that $p_{eff}<p_{dir}$. In particular, if a large 
fraction of  
pions arise from resonance decays, 
$p_{eff}\rightarrow 1$ and it 
will take very precise measurements of the two-particle 
correlation 
function at small ${\bf q}$ to determine the true chaoticity, 
$p_{dir}$. 

A further complication arises if a fraction of particles 
are the
decay products of long-lived resonances \footnote{In some papers \cite{Cs1996} 
pions
originating from long lived
resonances are associated with a ``halo"  while those coming form short 
lived resonances or directly produced are related to a
``core". Then it is claimed among other things that 
the ``core" parameters of
the source (like radius and $\lambda$ factor) can be
obtained from the data just by eliminating the small $Q$ points 
and fitting only the remaining points. Even if such a
separation would be clear cut (there are doubts about this
because of the $\omega$ resonance), it would be of course dependent on the 
resolution of the detector.}. 
This topic as well as the problem of misidentification are
discussed in \cite{Bolz}.

\subsection{Dynamical models of multiparticle production and 
event generators}
Due to the lack of a full fledged theory of multiparticle production in
strong interactions different
models of multiparticle dynamics were proposed. 
 Bose-Einstein 
correlations measurements have been used
 either to test a particular model or/and to 
determine some of its parameters.
Among other things these models might be used to predict the
dependence of the chaoticity on the type of reaction. 
In the following we will sketch the main theoretical
ideas on which these models are based and 
mention briefly their relation to
data. 

One of the first models of particle production from which  
definite predictions on BEC can be derived is the Schwinger
model \cite{Schwinger} for $e^{+}-e^{-}$ reactions. It visualizes
the source as an one-dimensional string in a coherent state
and thus predicts the absence of any bunching effect.
A similar prediction follows from the bremsstrahlung
model \cite{Gunion}. Recoilless bremsstrahlung can be
described by a classical current which also corresponds to 
a coherent state. Given the fact that in all hadron
production processes BEC have been seen, it follows that
the above two models are ruled out by experiment.

More complex predictions follow from a {\em dual
topological} model due to Giovannini and Veneziano \cite{GV} 
 which associates the processes $e^{+}-e^{-}
\rightarrow hadrons$ to a unitarity cut in one plane,
reactions due to Pomeron exchange to a cut in two
planes, and annihilation reactions $\bar{p}-p$ to a cut 
in three planes. This model predicts then among other things   
that for $\pi^-\pi^-$ BEC the  
intercepts $C_2(k,k)$ of 
the second order correlation functions
for the above reactions should satisfy the following
relation:
\be
[C_2^{e^+e^-}(k,k)-2]/[C_2^{\pi
p}(k,k)-2]/[C_2^{ann}(k,k)-2]=1/\frac{1}{2}/\frac{1}{3}.
\label{eq:GV}
\ee  
    (A similar, but quantitatively different
relationship is predicted for $\pi^+\pi^-$ correlations.) 

Despite the fact that since the publication of this paper in
1977 many experimental BEC studies of these reactions
were performed, the above predictions could not yet be tested
quantitatively in a convincing manner. This is due among
other things to experimental difficulties (cf. below, in
particular 
factors (a),(b)) and illustrates the
unsatisfactory status of  
experimental BEC investigations.
A qualitative remark can however be made:
The expectation that the annihilation reaction leads to
more bunching than other reactions is apparently confirmed
(cf. e.g. ref.\cite{Angel} and section 2.1.1). As to the
difference between $e^{+}-e^{-}$ reactions and hadronic reactions the
experimental situation is rather confused (cf. also below).  

A somewhat related dynamical model based on Reggeon theory,
was proposed already in \cite{LRT}. A straightforward 
extension of
of this formalism to heavy
ion reactions does not work as it predicts that
the longitudinal radius is of ``hadronic" size \cite{CK}.

A different approach to BEC based on the classical current
formalism is proposed in \cite{Casado}. The currents are
associated with the chains of the dual parton model, and
contrary to what is assumed in other applications of the
classical current formalism, all the phases of these
elementary currents
 are fixed, so that the source is essentially
coherent. This is a special case of the classical current
approach presented in sections 4.2, 4.3 and 4.8 where 
allowance is made both for a chaotic and coherent component.
The model is intended to work for $p-p$ reactions where the 
authors state that resonances do not play an important role. It explains,
according to the authors, the dependence
of the $\lambda$ parameter in the empirical formula for the
second order correlation function
\be
C_2= 1 + \lambda \exp (-R^2 q^2) 
\label{eq:lambda1}
\ee
 on the multiplicity and energy.
Unfortunately the claim that in $p-p$ reactions 
 pions are only directly produced is unfounded.
Furthermore there are other factors which influence
the multiplicity dependence of $\lambda$ (cf. section 6.2) 
which are not considered in\cite{Casado} and which
are of more general nature. 

An orthogonal point of view for the interpetation of the
same $\lambda$ factor (also for directly produced pions,
only,) is due to \cite{BSWW}. In this approach 
the source is made up of totally chaotic elementary emitting
cells which are occupied by identical particles subject to
Bose-Einstein statistics. Different cells are independent so
that correlations between particles in different cells lead
to $\lambda=0$, while correlations between particles in the
same cell are characterized by $\lambda=1$. From the
interplay of these two types of correlations, one obtains    
with an appropiate weighting, large $\lambda$ values in
$e^{+}-e^{-}$ reactions and small $\lambda$ values in $p-p$
reactions, as in \cite{Casado}, but within a completely
different approach. 

We conclude the discussion of these two approaches by the
following remarks.
Besides the reservations about the role of
directly produced pions in BEC expressed above and which 
presents the two approaches in a rather academic
light, it is
unclear whether the $\lambda$ factor in $e^{+}-e^{-}$ reactions is
larger than in $p-p$ reactions as assumed in \cite{BSWW}. This issue awaits a critical
analysis of the specific experimental set-ups. 
  The fact that
quite different approaches lead to similar
conclusions about the $\lambda$ factor confirms that the 
parametrization of the second order correlation
function in the form (\ref{eq:lambda1}) is, as 
pointed out already in section 2.2, inadequate.  

We discuss now other two, closely related, approaches, which
make more detailed predictions about the form of the
correlation function in $e^{+}-e^{-}$ reactions:
 \cite{stringbowl},
\cite{bowler} on the one hand, and
\cite{stringander},\cite{AR1},
\cite{AR2}
on the other. Both approaches are
based on a variant of the string model (for a more extended 
review of this topic cf. e.g. \cite{Bow0}). Such a string
represents a coloured field 
formed between a quark $q$ and an antiquark $\bar{q}$, 
which tend to
separate. Because of confinement the breakup of the string
can be materialized only through creation of new 
$q\bar{q}$ pairs, which are the mesons produced in the
reaction. The difference between the Schwinger model of
confinement on the one hand and the models of Bowler and
Andersson-Hofmann-Ringn\'er on the other is that in the former the field
couples directly and locally to a meson, while in the latter
ones the quarks, which constitute the meson, are created at
different points. This feature destroys the coherence
inherent in the Schwinger model and makes possible the
Bose-Einstein bunching effect. 
  For massless quarks the second order
correlation function can be approximated by 
the relation \cite{stringander}
\be
C_2= 1+<\cos(\kappa\Delta A)/\cosh(b\Delta A/2)>
\label{eq:AH}
\ee
where $\Delta A$ denotes the difference between the
space-time areas of coloured fields spanned by the two 
particles,
$\kappa$ is the string tension and $b$ a
parameter characterizing the decay probability of the string.   
For massive quarks the formulae become more involved and
were approximated analytically in \cite{stringbowl} or 
calculated numerically in \cite{stringander},\cite{AR1},
\cite{AR2}. In this model the
correlation function depends both on the difference of
momenta $k_1-k_2$ as well as on their sum $k_1+k_2$,
reflecting the correlation between the momentum of the
particle and the coordinate of its production point. This
is a consequence of the fact that string models use a 
Wigner function type approach. 
   From the above equations follows
that there are two
length scales in the problem, one associated with
$\kappa$ and another with $b$. Phenomenologically these
correspond to $q_{\parallel}$ and $q_{\bot}$. Both these
lengths are correlation lengths rather than geometrical
radii. (As a matter of fact there is no geometrical radius 
in the string model.) Their magnitudes are quite different.

In both string approaches \cite{bowler},
\cite{stringander} one obtains
 a difference between BEC for identically charged and
neutral pions as found in \cite{apw}. However 
while in \cite{stringbowl} there is room for coherence, this is
apparently not the case for \cite{stringander},\cite{AR1},
\cite{AR2}, which
predict a totally chaotic source. Furthermore in 
\cite{stringbowl}
an energy dependence of the BEC is predicted (the correlation
function is expected to shrink with increasing energy),
while in \cite{stringander},\cite{AR1},\cite{AR2} the correlation function does
not depend on energy. 
 \footnote{To make the model more realistic in 
ref.\cite{stringander} resonances
were included according to the variant of the Lund model
(JETSET)
 in use at that time (1986) and agreement 
with $e^{+}-e^{-}$ data was found. Subsequently however it
was pointed out in \cite{Verb} that some resonance
weights used in \cite{stringander} were incorrect, so that the
agreement mentioned above is probably accidental.}

A rather discordant note in this string concerto 
\cite{stringbowl},
\cite{stringander} is represented by the paper by Scholten and Wu
\cite{Scholten}. These authors, using a different 
hadronization mechanism conclude that dynamical correlations,
at least in $e^{+}-e^{-}$ reactions dominate over 
BEC correlations so that BEC cannot be used to infer
information about the size and lifetime of the source
\footnote{What concerns $e^{+}-e^{-}$ reactions similar skepticism was
expressed by Haywood \cite{Hay}.}.   
This point of view seems too extreme, as it is
contradicted by some simple empirical observations: in
$e^{+}-e^{-}$ reactions, as well as in all other reactions,
correlations between identical particle are observed which
are much stronger than  
those of non-identical ones, the correlation functions are
(in general) monotoneously decreasing functions of the
momentum difference $q$ and in nuclear reactions the
``radii" obtained from identical particole correlations
increase with the mass number of the participating nuclei.
All these observations are in agreement with what one would
expect from BEC, which suggests that dynamical correlations  
cannot distort too much this picture. However a thorough
comparison of BEC in different reactions, using the same
experimental techniques, appears highly desirable. 
     
\paragraph{Event generators}

The model \cite{stringander} was implemented by 
Sjostrand \cite{Sjos} into JETSET under the name LUBOEI by
modifying a posteriori the momenta of produced pions so that
identical pairs of pions are bunched according to 
\cite{stringander}. This
manipulation ``by hand" violates energy-momentum
conservation which was imposed at the beginning in JETSET.
To compenate for this, the momenta are rescaled so that
energy-momentum conservation is restored. However this
rescaling introduces spurious long-range correlations, which
bias the BEC. Nevertheless, 
in general
this program leads to a reasonable description of the
bunching effect in the second order correlation function
\footnote{Cf. however ref. \cite{FW97}}.
More refined features of BEC which reflect the quantum
mechanical essence of the effect, cannot be 
obtained of course. One reason for this, of rather technical
nature, is
due to the fact that the ad-hoc modification of two-particle
correlations does not yet include many-body correlations,
reflected in the symmetrization (or antisymmetrization) 
of the entire wave-function.

Another reason of fundamental character is the fact that
event generators like any Monte Carlo algorithm deal in
general with probabilities
\footnote{For heavy ion reactions the "Quantum Molecular
Dynamics" (QMD) model \cite{QMD} attempts to surpass this
deficiency by using wave
functions rather than probabilities as input. However this
model
also neglects (anti)symmetrization effects and cannot be used for
interferometry studies.}
 and therefore cannot account for quantum effects, which are
based on phases of amplitudes
 \footnote{It is interesting to mention that for {\em one} 
string Andersson and Hofmann (\cite{stringander})proposed a 
formulation of the BEC effect in terms of {\em amplitudes}. 
However this procedure cannot be used to generate events.}.    
The event generator JETSET was 
further developped by L\"onnblad and Sj\"ostrand , 
(\cite{SL1}, \cite{SL2}) and used to estimate the influence
of BEC on the determination of the mass of W in $e^{+}-e^{-}$
reactions, a subject of high current interest for the
standard model and in particular for the search of the Higgs
particle.
This effect was also studied using different
event generators in \cite{Jadach} and \cite{Kart}. The
argument of L\"onnblad and Sj\"ostrand is the following.
Consider the reaction $e^{+}-e^{-}\rightarrow W^+W^- \rightarrow
q_1\bar{q}_2q_3\bar{q}_4$ when both $W$-s decay into
hadrons. Then according to \cite{SK} the typical space-time
separation of the decay vertices of the $W^+$ and the $W^-$
is less than 0.1 fm (at LEP 2 energies) and thus much
smaller than a typical hadronic radius ($\sim 0.5$ fm. There
will thus be a Bose Einstein interference between a pion
from the $W^+$ and a pion (with the same charge) from the
$W^-$ and one cannot establish unambigously the
``parenthood" of these pions. This prevents then in this
model a precise
determination of the invariant mass of the $W$-s. In
\cite{SL1} algorithms for the inclusion of this effect into
the determination of the mass of the $W$ are proposed and
for certain scenarios mass corrections   
 of the order of 100 MeV at 170 GeV c.m. energy are 
obtained. However, as
emphasized in \cite{SL1} other scenarios with less or no
effect of BEC on the mass determination of the $W$ are
possible. Thus in \cite{Jadach} and \cite{Kart} effects of
the order of only 20 MeV are found. For more details we 
refer the reader to the original literature. 

This aspect of BEC is interesting in itself as it
illustrates the possible applications of this effect in 
electroweak interactions, a domain which is beyond
the usual application domain of BEC, i.e. that of 
strong interactions.     

The Lund model was applied also to heavy ion reactions 
and then extended to include BEC (e.g. the SPACER
version \cite{Pra90} of the Lund model).
The topic of event generators for heavy ion reactions is of
current interest because of the ongoing search for quark
gluon plasma.   
Padula, Gyulassy and Gavin \cite{PaGyGa}
suggested
to use for this purpose the Wigner function formalism
 in order to take into
account explicitly the correlation between momenta and
coordinates, as implied by the inside-outside cascade
approach. This is evidently another way of expressing the
non-stationarity of the correlation function mentioned above.
(An explicit introducton of momentum-coordinate
correlations in particle physics event generators like 
JETSET/LUBOEI
  is not necessary,
because the
non-stationarity is delivered ``free
house" by the string model used in the LUND generator.) 
 
On the other hand the Wigner formalism may present also another
 advantage as emphasized more recently by Bialas
and Krzywicki \cite{BK}. This has to do with
 the important difficulty mentioned also above and which is 
inherent in all event
generators, namely the probabilistic nature
of Monte Carlo methods. 
The Wigner function has in
certain limits the meaning of a wave function and thus
provides quantum amplitudes. The proposal of
Bialas and Krzywicki consists then 
in starting from the single particle
distribution  $\Omega_0({\bf k})$ constructed from
non-symmetrized particle wave functions as produced 
by conventional event generators and
writing the Wigner function 
\be
g({\bf k};{\bf x}) =  \Omega_0({\bf k}) w({\bf k};{\bf x}) 
\label{eq:BK8}
\ee
where  w({\bf k};{\bf x}) is the conditional probability
that given that the particles with momenta $k_1,, k_2,
...k_n$
are present in the final state, they are produced at the
points $x_1, ...x_n$. Then the art of the model builder
consists in guessing the probability $w({\bf k};{\bf x})$.
This may be easier than guessing from the beginning the 
exact Wigner function. For example
a simple ansatz would be to assume that the
likelyhood to produce a particle from a given space point is
statistically independent of what happens to other
particles.
This means that $w({\bf k};{\bf x})$ can be factorized in
terms of the individual particles.  
 Implementations of this scheme were discussed in
\cite{Wosiek}, \cite{FWIT}
  
When using the Wigner formalism or any model (like 
those used in event generators) which specifies momenta and
coordinates simultaneously, one must of course watch
 that the 
correlations between momenta and coordinates do not become
too strong. This apparently has not always been
done.\footnote{E.g. it
is unclear to us whether the ``pure" multiple
scattering approach 
of \cite{Humanic} satisfies the above
constraint.} 
  
 That such a procedure is dangerous since it can lead to 
unphysical {\em antibunching} effects, i.e. to violation of 
unitarity  was already mentioned in \cite{Axel} (cf. section
4.6). This point has been reiterated recently e.g. in 
\cite{Pratt97},\cite{WFKMSZ},\cite{Martin}.  

Concluding this section one should
emphasize that event generators are just an
experimental tool,
sometimes useful in the design of detectors or for getting
rudimentary information about experimentally 
unaccessible phase. Often they are however ab-used, e.g.  
to search for
``new" phenomena: if agreement between data and event
generators is found, one states that no ``new "
physics was found. Such a procedure is injustified, because
agreement with a model or an event generator is often
accidental. Furthermore, for the reasons mentioned above 
event generators cannot be used to obtain
the ``true" correlation function, i.e. they are no
substitute for a bona fide HBT experiment (cf. also
\cite{Aich} for a critical analysis of transport models
from the point of view of interferometry).

\subsection{Experimental problems}
 The
confrontation of model predictions with experimental BEC
data has been hampered by two major facts: (i) most models
are idealizations, i.e. they use assumptions which are too 
strong. Examples of such assumptions are: neglect of final 
state interactions, boost invariance, particular analytical 
forms of the correlation functions; (ii) for various reasons
in almost no BEC experiment
so far a ``true" or ``complete"
correlation function was measured, i.e. a correlation
function as is defined by  
\be
C_2({\bf k}_1,{\bf k}_2) = \frac{P_2({\bf k}_1,{\bf
k}_2)}{P_1({\bf k}_1)P_1({\bf k}_2)}.
\label{eq:defC2}
\ee 
Here $P_2$ and $P_1$ are the double and single inclusive
cross sections respectively. Note in particular that the
single inclusive cross sections in the denominator are defined in terms of the
same density matrix as the double inclusive in the
numerator. This means that these cross sections and the
corresponding ratios have to be
measured for each event separately and only after that one
is allowed to average over events.
What is measured usually instead is a function which  
differs from eq.(\ref{eq:defC2}) in two respects: the
normalization of $C_2$ is not done in terms the product of
single inclusive cross sections $P_1$ but in terms of a
``background" double inclusive cross section, which is obtained
either  by considering pairs of (identically charged) 
particles which come from different
events,
 or by considering oppositely charged particles, or by simulating $P_2$
with an event generator which does not contain BEC. This
means that the structure of the event is falsified and
therefore this kind of normalization biases the results.
%\footnote{With the exception of certain detectors like that
%of the Na44 collaboration for which
%the definition of an event may be difficult, the author
%could not find a serious justification why in experiments
% one should not use the correct procedure for normalization.
%I am indebted to E.M. Friedlander, W. Kittel and B. R. Schlei
%for a clarifying corespondence along these lines.}. 
Furthermore, one
does not (yet) measure the full correlation function $C_2$
in terms of its six independent variables, but rather
projections of it in terms of single variables like
the momentum difference $q$, rapidity difference $y_1-y_2$
etc. Last but not least, the intercept of the correlation
function, which contains important information about the
amount of coherence, cannot be really be measured at present
because (a) one does not yet control sufficiently well the 
final state interactions which contribute to the intercept; 
(b) its experimental determination implies an extrapolation
to $q=0$. Such an extrapolation can be performed only if the
analytical form of the correlation function at $q\geq 0$
is known, which is not the case.

For these reasons at present it is difficult
to test quantitatively a given model, except when its
predictions are very clear-cut. 
This circumstance limits certainly the usefulness of 
BEC as a tool in determining the exact dynamics of a 
reaction.

\setcounter{equation}{0}

\section{Applications to ultrarelativistic 
nucleus-nucleus collisions}
\subsection{BEC, hydrodynamics and the search for
quark-gluon plasma}

The use of BEC in the search for quark-gluon plasma is in
most cases based on hydrodynamics. This is so because
the space-time evolution of the system 
is given by the equations of hydrodynamics the solutions of which are
different depending whether a QGP is formed or not.
In this way 
hydrodynamics also provides information about the equation of
state (EOS).
QGP being a (new) {\em phase} it is described by a specific
 EOS which is different from that of
ordinary hadronic matter. The proof that this phase has been
seen must include information about its EOS and thus the
combination of hydrodynamics with BEC constitutes the only
 consistent way through which the formation of QGP can be
tested.   

QCD predicts 
that the phase transition from
hadronic matter to QGP takes place only when a critical
energy density is exceeded. To measure this density 
we need
to know the initial volume of the system.
While via photon interferometry (cf. section 4.5 and 4.6) 
one can in principle measure the dimensions (and thus the
energy density) of the initial state, hadron interferometry
 yields information only about the final freeze-out stage
when hadrons are created. To obtain information about the
initial state with hadronic probes, 
hydrodynamical models have to be used in order to 
extrapolate backwards 
from the final freeze-out stage where hadrons are created, 
to the interesting initial stage. 
 The lifetime of the system as given by BEC
is also an important piece of information for QGP search. 
Indeed, in order to decide whether we have seen the  new 
phase, we have to measure
the lifetime of the system. Only lifetimes exceeding
significantly typical hadronic lifetimes ($10^{-23} sec.$)
could prove the establishment of QGP (cf. below)

\subsubsection{General remarks about the hydrodynamical 
approach}

Besides the main advantages of hydrodynamics related to the
information about initial conditions, freeze-out
and equation of state,
for the study of BEC in particular hydrodynamics is very
useful because it provides also 
the single
inclusive distributions which are intimately connected with 
higher order
distributions as well as the weights and 
the 
space-time
and momentum distributions of resonances, which influence
strongly the correlations.
The phenomenological applications of the hydrodynamical
approach to data  
are however hampered by two circumstances.  

(i) While the ultimate goal of BEC is the extraction of 
the minimum
set of parameters which include radii and coherence lengths
both for the chaotic and coherent components of the source,
in practice, mainly because of limited statistics
(but also because of an inadequate analysis of the data) 
 one had to limit oneself to the determination of 
a reduced number of parameters, which we will call in the
following ``effective radii" $R_{eff}$ 
and ``effective chaoticity" $p_{eff}$. 
In reality $R_{eff}$ is a combination of  
correlation lengths\footnote{In the following the concept of
length refers to space-time.} $L$ and geometrical lengths $R$
as introduced
in sections 4.3 and 4.8. Only in the particular case that one length
scale is much smaller than the other, one can assume that
one measures a ``pure" radius or a ``pure"  correlation
length. For simplicity in the following we shall assume that
this is the case and in particular we will assume 
 that $R \gg L$ so that,
$R_{eff}$ reduces to $L$. This limit might perhaps 
correspond to
what is seen in experiment, if one considers the expansion
of the system in the hadronic phase. (In the high temperature
limit $L\approx T^{-1}$ (cf. section 4.8).

(ii) The presentation of the data is yet biased by theoretical
prejudices. Instead of a consistent hydrodynamical analysis,
much simplified models are used (cf. section 5.1.5 where
these models are presented under the generic name of 
pseudo-hydrodynamics) for this presentation and therefore
 to obtain the real physical quantities, one would have to
solve a complicated mathematical ``inverse" problem, i.e.
one would have to reconstruct the raw data from those
presented in the experimental papers and then apply the
correct theoretical analysis to these. This has not been
done so far and even if the statistics would be sufficient   
for this purpose, the outcome is questionable because of the
difficulties implied by the numerics. (That is why it 
would be desirable that experimentalists and theorists
perform a joint analysis of the data or at least that 
 the data should be presented in ``raw" form.) 

The nearest approximation to the solution of the ``inverse"
problem found in the literature, is that of \cite{Schlei97} 
based on the application of the HYLANDER code by the Marburg
group: It
 consists in fitting the 
 results of the hydrodynamical calculation to the 
Gaussian form used by experimentalists,
\be
C_2( {\bf k}_1 , {\bf k}_2 ) = 1 +\lambda
\exp [ -\textstyle{\frac{1}{2}} q_\parallel^2 R_\parallel^2
-\textstyle{\frac{1}{2}} q_{out}^2 R_{out}^2
-\textstyle{\frac{1}{2}} q_{side}^2 R_{side}^2 ]
\label{eq:015a}
\ee
and comparing with the inverse width of the correlation
function as presented in the experimental papers
\footnote{In \cite{chapman} it was recommended that
experimentalists should use the more complete formula
\be
C_2( {\bf k}_1 , {\bf k}_2 ) = 1 +\lambda
\exp [ -\textstyle{\frac{1}{2}} q_\parallel^2 R_\parallel^2
-\textstyle{\frac{1}{2}} q_{out}^2 R_{out}^2
-\textstyle{\frac{1}{2}} q_{side}^2
R_{side}^2-2q_{out}q_{\parallel}R^2_{{out\parallel}}]
\label{eq:015bis}
\ee
where $2q_{out}q_{\parallel}R^2_{{out\parallel}}$
is called  ``cross" term. For a more detailed discussion of
its meaning and dependence on the coordinate system 
cf. ref. \cite{SSX}. Of course, in view of 
the defficiencies of the entire phenomenological procedure
outlined above and discussed in greater length in section 
5.1.5, these details are of limited importance. In
particular they do not affect the conclusions discussed here.}. 
Here  
$q^\mu \equiv p_1{}^\mu - p_2{}^\mu $, 
$K^\mu \equiv \frac{1}{2} (p_1{}^\mu + p_2{}^\mu)$ and
 $q_\parallel$ and $K_\parallel$ denote the components of  
${\bf q}$ and ${\bf K}$ in beam direction, and $q_\perp$ and
$K_\perp$ the components transverse to that direction;
 $q_{out}$ is  
the projection of the transverse momentum difference, 
${\bf q}_{\perp}$ 
on the transverse momen\-tum of the pair, $2 {\bf K}_\perp$, 
and $q_{side}$ the component perpendicular to ${\bf K}_\perp$. 
(For a source with cylindrical symmetry, the two-particle 
correlation 
function can be expressed in terms of the 5 quantities 
$K_\parallel$, $K_\perp$, $q_\parallel$, 
$q_{side}$ and $q_{out}$.)
$R_\parallel$, $R_{side}$,
$R_{out}$ are effective parameters,
associated via eq.(\ref{eq:015a}) to the corresponding 
$q$ components. 
%These ``effective radii" are lengths
%of homogeneity rather than radii.

 Eq.(\ref{eq:015a}) is equivalent to an expansion of
the correlation function $C({\bf q},{\bf K})$ for small $q$.  
The use of eq.(\ref{eq:015a}) for the representation of
correlations data implies then that one does not 
measure the geometrical radius of the system but the length
of homogeneity, which means that energy density
determinations based on BEC are an overestimate 
\footnote{In \cite{SinHot}, \cite{AkkSi} a distinction is made between the
``local" length of homogeneity $L_{h}(x,k)$ and the 
``hydrodynamical"
length $\L_{h}(x)$ which is the ensemble average of the
former.}. To take into
account the fact that the correlation function depends in
general not
only on the momentum difference $q=k_1-k_2$ but also on the 
sum $K=\frac{1}{2}(k_1+k_2)$ , the parameters $R$ and 
$\lambda$ are assumed to be functions of $K$ and rapidity 
$\frac{1}{2}(y_1+y_2)$.

Hadron BEC refer to the freeze-out stage. This stage is
usually described 
by the Cooper-Frye formula
\cite{cfrye}:

\begin{equation}
E\frac{dN}{d{\bf k}}=\frac{g_i}{(2\pi)^3} \int_\sigma  
\frac{p_{\mu} d\sigma^{\mu}}{\exp(\frac{p_{\mu} u^{\mu} - 
\mu_s-\mu_b}
{T_f}) - 1},
\label{eq:single}
\end{equation}
\\
which describes the distribution of particles with 
degeneracy
factor $g_i$ and 4-momentum $p^{\mu}$ emitted from a 
hypersurface element $d\sigma^\mu$ with 
4-velocity $u^{\mu}$
\footnote
{In most applications particles produced
with momenta $p_\mu$ pointing into the interior of the
emitting isotherm $(p_\mu d\sigma^\mu < 0)$ were assumed
to be absorbed and therefore their contribution
to the total particle number
was neglected. In ref \cite{latp} this effect was indeed 
estimated
to be negligible and recent attempts to reconsider it could
not change this conclusion.
Another effect is 
the interaction of 
the freeze-out system with the rest of the fluid.
This effect can be estimated by comparing the evolution of the
fluid with and without the frozen-out part. This is done
by equating the frozen-out part with that corresponding
in the equation of hydrodynamics to the case $p_\mu=0$.
The fluid parameters are modified by this procedure
at a level not exceeding $10\%$ \cite{ornik}.
 The influence of the freeze-out mechanism
on the determination of radii via BEC has been
discussed recently in several papers; cf.e.g. \cite{HV} 
and references quoted there.}. 
After the cascading
of the resonances we obtain the final observable spectra.

\subsubsection{Transverse and longitudinal
expansion}

The equations of hydrodynamics are non-linear and therefore
good for surprises. 
An illustration of this situation is
represented by the realization, to be described in more 
detail below, that the naive intuition about the role of 
transverse expansion in the determination of the transverse
and longitudinal radius may be completely misleading. Only a
systematic analysis based on 3+1 dimensional hydrodynamics
could clarify this issue. 

In the present section we will discuss Bose-Einstein 
correlations of pions and kaons produced in nuclear 
collisions at SPS
energies in the framework of relativistic
hydrodynamics.
Concrete applications were done for the symmetric reactions 
$S+S$ and $Pb+Pb$ at 200 AGeV. Many of the theoretical 
results were predictions at the time they were performed.
These predictions were subsequently confirmed in experiment.
In \cite{Hum} it was found  
that the transverse radius
extracted from data on Bose-Einstein correlations (BEC) 
for O$+$Au at $200$ AGev reached in the central rapidity 
region a value of about $8 fm$. 
It was then natural to conjecture that this could be an 
indication of transverse flow\cite{Hei},\cite{Lee},\cite{Alb}.
In the meantime the experimental observation 
in itself has been qualified \cite{Fer} and it now appears 
that the transverse radius obtained from the BEC data does 
not exceed a value of $4$-$5 fm$ (see however ref.\cite{Alb}).
Motivated by this situation in ref. \cite{Schlei97} an
investigation \footnote{In ref.\cite{KagMin} the 
dependence of the effective transverse
radius on the transverse velocity field was investigated 
for a fixed
freeze-out hypersurface.
The important effects of transverse expansion on the shape and position
of the hypersurface are not considered there.}
of the role of $3$-dimensional hydrodynamical expansion on the 
space-time extension of the source was performed and
compared with a 1+1 d calculation.

Contrary to what one might have expected it was found that 
{\em transverse flow does not
increase the transverse radius}. 
On the other hand, a strong dependence of the longitudinal 
radius on the transverse expansion was estblished.

Fig.10 shows two typical examples of the
Bose-Einstein correlations as functions
of $q_\parallel$ and $q_\perp$ for the $1$- and the 
$3$-dimensional hydrodynamical solution. 
The dependence of $C_2 (q_\parallel)$ on transverse expansion 
 agrees qualitatively with what one would expect. 
For a purely longi\-tu\-dinal expansion,
the effective longi\-tu\-di\-nal radius of the source is
 larger
than in the case of 3-dimensional expansion, which is 
reflected in a decrease of the width of the correlation 
function (see also in Fig.11 below).

On the other hand, the results for $C_2(q_\perp)$ were
at a first glance rather surprising.
Naively one might have expected that the transverse
flow would lead to an increase of the transverse radius, 
i.e., to a narrower correlation function $C_2(q_\perp)$.
However, in Fig.10 the curves that describe the 1-dimensional
and the 3-dimensional results are almost identical.
If anything, one would conclude that the effective 
transverse radius is {\it smaller} in the presence of 
transverse expansion.

This effect can however be explained if one takes
a closer look at the details of the hydrodynamic expansion 
process as investigated in ref.\cite{Schlei97}.
 Due to the strong 
correlation between the space-time point where a particle is
emitted, and its energy-momentum, 
the effective radii 
obtained from Bose-Einstein correlation data  present a
characteristic dependence on the average momentum of the 
pair, $K^\mu$.
Fig.11 shows the dependence of the effective radii
$R_\parallel$, $R_{side}$ and $R_{out}$ on the rapidity $y_K$ 
and the mean transverse momentum of the pair $K_\perp$, both
for the 1-dimensional and for the 3-dimensional calculation. 
The longitudinal radius
$R_\parallel$ becomes considerably smaller (by a factor of 
2-3) if transverse expansion is taken into account. 
For the 1-dimensional case, an approximate analytic
expression has been derived for the
$y_K$- and the $K_\perp$-dependence of the longitudinal
`radii`"
in Refs.\cite{LorSin})
\footnote{In this reference eq.(\ref{eq:016}) is used to
disprove the applicability of hydrodynamics to $p-p$ reactions in the 
ISR energy range ($\sqrt{s}=53 GeV$). Such a conclusion seems
dangerous given the approximations involved both in the
derivation of this formula as well as in the interpretation
of the BEC measurements in the above reactions.},\cite{Sin} 
(cf. also \cite{PaGyGa}):
\be
R_\parallel = \sqrt{\frac{2T_f}{m_\perp}} \frac{\tau_o}
{\cosh(y_K)}
\label{eq:016}
\ee
where $m_\perp = (m_{\perp 1}+m_{\perp 2})/2$ is the average
transverse mass of the two particles, $T_f$ is the freeze-out
temperature and
$\tau_o =(\partial u_\parallel / \partial x)^{-1}$ is the 
inverse gradient
of the longitudinal component of the $4$-velocity in the
center (at $x=0$)
\footnote{The expression (\ref{eq:016}) for $R_\parallel$ 
denotes in
fact the
length of homogeneity $L_h$ mentioned above.
 It refers to the region within which the
variation of Wigner function is small. By definition
$L_h \leq R$ where $R$ is the geometrical radius.}. 
In \cite{Schlei97}
 one finds 
that this approximate
expression describes $R_\parallel (K_\perp,y_K)$ for $S+S$
reactions quite 
well, both for
the 1-dimensional and the 3-dimensional case (see Fig.11).
However for $Pb+Pb$ reactions the same
 formula fails to account for the data\cite{Pbpl}. This is not
surprising, because eq.(\ref{eq:016}) is based on the
assumption of boost invariance, i.e. no stopping. 
 This assumption is not justified at SPS energies where
there is considerable stopping. 
The inelasticity increases
with atomic number and this may explain the breakdown of
the above formula.
This exemplifies the limitations of the
boost-invariance assumption, an assumption which must not be
taken for granted but in special circumstances.
 
 In \cite{jan} it was proposed to use
the information obtained from fitting the single inclusive 
distribution to constrain the parameters that enter into the
hydrodynamic description, and then to calculate the 
transverse radius directly. 
Indeed let $\Sigma $ denote the hypersurface in Minkowski
space on which hadrons are produced. Then one can define
e.g. a transverse radius 
\be
R_{\perp}=\frac{\int_{\Sigma}R\frac{p^{\mu}d\sigma_{\mu}}
{\exp[(p^{\mu}u_{\mu}-\mu)/T]-1}}
{\int_{\Sigma
}\frac{p^{\mu}d\sigma_{\mu}}
{\exp[(p^{\mu}u_{\mu}-\mu)/T]-1}}
\label{eq:BOW10}
\ee
where $u_{\mu}, T$ and $\mu$ denote the four-velocity,
temperature, and chemical potential on the hypersurface
$\Sigma$, respectively as in eq.(\ref{eq:single}).     

It is interesting to note that this method for the
determination of transverse radii based on the single
inclusive cross sections provides a geometrical
radius while the use of the second order correlation
function provides a coherence length (length of homogeneity).
  \begin{footnotesize} Comparing the effective transverse radius $R_\perp$ extracted from 
the Bose-Einstein correlation function to the mean 
transverse radius 
as calculated directly in \cite{jan} according to eq.
(\ref{eq:single}) , one finds in
\cite{Schlei97} that the two results 
agree to an accuracy of about 10$\%$. This conclusion is
confirmed and strengthened in a more recent study by Schlei
\cite{Schlei97} for kaon correlations.\end{footnotesize}

Of course, this approach can be used only if a solution of
the equations of hydrodynamics is available;
with pseudo-hydodynamical methods this is not possible.

\subsubsection{Role of resonances and coherence in the 
hydrodynamical approach to BEC}

This problem was investigated using an exact 3+1 d numerical
solution of hydrodynamics in \cite{bernd3}. 

The source 
distribution
$g(x,k)$ was determined from a 3-dimensional solution of the
relativistic hydrodynamic equations. 
Fig. 12 illustrates the effect of successively 
adding the contributions 
from $\rho$, $\omega$, $\Delta$ and $\eta$ decays to the
the BEC correlation functions of directly produced (thermal) 
$\pi^-$ (dotted lines), in longitudinal 
and in transverse direction.      
The width of the correlation progressively 
decreases as the decays 
of resonances with longer lifetimes are 
taken into account, and 
the correlation looses its Gaussian shape. 
The long-lived $\eta$ leads to a decrease of the intercept.

\paragraph{Pion versus kaon interferometry}
 
Ideally, a comparison   
of pion and kaon interferometry should lead to conclusions 
concerning possible differences in the space-time regions 
where these particle decouple from the hot and dense matter. 
 It was proposed that kaons may decouple 
(freeze out)
at earlier times and higher temperatures than 
pions \cite{heinz}.
Indeed, preliminary results had indicated that the effective 
longitudinal and transverse source radii 
extracted from $\pi\pi$ 
correlations
were significantly larger than those obtained 
from $KK$ correlations
\cite{NA44}. 
However, as we have seen in fig. 12, the BEC of pions are
strongly distorted by the contributions from resonance decay.
It was pointed out in ref. \cite{gpkaon}
in a study based on the Lund string model that such
distortions are not present for the BEC of kaons, 
and that consequently  
for the effective transverse radii one expects  
$R_\perp(K^\pm) < R_\perp(\pi^\pm)$, even in the 
absence of any 
difference
in the freeze-out geometry of {\it directly produced} 
pions and kaons.
These conclusions were confirmed in \cite{bernd3} within the
hydrodynamical approach and one
found furthermore that
this effect is even more pronounced if one considers 
longitudinal
rather than transverse radii.
Furthermore the interplay between coherence and resonance 
production which was 
not considered in \cite{gpkaon} was 
studied in \cite{bernd3}.
 There are also some striking differences between 
\cite{gpkaon} and \cite{bernd3} in the 
resonance production cross sections used.  

In fig. 13, BEC functions of $\pi^-$
(solid lines)  and of $K^-$ (dashed lines) are compared, 
at $k_\perp=0$ and $k_\perp=1$ GeV/c, respectively. The dotted  
lines correspond to the BEC function of thermally produced 
$\pi^-$. 
It can be seen that the distortion due to the decay 
contributions from long-lived resonances disappear only at
large $k_\perp$.
Fig. 14 shows the effective radii $R_{||}$, 
$R_{side}$ and $R_{out}$
as functions of rapidity and transverse momentum 
of the pair, both 
for $\pi^-\pi^-$ (solid lines) and for $K^-K^-$ 
pairs (dashed lines).
For comparison, the curves for thermally 
produced pions (dotted lines) are also included.  
 The effective longitudinal radii 
extracted from $\pi^-\pi^-$ correlations are 
considerably larger than 
those obtained from  $K^-K^-$ correlations. 
In the central region
the two values for $R_{||}$ differ by a factor of $\sim 2$. 
For the 
transverse radii, the factor is $\sim 1.3$. 
A comparison between 
results for $K^-$ and thermal $\pi^-$ shows that part of this 
effect can be accounted for by kinematics (the pion mass being 
smaller 
than the kaon mass; see also eq. (\ref{eq:016})). 
Nevertheless, the 
large difference between the widths of pion and 
kaon correlation 
functions is mainly due to the fact that pion correlations are 
strongly affected by resonance decays, 
which is not the case for 
the kaon correlations. In the hydrodynamic scenario of 
ref.\cite{Bolz}, 
about $50\%$ of the 
pions 
in the central rapidity region are the 
decay products of resonances 
\cite{jan}, while less than $10\%$ of the 
kaons are created in resonance 
decays ($K^\star \rightarrow K\pi$ dominates, contributing 
with about $5\%$).

In ref.\cite{bernd3} the problem of coherence within the
hydrodynamical approach to BEC was also investigated.
Fig. 15 shows the $\pi^-\pi^-$ correlation 
functions in the presence of partial
coherence.
In order to extract effective radii from Bose-Einstein
correlation functions in the presence of partial coherence, 
eq.(\ref{eq:015a}) must be replaced by the more general form
\begin{eqnarray}
C_2({\bf k}_1,{\bf k}_2)\: = &1& +\:\:
\lambda \cdot 
\:p_{eff}^2\:
\exp \left[-\: \frac{1}{2}\:
\sum (qR)^2\: \right]
\nonumber\\
&&+ \:\:
\sqrt{\lambda} \cdot 
\: 2\:p_{eff}\:(1\:-\:p_{eff})\:
\exp \left[-\: \frac{1}{4}\:
\sum (qR)^2\: \right]
\label{eq:024}
\end{eqnarray}

\subsubsection{Comparison with experimental data}

Some of the predictions made in \cite{Bolz} for $S+S$ 
reactions  could be checked
experimentally in refs.\cite{Alber}\cite{Ferenc}, 
in particular the rapidity and transverse momentum
dependence of radii and remarkable agreement was found. 
%(cf.Fig. 16,17). 
 In \cite{OrnikPR96} the
hydrodynamical calculations were extended to $Pb+Pb$
reactions and compared with $S+S$ reactions and, where data
were available, with experiment. 
The calculation of Bose-Einstein correlations (BEC) was
performed using the formalism outlined in refs.
\cite{bernd2,bernd3,Schlei97} including the decay of 
resonances. The
hadron source was assumed to be fully chaotic.  

Figs.16 and 17  show the calculations for the effective radii
$R_\parallel$, $R_{side}$ and $R_{out}$ as functions of 
rapidity $y_K$
and transverse momentum $K_\perp$ of the pion pair compared 
to the corresponding NA35 and preliminary NA49 data 
\cite{alber,QM95}, respectively. All these
calculations, which in the case of $S+S$ had been true 
predictions, agree surprisingly well with the data\footnote
{The EOS used
in the hydrodynamical studies quoted above included a phase
transition from QGP to hadronic matter. How critical this
assumption is for the agreement with data is yet unclear and
deserves a more detailed investigation. On the other hand
the very use of hydrodynamics is based on local equilibrium,
and this equilibrium is favoured by the large member of
degrees of freedom due to a QGP.}. 
This suggests that our understanding of BEC in
heavy ion reactions has made progress and confirms the
usefulness of the Wigner approach when coupled with full
fledged hydrodynamics.
An important issue in comparing data with theory is the
detector acceptance of a given experiment. This is also 
discussed in detail in \cite{OrnikPR96}.

{\footnotesize Another application of hydrodynamics to the QGP search in 
heavy ion reactions is due to Rischke and Gyulassy 
\cite{Gyu96} who investigate the ratio $r=R_{out}/R_{side}$.
 Based
on considerations due to Pratt \cite{Pratt}, this quantitiy
 had been proposed by Bertsch and
collaborators\cite{Bertsch} as a signal of QGP. Under
certain circumstances
one could expect that for a long lived QGP phase
$r$ should exceed unity, while for a hadronic system, due to
final state interactions, the out and side sizes should be
comparable.
 
 The authors
of \cite{Gyu96} performed a quantitative hydrodynamical
study of $r$ in order to check whether this signal survives
a more realistic investigation, albeit they did not take
into account resonances. For directly produced pions it is
found that $r$ indeed reflects the lifetime of an
intermediate (QGP) phase. However we have seen from 
\cite{Bolz}, \cite{OrnikPR96} that for pion
BEC, when resonances are considered,
 hydrodynamics with an
EOS containing a long lived QGP phase, leads (in agreement
with experiment) to values of $r$ of order unity. 

 To avoid this complication, in \cite{Risch} it was
proposed to consider kaons at large $k_{\bot}$.
However even this proposal may have to be qualified, besides
the fact that it will be very difficult to do kaon BEC at
large $k_{\perp}$.
Firstly one has to recall that the entire formalism on which
the $r$ signal is based and in particular the parametrization 
(\ref{eq:015a}) is questionable. Secondly    
 it remains to be proven that this signal survives if
one imposes simultanoeusly the essential constraint due to the single
inclusive distribution\footnote{I am indebted to B. R.
Schlei for this remark.}. Furthermore it is unclear up to
what values of  $k_{\perp}$ the
Wigner formalism, (which is a particular case of the
classical current formalism) on which the theory is based, 
is applicable. For these reasons the determination of the 
lifetime of the
system via ``pure" hydrodynamical considerations is
certainly an alternative which deserves to be considered
seriously, despite its own difficulties.}

\subsubsection{Bose-Einstein correlations and
pseudohydrodynamics}

As mentioned already, the initial motivation for proposing
the Wigner function formalism for BEC was to explain why the
experimentally observed second order correlation
functions $C_2$ were depending not only on the momentum 
difference 
$q=k_1-k_2$ but also on their sum $K=k_1+k_2$. However in
the mean time it was shown \cite{aw},\cite{APW} that this
feature follows from the proper application of the
space-time approach in the 
  current formalism even without assuming expansion. 
Furthermore the Wigner formalism is useful only at small $q$
 and cannot be applied in the case of strong correlations 
between positions and momenta
while the current formalism is not limited by these
constraints. As
explained above the use of the Wigner formalism can
be defended if combined with bona-fide hydrodynamics and an
equation of state. 
 
This notwithstanding, besides a few real, 
albeit numerical, hydrodynamical 
calculations, most phenomenological papers on BEC in heavy 
ion reactions (cf. e.g.
\cite{MS},\cite{Bert1},\cite{Pratt3}, \\ 
\cite{Bertsch94},      
 \cite{Chao94}-\cite{Bert2},\cite{Akk},\cite{csorlor},        
 \cite{Csorg4}-\cite{Wu98}
have used the Wigner formalism without a proper
hydrodynamical treatment, i.e. without solving the equations
of hydrodynamics; hydrodynamical concepts
like velocity and temperature were used just to parametrize 
the Wigner source function. While such a procedure may be 
acceptable as a theoretical exercise, it is certainly no
substitute for a professional analysis of heavy ion
reactions. This is a fortiori true when real data have to be
interpreted\footnote{A recent experimental paper \cite{2Na49}
where such
a procedure is used is a good illustration of the limits of
 pseudo-hydrodynamical models. Despite the fact that the
statistics are so rich that ``the statistical errors on the
correlation functions are negligible", the outcome of the
analysis is merely the resolution of the ambiguity between
temperature and transverse expansion velocity of the source.  
It is clear that such an
ambiguity is specific to pseudo-hydrodynamics and is from
the beginning absent in a correct hydrdynamical traetment. 
Moreover even this result is questionable given some
 doubtful assumptions
which underly this analysis. To quote just two: 
(i) The assumption of boost
invariance made in \cite{2Na49} decouples the longitudinal 
expansion from the
transverse one. This not only affects the conclusions drawn 
in this analysis but prevents the (simultaneous) 
interpretation of the
experimental rapidity distribution.
(ii) The neglect of long lived resonances which strongly
influences the $\lambda$ factor and thus also the extracted
radii. 
Of course, despite the claimed richness of the data, no
attempt to relate the observations to an equation of state
 can be made within this approach.}.

As exemplified in previous sections such a
procedure is unsatisfactory, among other things because it
can lead to wrong results.

The use of this ``pseudo-hydrodynamical"
approach is even
more surprising if one realizes the fact that the Wigner
formalism not only is not simpler 
than the more general current
\footnote{We remind that the classical current formalism, in
opposition to the Wigner function formalism, does not apply
only to small $q$ and to semiclassical situations.
Furthermore it
allows also for coherence and new phenomena like
particle-antiparticle correlations.}   
formalism but it is also less economical. The number of 
independent 
parameters necessary to characterize the BEC within the
Wigner formalism of ref.\cite{Nix} is 
\footnote{For each value of ${\bf K}$ a Gaussian ellipsoid is
described by three spatial extensions, one temporal
extension, three components of the velocity in the local
rest frame and the three Euler angles of orientation.}
$10$, i.e. it is as 
large as that in the current formalism. However the $10$
parameters of \cite{Nix} describe a very particular source
\footnote{To consider such an approach as ``model independent"
\cite{Heavy}, \cite{HeinzNP} is misleading.},
as compared with that of the current formalism: besides the
fact that the correlation function 
source is assumed to be
Gaussian, it is completely chaotic and it can provide
only the length of homogeneity $L_h$ \cite{MS}. 
For the search of quark gluon plasma, however, the
geometrical radius $R$ is relevant, because the energy 
density is defined in terms of $R$ and the use
of $L_h$ instead of $R$ leads to an overestimate of the
energy density \footnote{One may argue that the ``length of 
homogeneity"
\cite{MS} $L_h$ defined in terms of the Wigner function 
is a particular case of the
correlation length $L$ defined in terms of the correlator
(cf. section 4.3).
While $L_h$ is always limited from above by $R$, $L$ can
be either smaller or larger than $R$.}.   
Furthermore the physical significance of the parameters
of the Wigner source is unclear if the Gaussian assumption 
does not hold\footnote{Even if a Gaussian form would hold 
for directly
produced pions, resonances would spoil it \cite{Bolz}, 
\cite{OrnikPR96},
\cite{Schlei97}.}. 
Not only is there no a priori reason for a Gaussian 
form,  but on the contrary, 
 both in nuclear and particle physics
 as well as
in quantum optics, there exists experimental evidence that
in many cases an exponential function in $|q|$  is at small
$q$ a better
approximation for $C_2$ than a Gaussian. Furthermore, in the
presence of coherence, no single simple analytical 
function, and in particular no single Gaussian is expected 
to describe $C_2$. This is a straightforward consequence of 
quantum statistics.

Given the fact that good experimental BEC data are expensive
both in terms of accelerator running time and man-power, 
the use of inappropiate theoretical tools, when more
appropiate ones are available, is a waste which has
to be avoided.    
For the reasons quoted above we will not
discuss in more detail the numerous and sometimes
unnecessarily long
papers which use pseudo-hydrodynamical methods.

\paragraph{Concluding remarks.}
In a consistent treatment of single and double inclusive 
cross sections for
identical pions via a realistic hydrodynamical model, 
resonances 
play a major
role leading to an increase of effective radii of sources. 
Effective longitudinal radii are more sensitive
to the presence of resonances than transverse ones.
>From the hydrodynamical treatment we
learn that the hadron source (the real fireball) is 
represented
by a very complex freeze-out hypersurface (cf. ref.
\cite{OrnikPR96}). The
longitudinal and transverse extensions of the fireball 
change dynamically as a function of time, rather than show 
up in static effective radii.
Thus, the interpretation of BEC measurements is also 
complicated.  
This is a theoretical task and if only for this
reason experimentalists
would be well advised to include in their teams also
qualified theorists.

For heuristic applications, 
when no
quantitative 
comparison with data is intended, besides 
the
current formalism, analytical approximations of the
equations of hydrodynamics can be useful, because they allow
a better qualitative understanding of hydrodynamical
expansion.  However when a
quantitative interpretaion of experiments is intended and 
in particular a connection with the equation of state is
looked for,
the only recommendable method is full-fledged hydrodynamics.

\subsubsection{Photon correlations, pseudo-hydrodynamics and
pseudo-Wigner formalism}

 Photon correlations
have been investigated within the context of quark-gluon
plasma search, since they present certain methodological
advantages as compared with hadrons. While experimentally
genuine photon BEC in high energy heavy ion reactions have
not yet been unambigously identified, because of the strong 
$\pi{^0}$ background and
the small cross sections for photon production,
there are several theoretical studies devoted to this topic.
The
advantage of photon BEC resides in the fact that, while 
correlations between
hadrons are influenced by final state interactions, photon
 correlations are ``clean" from this point of view. 
For high energy physics photons
present another important advantage due to the fact that
they can provide direct information about the early stages 
of the interaction where quarks and gluons dominate and 
hadrons have not yet been created.
 In particular photon BEC contain
information about the lifetime of the quark-gluon plasma 
\cite{kap}-\cite{Sriv}, 
\cite{Sriv5}, \cite{Axel}.
Among other things it was argued e.g. in \cite{Sriv5} 
and 
 confirmed in \cite{Axel} using a more correct
formalism, (cf. below),
that the 
correlation function $C_2$ in the
transverse direction can serve as a signal for QGP as it is
sensitive to the existence of a mixed phase.

Unfortunately, some of these studies 
(refs.\cite{kap}-\cite{Sriv},
\cite{Sriv5}) besides suffering from
the more general disease of pseudo-hydrodynamics, use an
input formula for the second order BEC, which is essentially
incorrect even within the Wigner-formalism (cf. section 4.6).
Besides this, some approximations made in \cite{kap} e.g.
are inadequate.
This question was analyzed in more detail in \cite{Axel} 
where it was found  that only some of the results of ref.
(\cite{kap}) survive a more critical analysis. 

  The formula for the two-particle inclusive probability 
used in (\cite{kap}-\cite{Sriv},\cite{Sriv5})
reads

\begin{equation}
P_2({\bf k}_1,{\bf k}_2) \ =\ \int d^4x_1 \int d^4x_2\ 
g\left(x_1,k_1\right)\ 
g\left(x_2,k_2\right)\ [1\ +\ \cos((k_1-k_2)(x_1-x_2))],  
\label{eq:4}
\end{equation}

while ref.(\cite{Axel} uses the more correct formula

\begin{equation}
P_2({\bf k}_1,{\bf k}_2) \ =\ \int d^4x_1 \int d^4x_2\ 
g\left(x_1,\frac{k_1+k_2}{2}\right)\ 
g\left(x_2,\frac{k_1+k_2}{2}\right)\ [1\ +\ \cos((k_1-k_2)
(x_1-x_2))],  
\label{eq:3a}
\end{equation}

(From eq.(\ref{eq:3a}) one gets for the second order
correlation function eq.(\ref{eq:Heinz23})).  
It is found in \cite{Axel} that two rather
surprising properties of the two-photon correlation function
presented in \cite{kap} are artifacts of
inappropriate approximations in the evaluation of space-time
integrals. In \cite{kap}, it was claimed that the BEC
function in the longitudinal direction (a) oscillates and
(b) takes values below unity. As property (b) is
inconsistent with general statistical bounds, it was
important to clarify the origin of this discrepancy. 
On the other
hand, it was confirmed in \cite{Axel} that the 
correlation function in the
transverse direction does exibit oscillatory behaviour in
the out component of the momentum difference.
Furthermore, in this reference
a change of the BEC function in $\Delta y$ from Gaussian to 
a two-component shape with decreasing transverse photon 
momentum was found which may serve as evidence for the 
presence of a mixed phase and, hence, as a QGP signature.

However even after correcting the wrong input BEC formula
it is
questionanble whether the other approximations made in
refs.(\cite{kap}-\cite{Sriv},\cite{Sriv5},\cite{Axel}) may 
not invalidate the
above result. Besides the use of a simplified hydrodynamical
solution one has to recall that 
(i) the Wigner formalism like
the more general classical current formalism, is limited to
small momenta $k$ of produced particles (no recoil
approximation)
(ii) Besides this general limitation to small $k$ the
Wigner formalism is specifically limited to small differences
of momenta $q$.
(iii)  Although Eq. (\ref{eq:Heinz23}) does not suffer from
the violation of unitarity disease mentioned in section 4.6
of chapter 3, it is based on an approximation which 
makes it sometimes inapplicable for photons 
(cf. section 4.5). 
>From the above discussion one may conclude that the 
experimental problems of photon BEC are matched by
yet unsolved theoretical problems.

\subsection{Pion condensates}

One of the most interesting phenomena related to
Bose-Einstein correlations is the effect of Bose
condensates. The remarkable thing about this effect is that
it is not specific for particle or nuclear physics, but
occurs in various other chapters of physics, in particular
in condensed matter physics like superconductivity and 
superfluidity. Moreover, recently not only the condensation 
of a gas of atoms has been experimentally achieved 
\cite{Atomcond95} but the 
quantum statistical coherence of these systems has been
experimentally proven through Bose-Einstein
correlations \cite{Atomcond97}.  
The proposal to use BEC for the detection of condensates
was made a long time ago \cite{Stelte}. The more recent 
developments in heavy ion
reactions have made this subject of current interest.  
Already in
experiments at 
the SPS (e.g., $Pb+Pb$ at $E_{beam}=160$ AGeV) secondary 
particles are 
formed at high number densities in rapidity space
\cite{pbpb} and 
in future experiments at RHIC and 
the LHC one expects to obtain even higher multiplicities 
on the order
of a few thousand particles per unit rapidity. If local 
thermal (but not chemical) equilibrium is established and 
the number densities are 
sufficiently large, the pions may accumulate in their ground
state and 
a Bose condensate may be formed\footnote{Recently a different type of 
pion condensate, the disordered chiral condensate, has 
received much 
of attention in the literature. 
 It has been argued \cite{kogan} that
such
a condensate would lead to the creation of squeezed states.}. 
A specific scenario for the formation of a Bose condensate, 
namely, 
the decay of short-lived resonances, was discussed in 
Ref. \cite{bo-co} 
where conditions necessary for the formation of a Bose 
condensate in a heavy ion collision were investigated.  
In Ref. \cite{bo-co} it was found that if a pionic Bose 
condensate 
is formed at any stage of the collision, it can be expected 
to survive 
until pions decouple from the dense matter, and thus it can 
affect the spectra and correlations of final state pions.

 In ref.\cite{cond97} one investigated the influence of
such
a condensate
on the single inclusive cross section 
and on the second order correlation function of identically
charged pions (Bose-Einstein correlations BEC) in hadronic
reactions for {\em expanding} sources.
A hydrodynamical approach was used 
based on the HYLANDER routine.
The Bose-condensate affects the single inclusive 
momentum distributions $EdN/d^3k$, the 
momentum-dependent chaoticities $p$ and
the Bose-Einstein correlation functions $C_2$
 only over a limited momentum range.
This is due to the fact that in a condensate 
there exists a maximum velocity (which implies also a
maximum momentum difference $q_{max}$)
and leads to a 
very characteristic structure in single and double inclusive
spectra.

In Fig. 18 the results of the numerical 
evaluations of the
 Bose-Einstein correlation functions $C_2$
are shown for a spherically and for a longitudinally 
expanding source.
The presence of a Bose-condensate of only $1\%$ 
results in a decrease of the intercept by about 15\%.
 Furthermore  
due to a limited value of $q_{max}$ a part of the tail of
the two-particle correlation functions is not affected
by the pionic Bose-condensate and a peak appears.
To what extent such peaks 
can be observed in experimental data
depends among other things on
 the size of the source,
 the details of the freeze-out,  
 width of the momentum distribution in the bosonic
ground state, and detector acceptance.

\paragraph{Plasma droplets?}
If the phase transition from hadronic matter to QGP and
\\viceversa is of first order then one could expect  the
formation of a mixed phase, in which QGP and hadronic matter
coexist. Such a mixed phase manifests itself  in
the hydrodynamical evolution of the system \cite{mixed} 
and it influences among other things the transverse 
momentum distribution of photons as we have seen
in section 5.1.6. It was
suggested by Seibert \cite{Seibert} that the mixed phase could 
also lead to a granular structure which might be seen in the
fluctuations of the velocity distributions of secondaries
produced in the hadronization stage. Pratt, Siemens and
Vischer \cite{droplet1} (cf. also \cite{droplet2} proposed 
subsequently that in Bose-Einstein
correlations, too, 
one might see a signature of this granularity.

\setcounter{equation}{0}

\section{Correlations and 
Multiplicity Distributions}  
\subsection{From correlations to multiplicity distributions} 

An important physical observable in multiparticle production
is the multiplicity distribution\footnote{The multiplicity
distribution will be denoted sometimes in the following
also by MD} $P(n)$, i.e., the probability
to produce in a given event $n$ particles. The link 
between the 
multiplicity distribution $P(n)$ and BEC is
represented by the density matrix $\rho$ , since it is the
same $\rho$ which appears in the definition of $P(n)$
\be
P(n) \equiv <n|\rho|n> 
\label{eq:rho}
\ee
and the definition of correlation functions (cf.e.g.
eq.(\ref{eq:cln})). 
Usually one expresses $\rho$ in terms of the $\cal
P(\alpha)$ representation (cf. section 2.2 )
and by using for $\cal P(\alpha)$ simple analytical 
expressions 
one is able to derive the most characterisitc forms of
$P(n)$ in an analytical form,
like the Poisson or the negative binomial representation
(cf. e.g. \cite{book}).

Sometimes there exist however physically interesting cases
where no analytical expression for 
the multiplicity distribution $P(n)$ exists, but instead
the moments of $P$ are given.
>From the phenomenological point of view the approach to MD
via
moments presents sometimes important advantages because 
it allows the construction of an {\em effective} 
density matrix from the knwoledge of a few physical
quantities
like the correlation lengths and mean multiplicity, which in
turn
can be obtained from experiment (cf.e.g. \cite{Botke}). 
In the following we will address
this aspect of the problem, the more so that
in this way the link between correlations and MD 
becomes clearer. 

We start by recalling some definitions. Besides the normal
moments of the MD given by

\be
<n^q> \ \equiv \ \sum_n \ P(n) n^q
\label{eq:normal}
\ee

one uses frequently the factorial moments 

\begin{equation}
\Phi_q \ \equiv \ 
\sum_n \ P(n)\ \left(n(n-1)\cdot...\cdot(n-q+1)\right)
\label{eq:deffq}
\end{equation}

These can be expressed in terms of the inclusive 
correlation functions $\rho_q$ through the relation  

\begin{equation}
\Phi_q \ =\ \int_\Omega d\omega_1 ...\ \int_\Omega d\omega_q 
\ \ \rho_q({\bf k}_1 , ... , {\bf k}_q) \ = \ \langle
\frac{n!}{(n-q)!}  \rangle
\label{eq:facmom}
\end{equation}
Eqs. (\ref{eq:deffq}) and (\ref{eq:facmom}) illustrate
the fundamental fact that the inclusive cross sections
$\rho_q$ and thus
the correlation functions determine 
the moments of the MD, which are nothing else but 
the integrals of $\rho_q$. 
Although this relation between moments of MD and correlation
functions is a straightforward
aspect of multiparticle dynamics, the connection 
between MD and correlations was often overlooked. This is in
part due to the fact that
measurements of correlations are, for
reasons of statistics and other technical considerations,
frequently performed in different (narrower) regions of phase
space than measurements of MD
\footnote{This is e.g the case when MD are
measured with no proper identification of the particles,
while BEC refer of course to identical particles. Thus the
UA5 experiment \cite{UA5}, which discovered the violation of
KNO scaling in MD-s,  
measured only {\em charged} particles, without
distinguishing between positive and negative charges. However
the rapidity region accessible in this experiment was much
broader than the corresponding region in the UA-1 experiment
\cite{bu} where correlation measurements were performed and 
where a
distinction between positive and negative charges could be
made.}. However the importance of the use of the
relationship between MD and correlations can hardly be
overemphasized, just because of the different experimental
methods used in the investigation of these two observables. 
In the absence of a theory of multipaticle production, the
form of the correlators and the amount of chaoticity are 
unknown and have therefore to be parametrized and then
determined experimentally in correlation experiments. Both 
the parametrization and the measurements are affected by
errors. Similar considerations apply for MD, but because of
the different experimental conditions under which
correlation measurements and MD measurements are made, the
corresponding errors are different. 
 (Cf. also the discussion of the importance of higher order
correlation,  section 2.2.1).
Therefore from a phenomenological and practical point of
view, MD and correlations are rather complimentary and have 
to be interpreted together.     
In the following we will exemplify the usefulness of this
point of view.

\subsubsection{Rapidity dependence of MD in the stationary 
case}
The dependence of moments of multiplicity distributions 
$P(n)$ on the width of the bins in momentum or rapidity space
has been in the center of multiparticle production studies 
for the last 15 years. 
It got much attention after: 1) the experimental observation
\cite{UA5} 
by the UA5 collaboration that the normalized moments of 
$P(n)$ in the rapidity plateau region increase 
with the width of the rapidity window $\Delta y$; 2) the 
proposal by 
Bialas and Peschanski \cite{bipesh} that this behaviour,
which at a first look was power-like may reflect 
``intermittency", i.e. the absence
of a fixed scale in the problem, which could imply that 
self-similar phenomena play a role in multiparticle 
production.  

However soon after this proposal was published it was
pointed out in \cite{car} that the quantum statistical
approach, presented in section 2.2 and which implies a
{\em fixed} scale 
\footnote{This scale is the correlation length $\xi$ of eqs.
(\ref{eq:Lorentz}) or (\ref{eq:Gaussdef}).},
predicts a similar
functional relationship between the moments of MD and    
$\Delta y$. 
 In Fig. 19 from \cite{car} 
some examples of this behaviour are plotted and compared
with experimental data. For small $-\delta y$ the
(semilogarithmic) plot can be approximated by a power
function as indicated by the data. Recalling that the QS
formalism applies to 
identical particles, it follows that BEC could be at the
origin of the so called intermittency effect. This point of
view was corraborated subsequently also in \cite{gyul} and
\cite{cap} and was confirmed experimentally by the
observation that the ``intermittency" effect is strongly
enhanced when identical particles are considered
 \footnote{The UA5 data refer to a
mixture of equal numbers of positive and negative particles;
this dilutes the BEC effect.}
and /or when studied in more than one dimension
\footnote{This last
observation also supports the idea that BEC is the determining
factor in intermittency because the integration over
transverse momentum implied by a a one dimensional $y$
investigation diminishes the BEC.}. 
For a more recent very clear confirmation of this point of
view cf. the studies by Tannenbaum \cite{tan}. 
Further
developments related to ``intermittency" will be discussed in
section 6.3.

\subsubsection{Rapidity dependence of MD in the non-stationary
case}

The assumption of translational invariance in rapidity
permitted to apply the quantum optical formalism, in which
time has the analogous property, to MD and led to a simple
interpretation of the observed broadening of the MD with the
 decrease of the width of the rapidity window in high energy
 reactions.
However stationarity in rapidity
 is expected to hold only in the central region
(and only at high energies). Indeed experimental data on
proton(antiproton)-proton collisions in the energy range 
$52 < \sqrt{s}  <  540 GeV$ show
that if one considers shifted rapidity bins along the
rapidity axis the MD in these bins depend on the position 
of the bin: in the central region the MD is broader and can
be decribed by a negative binomial distribution while in the
fragmentation
region it is narrower and can be described by a Poisson
MD \cite{2comp}. 
 The mean multiplicity in the
central region increases faster (approximately like
$s^{1/4}$)
with the energy than that in the fragmentation region.
This was interpreted in \cite{2comp} as 
possible evidence for
 the existence of two sources, one of chaotic nature
localized in the central region and another coherent in the
fragmentation region. This interpretation is in line with 
the folklore that gluons
which interact stronger (than quarks) form a central blob
which may be equilibrated, while the fragmentation region
is populated by througoing quarks associated with 
the leading particles. 
\footnote{For a microscopic interpretation of this effect 
in terms of a partonic stochastic model, cf.
ref.\cite{stoch}.}.
 
A few years later the NA35-collaboration
\cite{Na35} measured BEC in $^{16}O-Au $ reactions at 200
GeV/nucleon in a relatively broad y region and found
evidence for a larger and more chaotic source in the central
rapidity region, and a smaller and more coherent source in
the fragmentation region. It was then natural to correlate
(cf. ref.\cite{revisit}) the two observations, i.e. that
refering to MD and that refering to BEC. Taken together the
credibility of the conjecture made in ref.\cite{2comp} is
strongly enhanced, just because we face here different
physical observables and different experiments, each with
its own specific corrections and biases.
Moreover it was
pointed out in \cite{revisit} that experiment
\cite{Na35}
did not necessarily imply that the two sources were
independent, but could also be interpeted as due to a single
partially coherent source. 

Indeed consider in a
simplified approach as used e.g. in quantum optics
a superposition $\pi$ of two fields
 one coherent
denoted by  $\pi_c$ and another chaotic denoted by
$\pi_{ch}$ so that 
$\pi = \pi_{c}(k_{\perp}^{(1)} +\pi_{ch}(k_{\perp}^{(2)})$. 
whre $Q_{\perp} =k_{\perp}^{(1)} -k_{\perp}^{(2)}$.
Assuming boost
invariance the correlator depends only on $k_{\perp}$ and we
have for the second order correlation function  
\be
C_2=1+2p(1-p)e^{-Q^2_{\perp}R^2_{\perp}/2}+p^2e^{-Q^2_
{\perp}R^2_{\perp}}
\label{eq:super1}
\ee
where the transverse ``radius" $R_{\perp}$ plays the role of
the correlation length and 
$p=\frac{<|\pi_{ch}>|^2>}{|\pi_c|^2 + <|\pi_{ch}>|^2>}$.

Assume now that the chaoticity $p$ is
rapidity dependent 
so that in one rapidity region, denoted
by (A), $p(A) \approx 1$. In that region
then the third term in the equation (\ref{eq:super1}) 
dominates, i.e.
\be
C_2  \approx  1 +  p^{2}(A)e^{-Q^{2}_{\perp}R^2_{\perp}(A)}
\label{eq:chaotic}  
\ee
In the parametrization used in \cite{Na35}, according to
which we have two independent sources, 
this suggests that the effective radius of source A is
$R^{A}_{\perp}$. Conversely, for the more coherent region 
denoted by B, $p(B)$ is small and $C_2$ reads
\be   
C_2 \approx 1+2p(B)(1-p(B))e^{-Q^{2}_{\perp}R^{2}_{\perp}(B)
/2}
\label{eq:coherent}
\ee
with an effective radius $R_{\perp}(B)/\sqrt{2}< R_{\perp}(A)$. 

This corresponds qualitatively to the observations made in 
ref.\cite{Na35}. Unfortunately these observations
have not yet been confirmed by another, independent experiment 
so that the reader should view these considerations with
prudence 
\footnote{For various caveats concerning the analysis of
BEC and MD data cf.
refs.(\cite{Na35},\cite{UA5},\cite{2comp},\cite{revisit},
\cite{Perugia}, \cite{QM88}).}.
In any case they prove
the usefulness of a global analysis which incorporates both
BEC and MD. On the other hand, a dedicated simultaneous 
investigation of the rapidity dependence of these two
observables appears very desirable.  
  
\subsubsection{Energy dependence of MD and its implications
for BEC; long range fluctuations in BEC and MD}

This subject has been discussed
recently in \cite{lrc}. 
Besides the rapidity dependence,
the dependence of MD on 
the center of mass energy of the collision, $\sqrt{s}$,
constitutes an important topic in the study of high energy  
multiparticle production processes. This energy 
dependence is usually discussed in terms of the violation 
\cite{UA5} of KNO scaling \cite{KNO}.
 KNO scaling implies 
that the
normalized moments $\langle n^m \rangle
/\langle n \rangle^m$ are constant as a function of $s$
(for high energies, i.e. large $\langle n \rangle$, 
these moments coincide with the normalized factorial moments). 
For charged particles it turned out that while KNO scaling 
is approximately satisfied over the range of ISR energies 
($20$ $GeV\leq \sqrt{s} \leq 60$ $GeV$), it is violated if 
one goes to SPS-Collider energies ($200$ $GeV\leq \sqrt{s} 
\leq 900$ $GeV$), i.e. one finds a 
considerable increase of multiplicity fluctuations with 
increasing energy. 
In the following we will show how BEC can be used to
to understand the origin of this 
$s$-dependence. 
To do this, one needs to distinguish between long range 
(dynamical) correlations (LRC)
and short range correlations (SRC)\footnote{The importance 
of making this 
distinction was pointed out, among other things, in
\cite{foa,capella}. In \cite{faes} rapidity correlations were
measured for events at fixed multiplicity in order to get 
rid of the effect of LRC.}. 
For identical bosons, one important type of SRC are  
 BEC which reflect quantum statistical interference. 
In addition, there exist dynamical SRC like 
final state interactions, which however are quite 
difficult to be separated from BEC.  Up to 1994 
one usually assumed that LRC do not play an important part in 
BEC measurements  
(see however \cite{gyul},\cite{cap})  
in the sense that for $Q>1 \ GeV$ the two-particle correlation 
functions do not significantly exceed unity. 

However in ref.\cite{lrc} evidence, 
based on observational data, was presented showing
that this is not the case and that new and important
information about LRC  
is contained in the BEC data obtained by the UA1-Minimum-Bias
Collaboration \cite{bu}. 
In principle, the observed increase of multiplicity 
fluctuations 
with $\sqrt{s}$ could be due to a change of the SRC as seen 
in BEC, i.e. of the chaoticity and radii/lifetimes.
This possibility was discussed in \cite{fowler} but could 
not be tested because of the lack of identical particle data
for multiplicity distributions at Collider 
energies.  At that time only the UA5 data \cite{UA5} 
for multiplicity distributions of 
charged particles were available.
Furthermore, up to this point, the effect of LRC
had only been studied in terms of two-particle correlations 
as a function of rapidity difference, i.e. in one dimension.
With the advent of the newly 
analyzed UA1 data \cite{bu} for identical particles in three
dimensions  (essentially in $Q^2=-(k_1-k_2)^2$) this situation 
 changed. The analysis of \cite{lrc} led to the conclusion
of the existence in BEC data of long range fluctuations 
in the momentum space density of secondaries 
and to the realization that the increase with energy  
of multiplicity fluctuations is to a great extent due to 
an increase of the asymptotic values of the 
$m$-particle correlation functions $C_m^{asympt.}$, i.e. 
their values in the limit of large 
momentum differences where BEC do not play a role 
\footnote{The effect of LRC on BEC was discussed 
in ref.\cite{gyu}, where several  models specific 
for nucleus-nucleus collisions were considered, but
at that time no
evidence for this effect could be found.}.

In ref.\cite{bu}, the UA1 collaboration presents
the two-particle correlation of negatively charged 
secondaries as a function of the invariant 
momentum difference squared $Q^2=({\bf k}_1-{\bf k}_2)^2-
(E_1-E_2)^2$.
The data (cf. Fig. 20) have   
two unusual features: (I) at large $Q^2$ the correlation 
function  saturates above unity, and (II) at small $Q^2$ it 
takes on values above $2$.  The higher order 
correlation functions also exhibit property (I)
\cite{bu}\footnote{For higher order correlations the 
equivalent of property (II) is $C_m>m!$. (cf. also below). 
The values of 
$C_m({\bf k},...,{\bf k})$ for $m>2$ are apparently not yet 
available.}.
By comparing
the asymptotic values of the correlation functions at large 
momentum differences 
$C_m^{asympt.}$ ($m=2,...,5$),
with the normalized factorial moments,
$\phi_m \equiv \langle n (n-1)\cdot ... (n-m+1)\rangle /
 \langle n \rangle^m$ in the momentum space region$
|y| \leq 3, k_{\perp} > 0.15$ $GeV$ 
one finds that the contribution of the BE interference peak 
to the 
moments is negligible for such large rapidity windows.
Herefrom one concludes in \cite{lrc} that (I) indicates 
the presence of LRC in the momentum space
density of secondary particles and that it is quite plausible (cf. below) 
that (II) has to a great extent the same explanation. We
sketch here the arguments of ref.\cite{lrc}.

In general, LRC may be related to fluctuations
in impact parameter or inelasticity, or fluctuations in the 
number of sources. In what follows, let us label these 
fluctuations by a parameter $\alpha$. 
The $m$-particle Bose-Einstein correlation function
at a fixed value of  $\alpha$ is 
given by
\be
C_m({\bf k}_1,...,{\bf k}_m|\alpha) \ =\ \frac
{\rho_m({\bf k}_1,...,{\bf k}_m|\alpha)}
{\rho_1({\bf k}_1|\alpha) \cdot ... 
\cdot \rho_1({\bf k}_m|\alpha)}
\ee
where $\rho_\ell({\bf k}_1,...,{\bf k}_\ell|\alpha)$
are the $\ell$-particle inclusive distributions. 

The fluctuations in $\alpha$ are described by
a probability distribution $h(\alpha )$ with
\be
\int d\alpha \ h(\alpha) \ = \ 1.
\ee

If the experiment does not select events at fixed $\alpha$,
the measured inclusive distributions are 
\be
\rho_m({\bf k}_1,...,{\bf k}_m) \ =\ 
\langle\rho_m({\bf k}_1,...,{\bf k}_m|\alpha)\rangle\:,
\ee
where the symbol $\langle ... \rangle$ denotes an average 
over the fluctuating parameter $\alpha$, i.e.
\be
\langle X(\alpha ) \rangle \ \equiv \ \int d\alpha \ h(\alpha) \ 
X(\alpha )
\ee

The $m$-particle correlation function at the intercept 
reads
\be
C_m({\bf k},...,{\bf k}) \ =\ m! \ \frac{\langle\alpha^m
\rangle}{\langle \alpha 
\rangle^m}, 
\ee
where the symbols $<>$ refer to averaging with respect to
$h(\alpha)$.
 At large momentum differences one has 
\ba
&&C_m({\bf k}_1,...,{\bf k}_m) \ \rightarrow \ 
\frac{\langle \alpha^m\rangle }{\langle \alpha 
\rangle^m} = C_m^{asympt.} 
\qquad
 {\rm for} \quad |{\bf k}_i-{\bf k}_j|\rightarrow \infty 
\nonumber\\
&&(i\neq j, \ i,j=1,...m),
\label{eq:cal}
\ea
i.e., the $m$-particle 
correlation functions can have intercepts above $m!$ and 
saturate 
at values above unity for large momentum differences.

The most obvious candidate for the function $h(\alpha)$ is 
the inelasticity
distribution which describes the event to event fluctuations
of the inelasticity $K$. With the identification 
$\alpha \equiv \langle n(K)\rangle$, where
$\langle n(K)\rangle$ is the mean multiplicity at 
inelasticity $K$, one obtains from the above
considerations a first ``experimental" information about this important physical
quantity at collider energies. Previous experimental 
information about this
distribution was derived in \cite{fow1} from the data of ref. \cite{brick} at
$\sqrt s \simeq 17$ $GeV$.

The conclusions obtained in \cite{lrc} 
about LRC are based among other things on the different
normalizations used in different experiments. 
 Some tests related to these conclusions 
 are proposed:
\begin{itemize}
\item Analysis of BEC at lower energies (NA22 range) as well
as at $\sqrt s =1800$ $GeV$ with the same normalization as 
that used by the UA1 Collaboration.
The values of $C_2$ at $Q^2 > 1$ $GeV^2$ and possibly also at
very small $Q^2$ obtained in this way should exceed those 
obtained with the fixed multiplicity normalization in the 
same experiments. The following inequalities for $C_2(Q^2 > 
1$ $(GeV)^2)$, should be observed  if this normalization is 
used:\\
$C_2({\rm NA22}) <C_2({\rm UA1})<C_2({\rm Tevatron})$.
\item Analysis of BEC at UA1 energies with the same 
normalization as that used
so far by the NA22 and Tevatron groups (fixed multiplicity).
The enhancement
of $C_2$ at large $Q$ (and possibly also at small $Q$) 
observed so far, 
should disappear to a great extent.\\
\end{itemize}

\subsection{Multiplicity dependence of Bose-Einstein
correlations}

 The operators for the field
(intensity) and number of particles do not commute.
This means that
measurements of ``ideal" BEC can be performed only when no
restriction on the multiplicity $n$, which fluctuates from
event to event, is made. In practice, however, very often
such restrictions are imposed, either because of technical
reasons or because of theoretical prejudices.
To the last category belong considerations
imposed by  
the search for QGP in high 
energy heavy ion reactions. Thus one expects that
 by selecting events with
$n\geq n_{min}$, where $n_{min}$ is in general an energy
dependent quantity, one gets information about
the interesting "central" collisions. 

Another reason why multiplicity constraints are of practical
importance for QGP experiments is the need 
to compare
various QGP signals in a given event and at the same time
determine for that event the radius, lifetime and chaoticity
of the source, among other things in order to be able to
estimate the energy density achieved in that event. This
means that for QGP search it is interesting to perform 
interferometry measurements for
single events, which of course have a given multiplicity.  

For these reasons the investigation of the multiplicity
dependence of BEC constitutes an important
enterprise which we shall address in the following section.

\subsubsection{The quantum statistical formalism}
Correlation functions defined in quantum
statistics and used in quantum optics refer to ensemble 
averages 
of intensities \footnote{Herefrom the name ``intensity
interferometry".} $I$ of fields $\pi$, where
\be
I({\bf k}_{\perp},y)=|\pi({\bf k}_{\perp},y)|^2.
\label{eq:Shih1}
\ee

The total multiplicity $n$ of {\em identical particles}
over a given phase space region is given by
\be
n=\int d{\bf k}^2_{\perp}\int dy|\pi({\bf k}_{\perp},y)|^2
\label{eq:Shih2}
\ee
Both the field $\pi({\bf k}_{\perp},y)$ and the intensity 
$I({\bf k}_{\perp},y)$ are stochastic variables. Averaging 
over an appropriate ensemble, we get the mean total
multiplicity
\be
<n>=\int d{\bf k}^2_{\perp}\int dy<|\pi({\bf k}_{\perp},y)|^2>.
\label{eq:Shih2a}
\ee

In \cite{Shih} the dependence
of BEC within the QS formalism on the total multiplicity $n$
and on $n_{min}$  
was investigated and it
was found that the size of this effect is (especially at low
$<n>$) surprisingly large and must not be ignored, as had
been done before. 
Both the $n_{min}$ constraint and the $<n>$ constraint lead
to a decrease of the correlation function $C_2$
at fixed $y_g = y_1-y_2$, i.e. to an antibunching effect. 
The last effect can be
approximated, except for very large $n$ and small $y_g$
by a simple analytical formula
\be
C^{(n)}_{(2)}(y_g)\approx C_{(2)}(y_g)\frac{n-1}{nf_2}
\label{eq:Shih424}
\ee
where $C_{2}^{n}$ denotes the correlation function
 at fixed $n$ and $f_2$ is the reduced factorial moment.

These results show that BEC parameters like radii, lifetimes,
and chaoticity do depend on the particular experimental
conditions under which the measurements are performed.  

{\footnotesize 
It is worth
mentioning that a multiplicity dependence of BEC was
observed experimentally  for $p-p$ and $\alpha-\alpha$
reactions at $E_{cm}= 53$ GeV and $31$ GeV respectively
already in \cite{Axial}. 
There it was found that the transverse radius increases with
the multiplicity of charged particles $n_{charged}$.   
This effect was interpreted by Barshay \cite{Barshay} to be 
a consequence of
the impact parameter dependence. The same effect was seen in
heavy ion reactions \cite{Na35} and got a similar
interpretation in \cite{Kag}
\footnote{The approach of \cite{Kag} combines
simplified (1-d) hydrodynamics with a
multiple scattering model, which also exploits the
impact parameter dependence.}. 
The interpretation in terms of impact parameter dependence
could be checked directly in heavy ion reactions since here 
 the impact parameter can be determined on an event by event
basis.  

Another mechanism for the increase
of radius with multiplicity was proposed
by Ryskin \cite{Ryskin88}. He pointed out
that in high multiplicity events, which 
are associated by many authors to large transverse momenta
of partons and thus to a regime where perturbative QCD
applies, one expects 
that the size of the hadronization region should
increase with the multiplicity like $\sqrt{n}$.}
    
A related topic is the dependence of BEC on the rapidity
density $d=\Delta n/\Delta y$, which has also been observed 
experimentally  in $\bar p-p$
reactions at the CERN SPS collider \cite{Wheeler} and at the
Fermilab tevatron \cite{Alex}. Using a
parametrization 
\be
C_2=1+\lambda \exp(-R^{2}Q^2)
\label{eq:simple}
\ee
where $Q$ is the invariant momentun transfer, it was found 
that $R$ increased with $d$ while $\lambda$
decreased with $d$. This last 
observation is compatible with the results from \cite{Shih};
a more quantitative comparison would be possible only if,
among other things,
  the data were parametrized 
in a way more consistent with quantum statistics. 
Thus the four momentum
difference $Q$ is not an ideal variable for BEC (cf. section
2.1.5) and the coherence effect
has to be taken into account as outlined in section 2.2 and 
not by the simple empirical $\lambda$ factor.

\subsubsection{The wave function formalism; ``pasers"?}

The dependence on total multiplicity of BEC
 was investigated also within the wave function
formalism.
In section 2.1 where the  GGLP theory was presented
it was pointed out that the wave function formalism  may be 
useful for exclusive processes or for event generators. 
Indeed, in a first approximation,
the wave function $\psi_n$ of a system of $n$ identical 
bosons e.g. can be obtained from the product of $n$ single 
particle wave functions $\psi_1$  by symmetrization. Then the
calculation of $C_{(2)}^{(n)}$ 
is in principle straightforward and follows the lines of 
GGLP.

However when the multiplicities $n$
become large (say $n>20$) the explicit
symmetrization of the wave function formalism
becomes difficult. This lead  
 Zajc \cite{Zajc}  
 to use numerical Monte
Carlo techniques for estimating $n$ particle symmetrized 
probabibilities, which he then applied to calculate
 two-particle BEC. He was thus able to study the
question of the dependence of BEC parameters 
on the multiplicity. For an application of this approach to 
Bevalac heavy ion reactions, cf. \cite{WNZhang}. 
Using as input a second order BEC function  paramerized in 
the form (\ref{eq:simple})
 Zajc found, 
(and we have seen above that this was confirmed in 
\cite{Shih},) that the
``incoherence" parameter
$\lambda$ decreased with increasing $n$ \footnote{In
\cite{Zajc} the clumping in phase space due to Bose symmetry
was also illustrated;}.
   
However Zajc did not consider that this
effect means that events with higher pion
multiplicities are denser and more coherent. On the
contrary he warned against such an 
interpretation  and concluded that his results
have to be used in order to eliminate the {\em bias} 
introduced by
this effect into experimental observations 
\footnote{The same
interpretation of the multiplicity dependence of BEC was
given in \cite{Shih}. In this reference the nature of the 
``fake"
coherence induced by fixing the multiplicity is even
clearer, as one studies there explicitly coherence 
in a consistent  quantum statistical formalism.}.

This warning apparently did not deter the authors of 
\cite{Pratt1} and 
\cite{Chao} to do just that. Ref.\cite{Pratt1} went even so 
far as
to deduce the possible existence of pionic lasers from 
considerations of this type 
\footnote{The philosophy
of ref.\cite{Pratt1} is reflected e.g. in statements like
``multiparticle interference can enhance the emission of
pions". This language is apparently motivated
by the practice of event generators where BEC (or in general
quantum statistical effects) are ``added" to the usual
schemes at the end of the calculation. That this does not
reflect the true sequence of the physical processes
considered is obvious.}.

Ref. \cite{Pratt1} starts by proposing an algorithm for
symmetrizing the wave functions which presents
the advantages that it reduces very much the computing time
when using numerical techniques, which is applicable also
for Wigner type source functions and not only plane wave 
functions,
and which for Gaussian sources provides even analytical 
results. 
Subsequently in ref.\cite{Zimanyi} wave packets were 
symmetrized and in
special cases the matrix density at fixed and arbitrary $n$ 
was derived in analytical form.
This algorithm was then applied to calculate the influence of
symmetrization on BEC and multiplicity distributions. As in
\cite{Zajc} it is found in \cite{Pratt1} that 
the symmetrization produces an effective
decrease of the radius of the source, a broadening of 
the multiplicity distribution $P(n)$ and an increase of the
mean multiplicity as compared to the non-symmetrized case.
What is new in \cite{Pratt1} is (besides the algorithm)
mainly the meaning the author attributes to these results. 

In a concrete example Pratt considers
a non-relativistic source distribution $S$ in the absence of
symmetrization effects:

\be
S(k,x)=\frac{1}{(2\pi R^2mT)^{3/2}}\exp\left(-\frac{k_0}{T}-
\frac{x^2}{2R^2}\right)\delta(x_0)
\label{eq:Pratt8}
\ee
where 
\ba
k_0/T=k^2/2{\Delta}^2
\ea
Here $T$ is an effective temperature, $R$ an effective
radius, $m$ the pion mass and $\Delta$ a constant with
dimensions of momentum.

Let $\eta_0$ and $\eta $ be the number densities before 
and after symmetrization, respectively. In terms of $S(k,x)$
we have
\be
\eta_0= \int S(k,x)d^4kd^4x
\label{eq:dens}
\ee
and a corresponding expression for $\eta $ with $S$ replaced
by the source function after symmetrization.
Then one finds \cite{Pratt1} that $\eta $ increases with 
$\eta_0$ and
above a certain crtitical density $\eta^{crit}_0 $, 
$\eta $ diverges. This is
interpreted by Pratt as ``pasing".

The reader may be rightly puzzled by the fact that while 
$\eta $ has a clear physical significance the number density
$\eta_0 $ and a fortiori its critical value have no physical
significance, because in nature
there does not exist a system of bosons the wave function of
which is not symmetrized.   
Thus contrary 
to what is alluded to in
ref.\cite{Pratt1}, this paper does not does address really 
the question how a condensate is reached.
Indeed the physical factors which induce condensation are, 
for systems in (local) thermal and chemical equilibrium
\footnote{For lasers the determining dynamical factor is
among other things the inversion.},
pressure and temperature and the symmetrization is contained
automatically in the distribution function
\be
f=\frac{1}{\exp[(E-\mu]/T]- 1}
\label{eq:f}
\ee 
in the term -1 in the denominator;
$E$ is the energy and $\mu$ the chemical potential.

To realize what is going on
 it is useful to observe that the increase of 
$\eta_0 $ can be achieved by decreasing $R$ and/or $T$. Thus
$\eta_0$ can be substituted by one or both of these two 
physical
quantities. Then the blow-up of the number density $\eta $
can be thought of as occuring due to a decrease of $T$ and/or
$R$. However this is nothing but the well known
Bose-Einstein condensation phenomenon and does not represent
anything new.
While from a purely mathematical point of view the
condensation effect can be achieved also by starting
with a non-symmetrized wave function and symmetrizing it 
afterwards ``by hand" , the causal i.e. physical relationship   
is different: one starts with a bosonic i.e symmetrized
system and obtains condensation by decreasing the
temperature or by increasing the density of this {\em bosonic} 
system. 

Another confusing interpretation in \cite{Pratt1} relates to
the observation
made also in \cite{Zajc} that the symmetrization produces a 
broadening of the
multiplicity distribution (MD). In particular starting with a
Poisson MD for the non-symmetrized wf one ends up
after symmetrization with a negative binomial. While Zajc
correctly considers this as a simple consequence of Bose
statistics, ref.\cite{Pratt1} goes further and associates
this with the so called pasing effect. That such an
interpretation is incorrect is obvious from the fact that
for true lasers
the opposite effect takes place. Before ``condensing" i.e.
below threshold their
MD is in general broad and of negative binomial form 
corresponding to
a chaotic (thermal) distribution while above threshold the
laser condensate is produced and as such
corresponds to a coherent state and therefore is
characterized by a Poisson MD. Last but not least the fact
that this broadening increases with $n$ is not due, as one
might be mislead to believe from \cite{Pratt1},\cite{Chao}, 
to the approach to ``lasing criticality", but simply to the
fact that the larger $n$ the larger the number of
independent emitters is and the better the central limit
theorem applies. This theorem states (cf. section 2.2)
that the field produced by a large number of independent
sources is chaotic.       

Finally a terminological remark appears necessary here.
We believe that names like ``paser" or pionic laser used in
the papers quoted above
are unjustified and misleading.
The only characteristic which the systems considerd in these
papers  possibly share with lasers is the condensate property
 i.e. the bunching of particles in a given (momentum) state. 
However lasers are much more than just condensates; one of
the main properties of lasers which distinguishes them 
from other condensates is the
directionality, a problem which is not even mentioned in 
the ``paser" literature.
\footnote{For a model of directional coherence, not 
necessarily related to pion condensates, 
cf.\cite{FW85}; experimental hints of this effect have 
possibly been seen in \cite{Zajc84}.}

To conclude this excursion into what in our opinion is a
surrealistic interpretation of multiparticle
correlations, it is hard to find new physics in the ``paser"
papers, although they probably represent progress in the
mathematical problem of symmetrization of wave functions or
wave packets\footnote{For another investigation of the effect
of symmetrization on the
single inclusive cross section cf.
\cite{Ray} and \cite{Wied88}. In \cite{Wied88} second
order correlation functions are also considered.(Cf. also
ref.\cite{RH} for this topic.)}.

\subsection{The invariant $Q$ variable in the space-time
approach; 
higher order correlations; ``intermittency" in BEC?}

The issue of apparent power-like rapidity dependence of 
moments of MD was discussed
in section 6.1.1 where it was pointed out that this dependence
could in principle be understood within the QS formalism
without invoking the idea of intermittency. However
this was not the end of the story because:1) it was observed
 \cite{bu} that the power like
behaviour extends also to BEC data (in the invariant variable
$Q$). This was surprising because up to that moment BEC data
could usually be fitted by a Gaussian or exponential
function, albeit these data did not extend to such small $Q$
values as those measured in \cite{bu}; 2) Bialas   
 \cite{bi} (cf. also \cite{bizi}, \cite{zi}) proposed that the
source itself has no fixed size, but is fluctuating from
event to event with a power distribution of sizes. Since the
measurements made in \cite{bu} were in the invariant variable     
$ Q$ rather than
$\Delta y$, one had to understand whether the $QS$ approach 
which
implies fixed scales does not
lead to a similar behaviour in $Q$.

Refs. \cite{intpl},
\cite{intpr} addressed this question and proved that,
indeed, by starting from a space-time correlator with a
fixed correlation length and a source distribution with a
fixed radius, one gets after integrations over the
unobserved variables a correlation function which is
power-like in a limited $Q$ range.   

In the following we shall sketch how this happens.
The two-particle Bose-Einstein correlation function is defined as 
\be
C_2({\bf k}_1,{\bf k}_2) = \frac{\rho_2({\bf k}_1,{\bf k}_2)}
{\rho_1({\bf k}_1)
\cdot \rho_1({\bf k}_2)}
\label{eq:c2rho}
\ee
where $\rho_1({\bf k}^{})$ and $\rho_2({\bf k}_1,{\bf k}_2)$
are the one- and two-particle particle inclusive spectra,
res\-pec\-tive\-ly.

The two-particle correlation function projected on $Q^2$ is
\be
C_2(Q^2) = 1 + \frac{{\cal I}_2(Q^2)}{{\cal I}_{11}(Q^2)}
\label{eq:c2rat}
\ee
with the integrals
\ba
{\cal I}_{11}(Q^2) &=& \int d\omega_1 \int d\omega_2 \: \  
\delta \left[ Q^2+(k_1^\mu-k_2^\mu)^2
\right] \rho_1({\bf k}_1) \rho_1({\bf k}_2) \/  \nonumber\\
{\cal I}_2(Q^2) &=& \int d\omega_1 \int d\omega_2 \: \ 
\delta \left[ Q^2+(k_1^\mu-k_2^\mu)^2
\right] \left(\rho_2({\bf k}_1,{\bf k}_2) \ - \ 
 \rho_1({\bf k}_1) \rho_1({\bf k}_2)\right)
\label{eq:i112}
\ea
where $d\omega \equiv d^3k/(2\pi)^32E$ is the invariant 
phase space volume element. 
Two types of sources, a static one and an expanding one, 
were considered.  
The parameters
employed are physically meaningful quantities in the sense 
that they give the lifetime, radii and correlation lengths 
of the source. This 
is not the case for the ad-hoc parametrization of  
$C_2(Q^2)$, e.g.,  with a Gaussian 
\begin{equation}
C_2(Q^2)\:=\:1\:+\:\lambda_Q\:e^{-R_Q^2 Q^2}
\label{eq:c2qmi1}
\end{equation}

To illustrate the behaviour of the correlation function 
as a function of $Q^2$,  one applied the formalism to 
describe
UA1 data in the phase space region $| y | \leq 3.0$ and 
$k_\perp \geq 150 MeV$.
In \cite{intpl},\cite{intpr} $C_2(Q^2)$ was calculated for 
the static and for the expanding source
by Monte Carlo integration. 
(for the static source, approximate analytical results could 
 be obtained only for $|y|>1.5$, where they agree with the 
numerical results \cite{intpr}). 
Fig. 21 shows the results of fits to the UA1  
data\footnote{The
$C_2(Q^2)$ data of \cite{bu}  were normalized so that at 
large $Q^2$, $C_2(Q^2) \approx 1$.
An explanation for the experimental observation of
\cite{bu} that $C_2(Q^2)$ exceeds by a multiplicative factor
of $\sim 1.3$ both the upper and lower
``conventional'' limits of 2 and 1, respectively,
is discussed in ref. \cite{lrc}.}
for the static source
(dashed line) and for the expanding source (solid line),   
which were obtained under the assumption of a purely
chaotic source, $p_0=1$
(the data are consistent with  
an amount of $\leq 10 \%$ coherence\footnote{The older 
UA-1 data \cite{UA1}
were limited to larger $Q^2$ values. 
Furthermore the quantum statistical
interpretation \cite{mad} of these data is 
different from that in \cite{intpl}, which
is based on space-time concepts. 
This explains the different values of
chaoticity obtained in \cite{intpl}, \cite{intpr} on the one
hand and in 
refs. \cite{UA1},\cite{mad} on te other hand}, 
but the sensitivity
to $p_0$ was not sufficient to further constrain the degree 
of coherence within these limits). 
For comparison, the result of 
a power law fit (dotted line) as suggested
in ref. \cite{bu} 
\begin{equation}
C_2(Q^2)\:=\:a\:+\:b\cdot \left(\frac{Q^2}
{1 GeV^2}\right)^{-\phi}
\label{eq:power}
\end{equation}
is also plotted.

One finds that the data can
be well described both with the static and the expanding 
source model with reasonable values for the radii, lifetime 
and correlation length.

The results of above show 
that the power-like behaviour of $C_2$ (the same holds for 
higher order correlations) can be reproduced by assuming a 
conventional space-time source with fixed parameters, 
i.e., without invoking ``intermittency"
\footnote{A similar point of view is expressed in
\cite{PPT1} for the particular case of bremsstrahlung 
photons.}. This conclusion of \cite{intpl} is strengthened
by an explicit consideration of resonances in \cite{intpr}.

The advantages of the QS formalism as compared with the wave
function 
formalism emerge clearly also in the problem of higher order
correlations. In the QS formalism 
 higher order
correlations are treated on the same footing as lower order
ones and emerge just as consequences of the form of the
density matrix. Therefore 
questions found sometimes in the literature like ``what is 
the influence of higher order
correlations on the lower one" do not even arise in QS and in
fact do not make sense.
We emphasized in section 2.2.1 the importance of higher order 
correlations  for the phenomenological determination 
of the form of the correlation function and this applies in
particular when a single variable like $Q$ is used 
 instead of the six
independent degrees of freedom inherent in the correlator.
For this reason the space-time integration at fixed $Q$
has been used recently \cite{Nelly}
also in the study of higher order correlations and applied
to
the NA22 data \cite{Na22}. It was found among other things
that an expanding source with fixed parameters 
as defined in chapter 4 can account for the data up to and
including the fourth order, confirming in a first
approximation the Gaussian form of the density matrix.
The fact that previous attempts in this direction
 like \cite{Na22} and \cite{UA1new} met with
 difficulties may be due to the fact
that in these two 
experimental studies the QO formalism in momentum space was
used and possibly also because no simultaneous fit of all
orders of correlations was performed, as was the case in
\cite{Nelly}.

\section{Critical discussion and outlook}

For historical reasons related to the fact that the GGLP
effect was observed for the first time in annihilation at
rest when (almost) exclusive reactions were studied, the
theory of BEC was initially based on the wave function 
formalism.
This formalism is not appropiate for inclusive reactions 
in high energy physics, among other things because it yields
a correlation function which depends only on the momentum
difference $q$ and not also on the sum of momenta, it does
not take into account isospin, and cannot treat adequately
coherence. This last property is essential as it leads to
one of the most important applications of BEC, i.e. to 
condensates. Furthermore this property strongly affects
another essential application of BEC, namely the
determination of sizes, lifetimes, correlation lengths and
correlation times of sources. The adhoc parametrization of
the correlation function under the form (\ref{eq:lambda}) 
where $\lambda$ is supposed to take into account (in)coherence
is unsatisfactory. An improvement based on an analogy with 
quantum optics
leads to an additional term (with the same number of
parameters). A more complete and more correct treatment of 
BEC is provided by the space-time formalism of classical
currents.

Almost all studies of BEC assume a Gaussian density
matrix. Possible deviations from this form could and should
be be looked for by studying higher order correlations.

The classical current approach is based on an exact solution of the equations
of quantum field theory and it  
can be considered at present the most advanced
and complete description of BEC. It introduces a primordial
correlator of currents which is characterized by finite 
correlation
lengths and correlation times. The geometry of the source
is an independent property of the system and it is here that
the traditional radii and lifetimes enter. The form of
the geometry and of the correlator is not given by the
theory and it is up to experiment to determine these.

For a  Gaussian density matrix     
 the space-time approach within the classical current formalism 
leads to a minimum of ten independent parameters; 
these include geometrical and dynamical scales as well
as the chaoticity. Phenomenologically these scales can be
separated only by considering simultaneously single and
double inclusive cross sections. Experimentally this 
separation as well as the determination of all parameters 
has not yet been performed and constitutes an
 important task for the future. {\em This should be done not only for particle reactions but also
for heavy ion reactions, as an alternative to the
pseudo-hydrodynamical approach which cannot provide this 
separation.} It would also be desirable to extend the 
parametrization for an expanding source by
dropping the assumption of boost invariance.

Besides the conventional
$++$ or $--$  pion correlations there exist also $+-$
correlations, which become important for sources of 
small lifetimes. These "surprising" field theoretical 
effects represent squeezed states, which unlike what happens
in optics,
appear in particle physics ``for free". 
These effects can be used to investigate the difference 
between classical
and quantum currents. Their detection
constitute one of the most important challenges for 
future experiments.

BEC are influenced by final state
interactions.   
 Coulomb final state interactions do not play a major part
 except at very small $q$. These small
values have apparently not yet been reached in experiment
and it is questionable whether they will be reached in the
foreseable future. In heavy ion reactions this is due to
the large values of radii which multiply $q$ and even for
typical scales of $1$ fm the present resolution of detectors is not
sufficient to make this effect very important. 
Nevertheless since Coulomb corrections have
been studied so far only  within the
wave function formalism and usually by applying the
Schr\"odinger equation, it would be desirable to extend
this study by considering the Klein-Gordon equation 
which is more appropiate for mesons. Even more interesting
would be to study Coulomb corrections within
the classical current formalism.  

Resonances 
 play an essential part in final 
state interactions and progress has been achieved in their
understanding, in particular in heavy ion reactions, where
their influence has been investigated using solutions of the
equations of hydrodynamics within the Wigner function formalism

BEC have been investigated  in $e^{+}-e^{-}$, hadron-hadron 
and heavy ion reactions, however a systematic comparison
of results using the same paramterization and normalization
awaits still to be done\footnote{At present even for the
same type of reaction different normalizations are
frequently used and this can lead to {\em apparently}
different  
results, as exemplified in the case of $NA22$ and $UA1$
data. Tests for a better understanding and elimination of
these discrepancies have been proposed (cf. section 6.1.3)}. 

Correlations are intimately related to multiplicity
distributions which can serve as complementary tools in the
determination of the parameters of sources. Therefore a
systematic 
investigation of BEC and multiplicity distributions in the
same phase space region is desirable. How useful this can be
has been shown by proving in this way the influence of long
range correlations on BEC.

BEC can be useful in heavy ion reactions and in particular 
for the search of quark-gluon plasma if
one of the two conditions are satisfied:

(A) the investigation is based on  full-fledged
hydrodynamics, impying the solution of the equations of
hydrodynamics with explicit consideration of the equation of 
state. 

(B) the investigation is based on the classical current 
space-time approach.

Case (A) has the advantage that the dependence of the 
equation of state on the 
 phase transition may be reflected
also in single inclusive cross sections and in the BEC.
(So far, however, the sensitivity of BEC on the equation of 
state could not yet 
been proven with present data \cite{SchleiHung}.) 
It has however the disadvantage that  
 its applicability is
restricted because it is based on
the Wigner function, which is a particular case of
the classical current formalism, 
for small $q$ and not too strong correlations between momenta
$k$ and coordinates $x$. 

Case (B) has the disadvantage that there is no contact with
the equation of state. However it has the advantage
it is not restricted to small $q$ and weak
correlations between $k$ and $x$.

Unfortunately many of the ``theoretical" papers on BEC in
heavy ion reactions do not satisfy either condition (A) or
condition (B). This is the case with most of the  
 pseudo-hydrodynamical papers which use
a parametrization of the source function based on
qualitative hydrodynamical 
considerations without the use of an equation of state and
without solving the equations of hydrodynamics. This
pseudo-hydrodynamical 
approach has the disadvantages of (A)
 and (B) but none of
their advantages\footnote{Pseudo-hydrodynamics as compared
with hydrodynamics has 
also the supplimentary disadvantage that it does not allow a
separation between
geometrical radii and correlation lengths. Moreover it
does not have  
even the excuse of
simplicity, since the number of free parameters in the
psudo-hydrodynamical approach is as large as that in the
classical current space-time approach.}.

>From the above considerations it is seen that there are many
usolved problems in the investigation of BEC, some of them
of theoretical, but most of them of experimental nature.   

While an analysis of the experimental BEC
 deserves a special review, some of the obvious
reasons for this unsatisfactory experimental situation are: 

1) Most of the BEC experiments performed so far use 
inadequate detectors, because they 
are not dedicated experiments but rather byproducts of  
experiments planned for other purposes. What is needed amng
other things is 
track by track detection and improved identification of
particles.

2) Insufficient statistics. An improvement of statistics 
especially at small $q$
by at least one order of magnitude is necessary to address 
some of the problems enumerated above.

3) Incorrect or incomplete parametrizations of the
correlation functions. Very often and in part because of 2)
not all six independent variables of $C_2$ are measured, but
projection of these. Very popular among these projections is  
the relativistically invariant variable $Q$.  This is not a 
good variable for BEC studies, because among other things it 
mixes the space and time
variables in an uncontrollable way. Furthermore, in most
parametrizations, coherence is not (or inadequately)
considered.
  
4) Inadequate normalizations. Practically 
 all
BEC experiments use a normalization procedure of the
correlation function which does not correspond to its
definiton. This definition refers to the single inclusive 
cross sections
obtained in the same event as the double inclusive cross 
section. The use of an ``uncorrelated background", instead,
biases the results. 

The solution of the problems mentioned above will make of boson
interferometry what it is supposed to be: a reliable method
for the determination of sizes, lifetimes, correlation
lengths, and coherence of sources in subatomic physics. 

A more pedagogical presentation of the theory of
Bose-Einstein correlations, which discusses also 
its quantum optical context, including a comparison between
the HBT and the GGLP effects and between photon and hadron
intensity interferometry can be found in the
book by the author \cite{book}. 
Some of the most representative theoretical and experimental
papers on BEC and which are frequently quoted within the 
present
review have been reprinted in a single volume in ref.  
 \cite{reprints}.  

{\bf Acknowledgements}
I am indebted to D. Strottman for a careful reading of the
manuscript and for many helpful comments.

\newpage

\section{Figure captions}

\begin{description}

\item[Fig. 1] The GGLP experiment schematically.

\item[Fig. 2]  Second order correlation function
$C_2(k,k)\equiv g^{(2)}$ 
as a function of the squeezing parameter $r$ for pure
squeezed states (from. ref.\cite{VW}).

\item[Fig. 3] Second order correlation function as given by 
the quantum statistical formalism of ref. \cite{Stelte} for 
various values of the coupling constant $g$ for $\kappa =0.5$
(from \cite{Stelte}). The parameter $\kappa$ is related to
the chaoticity $p$ via the relation  $\kappa =<n_c>/<n_{ch}>
=1/p-1$ where $p=<n_{ch}/n>$ and 
$n = n_{c} +n_{ch}$ is the total multiplicity. $Y$ is the
maximum rapidity.

\item[Fig. 4] The same for various values of $\kappa$ at 
$g = 0$ (from \cite{Stelte}).

\item[Fig. 5] Second order correlation function for various 
values
of the ``incoherence" parameter $\lambda$ as given by the
phenomenological equation (\ref{eq:Deutschmann}). 

\item[Fig. 6] Diagrams contributing to different terms of the two-particle
correlator in the point-like random source model 
(from ref.\cite{APW}). 

\item[Fig. 7] The linear polarization vectors
$\epsilon_{lambda}(k)$ for two photons with momenta ${\bf
k}_1$ and ${\bf k}_2$ (from ref.\cite{Feld}).

\item[Fig 8] Geometry of the boost-invariant source.

\item[Fig. 9] Intercept of two-particle correlation function
in the
presence of coherence and resonances (from ref.
\cite{Bolz}). 

\item[Fig. 10]  Bose-Einstein 
correlation functions in longitudinal
and transverse direction, for the  
 3-dimensional (solid lines) and the 1-dimensional 
calculations
(dashed lines) (from ref. \cite{Schlei97}).

\item[Fig. 11]  Dependence of 
the longitudinal and transverse radii
extracted from Bose-Einstein correlation functions on the 
rapidity 
$y_K$ and average momentum $K_\perp$ of the pair. As before,
solid
lines correspond to the 3-dimensional and dashed lines to the 
1-dimensional results.
The open circles indicate values of $R_\parallel$
obtained from \mbox{eq.(\ref{eq:016})} (from ref. 
\cite{Schlei97}).

\item[Fig. 12]  Bose-Einstein correlation functions of 
negatively 
charged pions, in longitudinal
and transverse direction. 
The separate contributions from resonances
are successively added to the 
correlation function of direct (thermal) 
$\pi^-$ (dotted line). 
The solid line describes the correlation 
function of all $\pi^-$ (from ref. \cite{bernd3}). 

\item[Fig. 13]  Correlation functions 
of all $\pi^-$ (solid lines),
thermal $\pi^-$ (dotted lines) and all $K^-$ (dashed lines), 
for
$k_\perp=0$ and for $k_\perp=1$ GeV/c (from ref. 
\cite{bernd3}). 

\item[Fig. 14]  Dependence of the longitudinal and 
transverse radii
extracted from Bose-Einstein correlation functions on the 
rapidity 
$y_k$ and average momentum $k_\perp$ of the pair, for 
all $\pi^-$ (solid lines), thermal $\pi^-$ (dotted lines) 
and all 
$K^-$ (dashed lines). The full circles were obtained
by substituting the value $<t_f(z=0,r_{\perp})> = 2$ fm/c 
for the average lifetime of the system 
(calculated directly from hydrodynamics by
averaging over the hypersurface) into eq. (\ref{eq:016}), 
with
\mbox{$T_f=0.139$ GeV} (from ref. \cite{bernd3}). 

\item[Fig. 15]  $\pi^-\pi^-$ Bose-Einstein correlation 
functions in
the presence of partial coherence (from ref. \cite{bernd3}). 

\item[Fig. 16] Effective radii extracted from Bose-Einstein
correlation functions as a function of the rapidity $y_k$ of
the pair and the transverse average momentum $K_{\perp}$ of
the pair for all pions (
from ref. \cite{OrnikPR96}).

\item[Fig. 17] Effective radii extracted from Bose-Einstein
correlation functions as a function of the transverse
average momentum $K_{\perp}$ of the pair for all pions
compared with data (from ref. \cite{OrnikPR96}).   

\item[Fig. 18] 
Two-particle BE correlation functions for a spherically
and a longitudinally expanding source.
The different line styles correspond to different
condensate densities $n_{co}$ compared to the 
thermal number densities $n_{th}$ (from ref. \cite{cond97}).

\item[Fig. 19] Normalized factorial moments $\Phi_q$ of
order $q$ in finite (pseudo) rapidity windows of width
$\delta y$ around $y_{CMS} = 0$, plotted against 
$\delta \eta $ (from ref. \cite{car}).

\item[Fig. 20] Second order correlation function for negative
particles at $\sqrt s = 630 GeV$ (from ref. \cite{bu}).

\item[Fig. 21] 
$C_2(Q^2)$ for the static source model (dashed line) and for
the expanding source model (solid line) compared
to the UA1 data \cite{bu}.
The dotted line shows a power law fit. 
\end{description}

Notes to be added in proof:
Erase footnote 32 on p. 59
Quotation at the end: ``For God's sake, stop researching for
a while and begin to think." 
\newpage

\begin{figure}[h]
\setlength{\unitlength}{1cm}
\begin{picture}(17.,25.)
\includegraphics{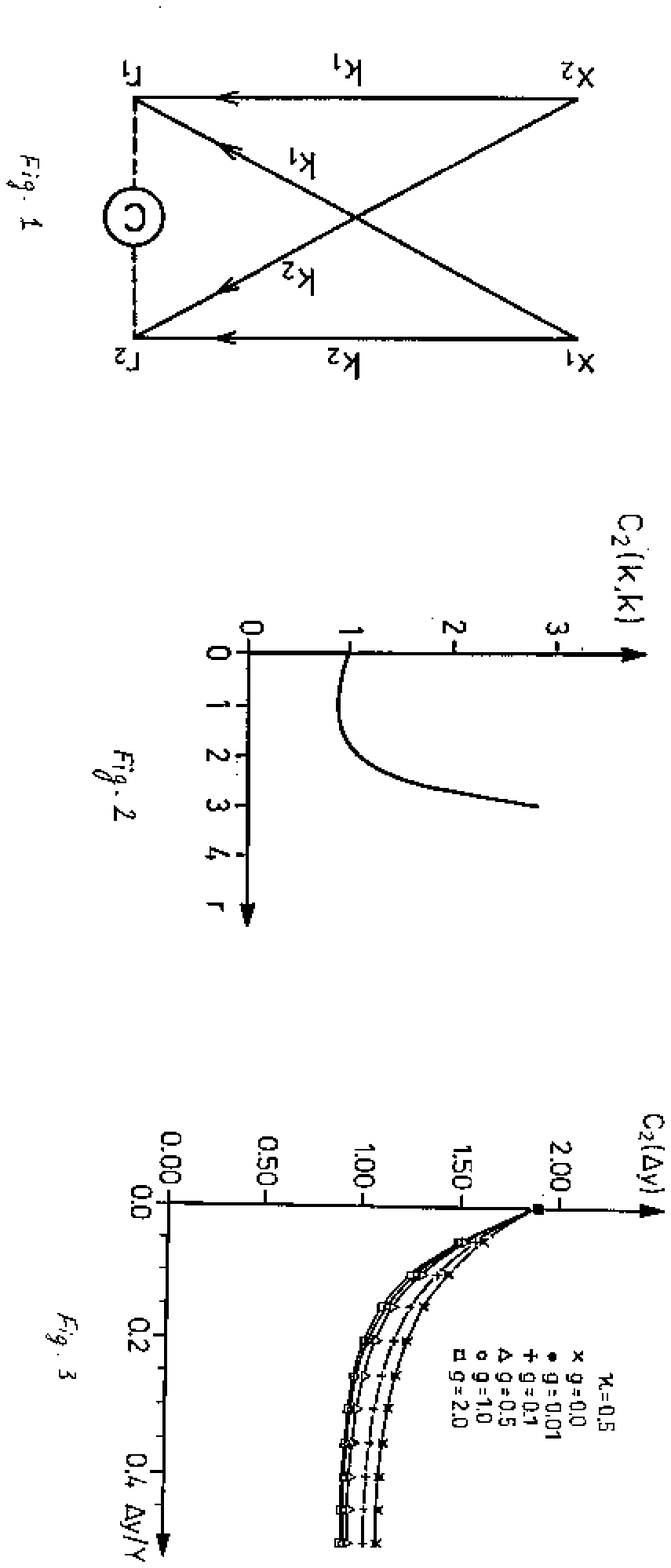}
\end{picture}
\end{figure}
\newpage
\begin{figure}[h]
\setlength{\unitlength}{1cm}
\begin{picture}(17.,25.)
\includegraphics{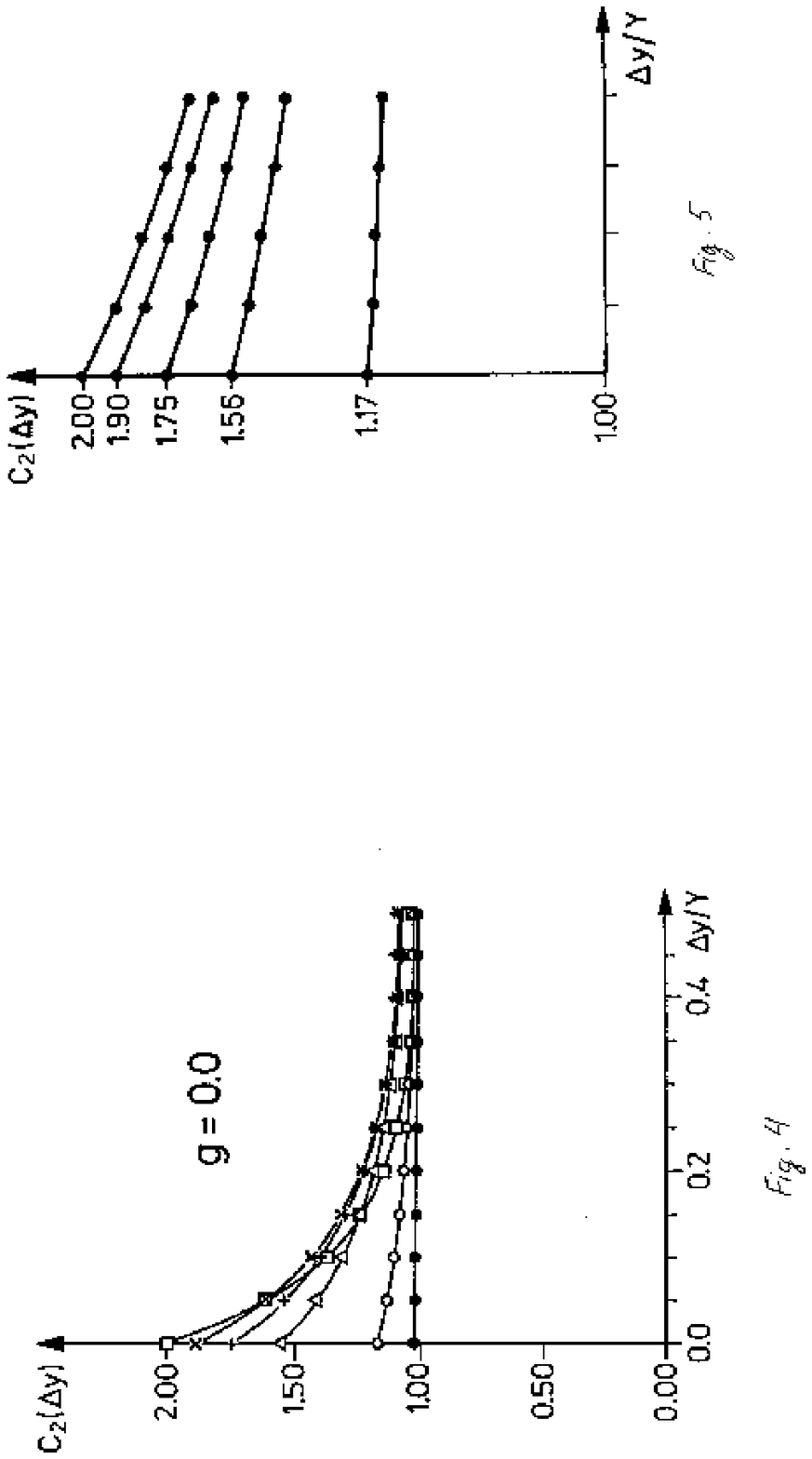}
\end{picture}
\end{figure}
\newpage
\begin{figure}[h]
\setlength{\unitlength}{1cm}
\begin{picture}(17.,25.)
\includegraphics{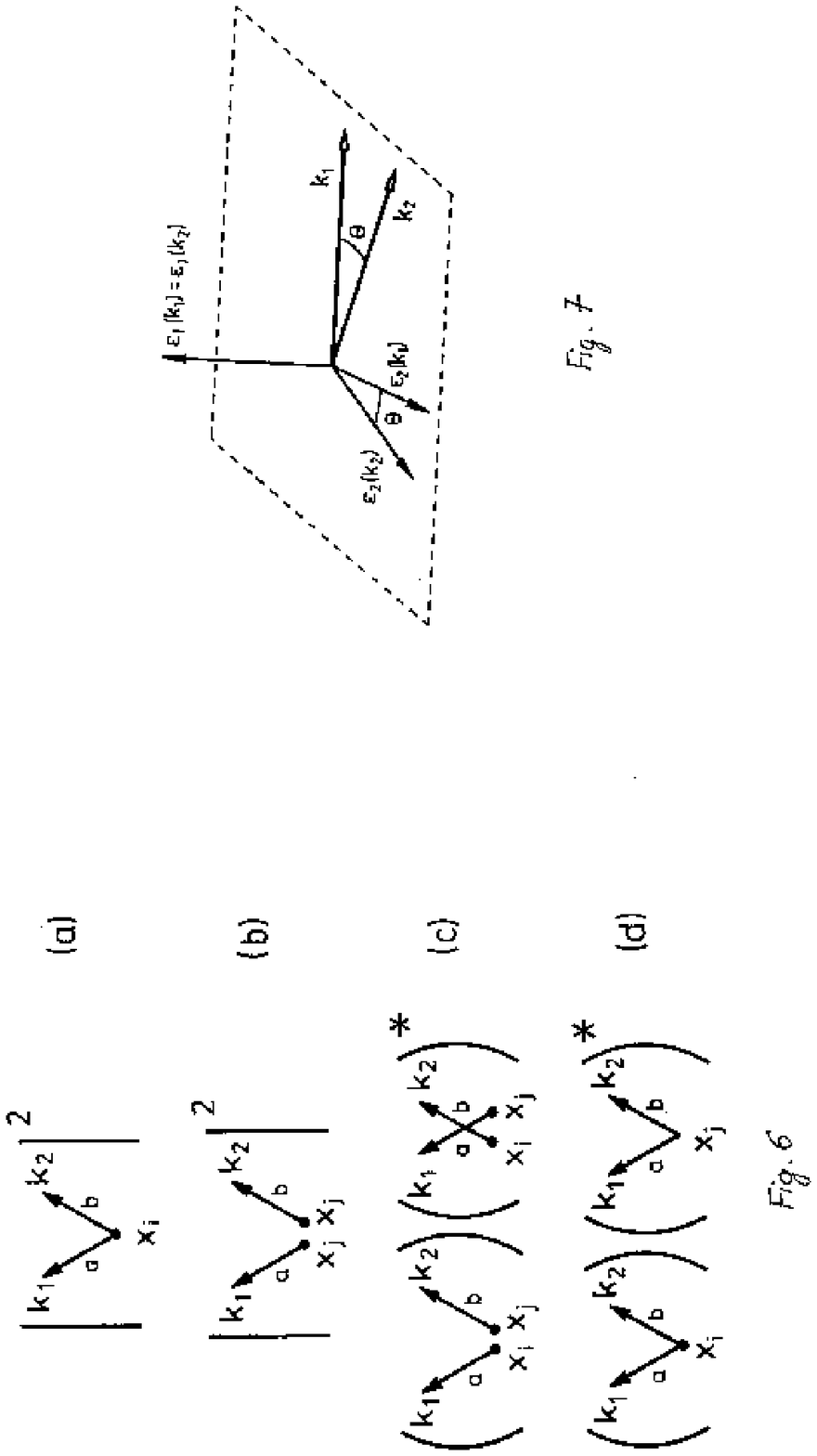}
\end{picture}
\end{figure}
\newpage
\begin{figure}[h]
\setlength{\unitlength}{1cm}
\begin{picture}(17.,25.)
\includegraphics{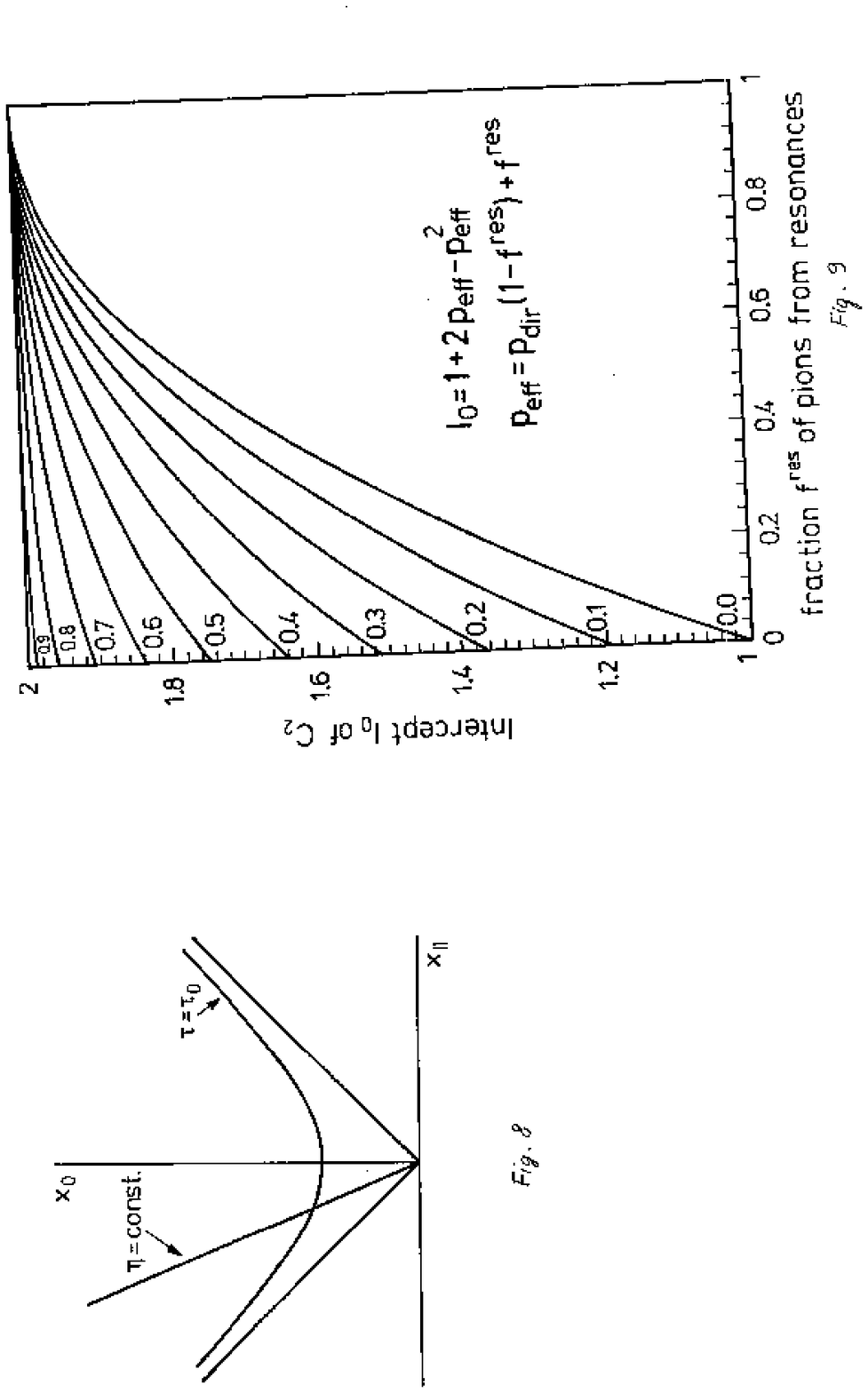}
\end{picture}
\end{figure}
\newpage
\begin{figure}[h]
\setlength{\unitlength}{1cm}
\begin{picture}(17.,25.)
\includegraphics{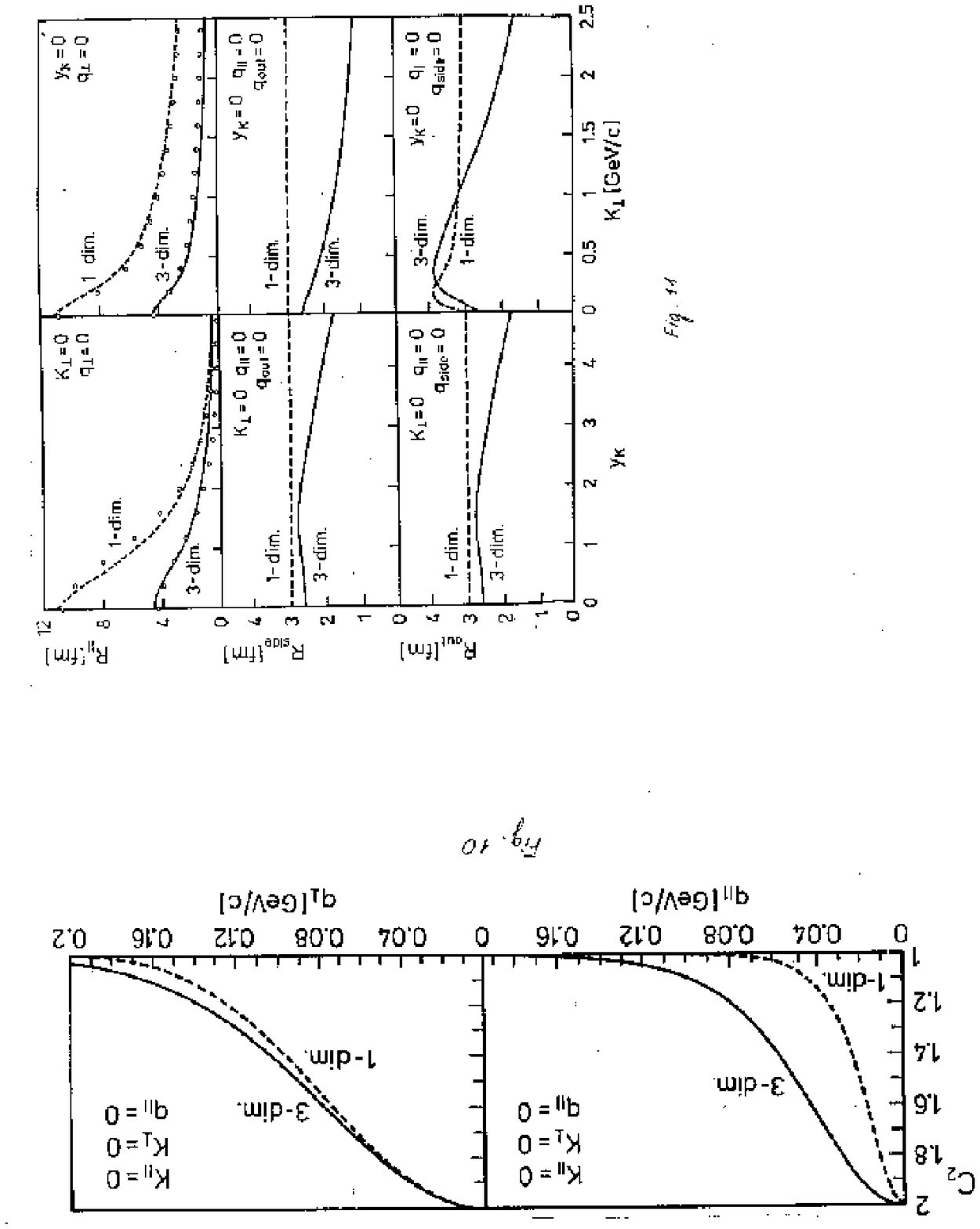}
\end{picture}
\end{figure}
\newpage
\begin{figure}[h]
\setlength{\unitlength}{1cm}
\begin{picture}(17.,25.)
\includegraphics{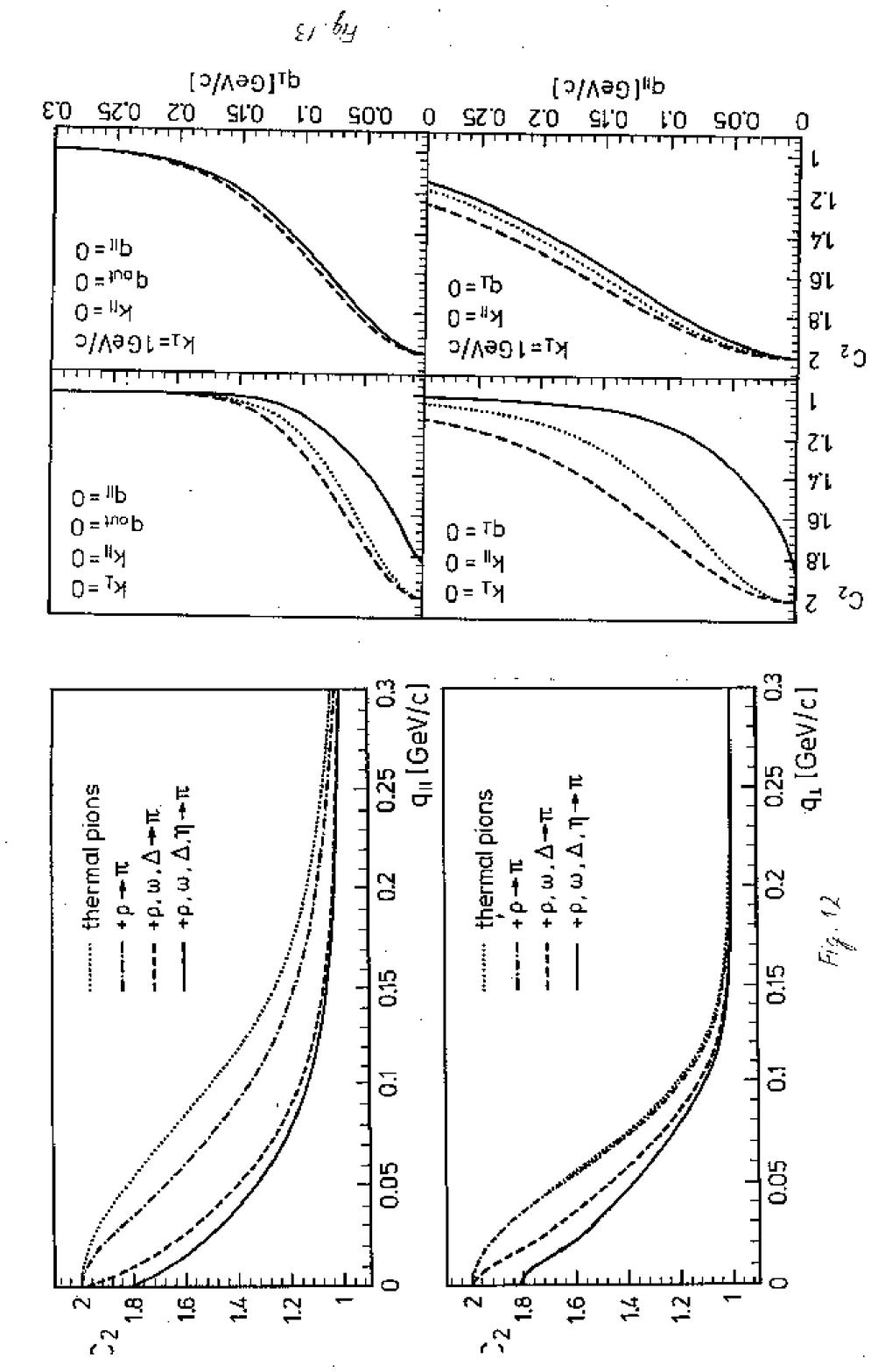}
\end{picture}
\end{figure}
\newpage
\begin{figure}[h]
\setlength{\unitlength}{1cm}
\begin{picture}(17.,25.)
\includegraphics{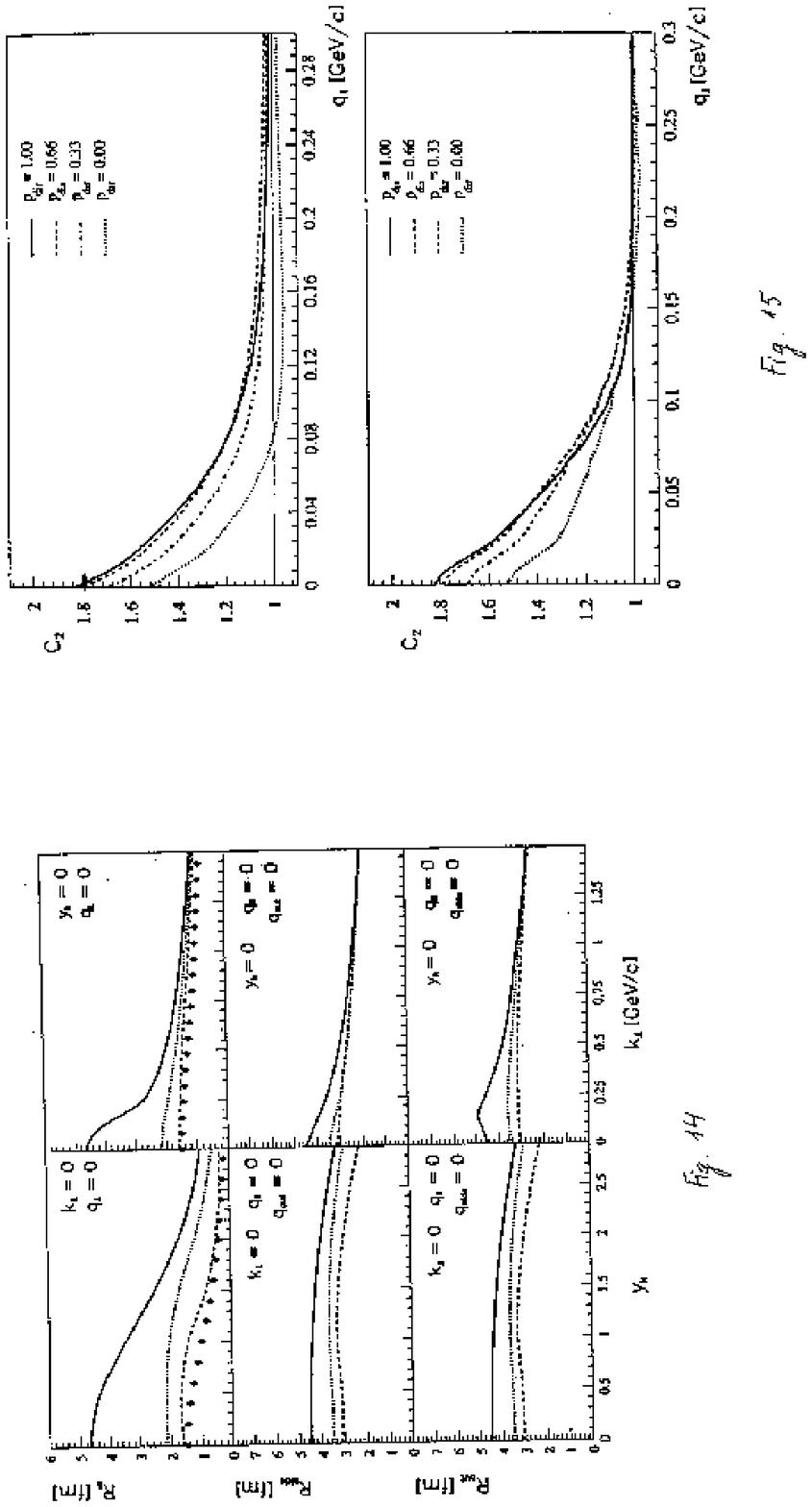}
\end{picture}
\end{figure}
\newpage
\begin{figure}[h]
\setlength{\unitlength}{1cm}
\begin{picture}(17.,25.)
\includegraphics{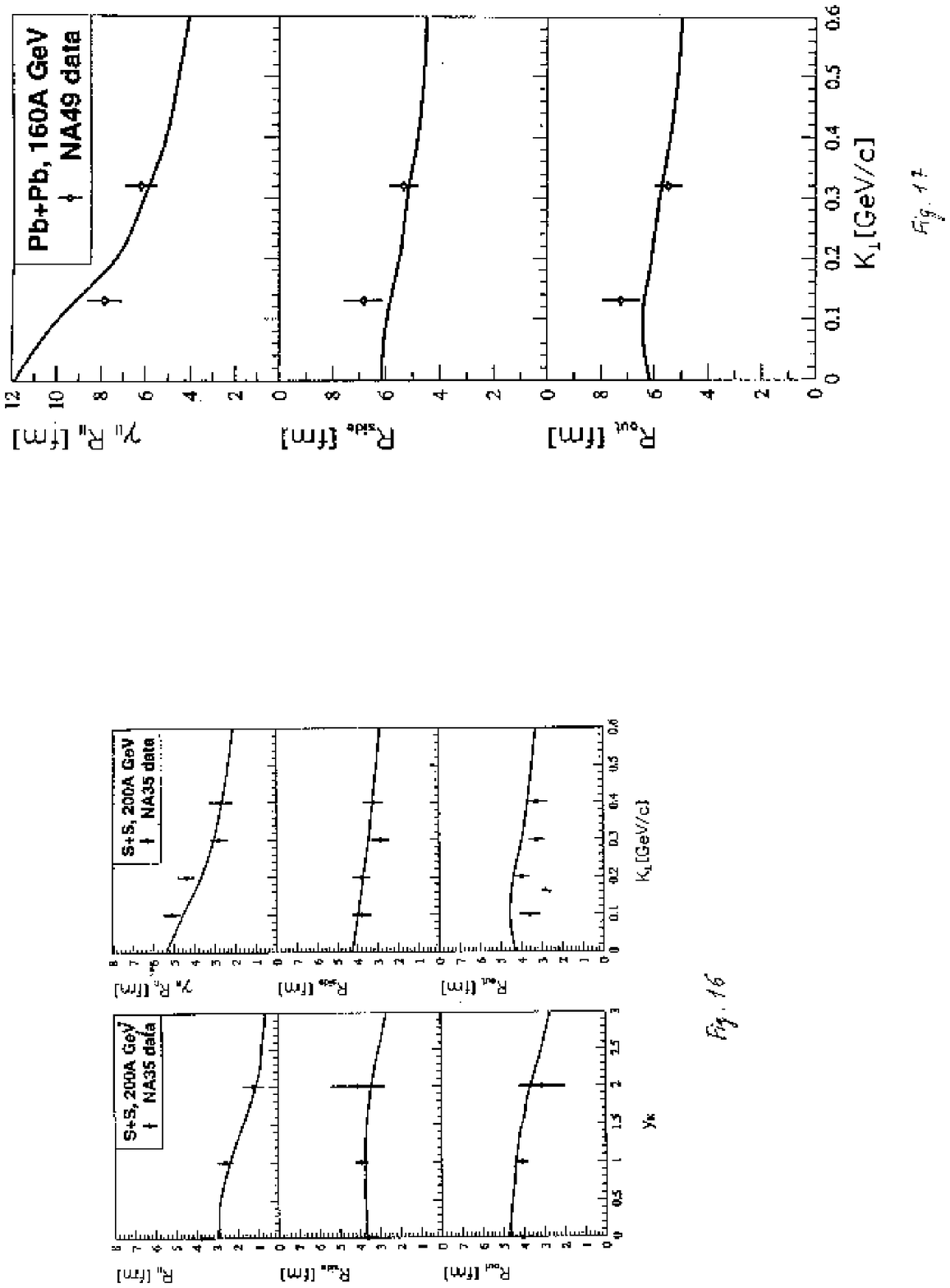}
\end{picture}
\end{figure}
\newpage
\begin{figure}[h]
\setlength{\unitlength}{1cm}
\begin{picture}(17.,25.)
\includegraphics{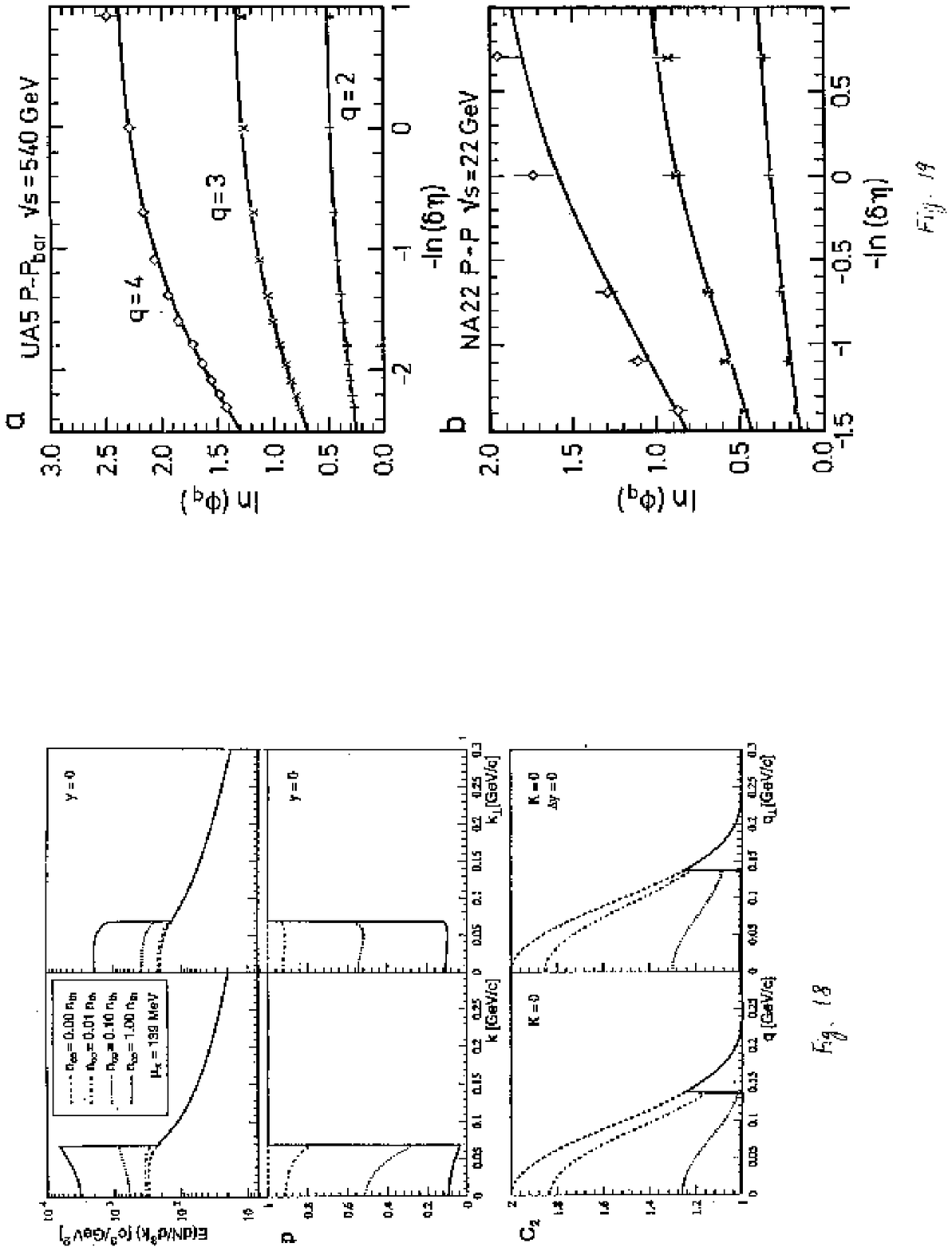}
\end{picture}
\end{figure}
\newpage
\begin{figure}[h]
\setlength{\unitlength}{1cm}
\begin{picture}(17.,25.)
\includegraphics{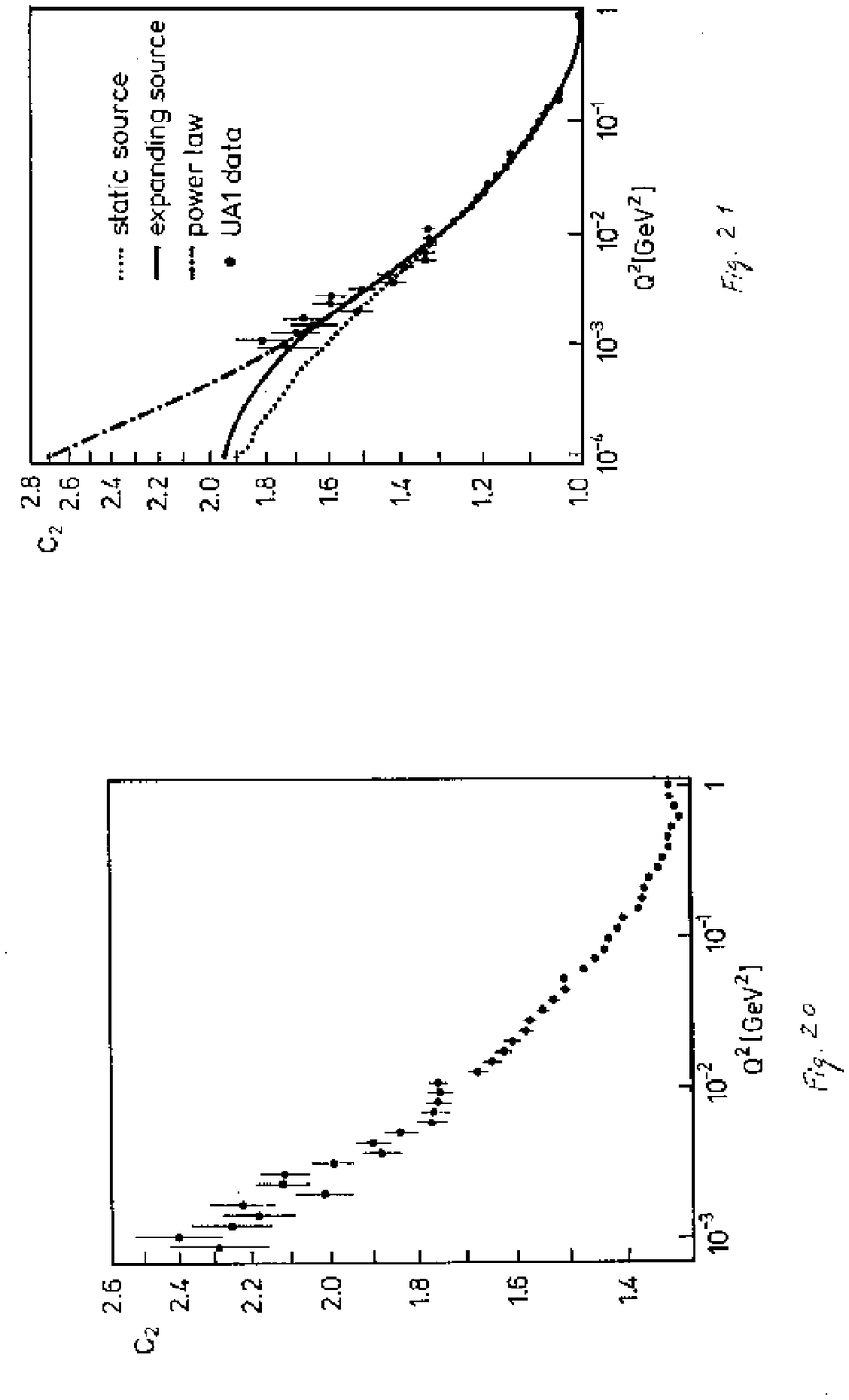}
\end{picture}
\end{figure}
\end{document}